%% file: main.tex
\documentclass[twocolumn]{aastex631}
\usepackage[utf8]{inputenc}
\usepackage{amsmath}	% Advanced maths commands
\usepackage{amssymb}
\usepackage{enumitem}
\begin{document}

\title{An Independent Search for Small Long-period Planets in \textit{Kepler} data \\I: Detection Pipeline}

%% The \author command is the same as before except it now takes an optional
%% argument which is the 16 digit ORCID. The syntax is:
%% \author[xxxx-xxxx-xxxx-xxxx]{Author Name}

\author[0000-0002-2805-9405]{Oryna Ivashtenko}
\affiliation{Department of Particle Physics and Astrophysics, Weizmann Institute of Science, Rehovot 7610001, Israel}
\footnote{oryna.ivashtenko@weizmann.ac.il}

\author[0000-0001-5162-9501]{Barak Zackay}
\affiliation{Department of Particle Physics and Astrophysics, Weizmann Institute of Science, Rehovot 7610001, Israel}
\footnote{barak.zackay@weizmann.ac.il}
%\collaboration{}{()}

\begin{abstract}
\textbf{The unprecedented photometric precision of \textit{Kepler} mission allows searching for Earth-like planets. However, it remains difficult to distinguish these low signal-to-noise planets from the false alarms originating from correlated and non-Gaussian noise. It reduces the resulting planetary catalog reliability and makes it hard to measure the occurrence rate of small long-period planets.
We aim to obtain a more reliable catalog of small long-period planet candidates from \textit{Kepler} data and use it to improve their occurrence rate estimate.
This work develops an independent search pipeline for small (\textit{Kepler}~Multiple-Event Statistic, MES$\lesssim$12) long-period (50-500~days) planets. It designs and implements a detection statistic that takes into account noise non-Gaussianity and physical prior. For every threshold-crossing event, it runs permutation and injection procedures to calculate the probability of it being caused by a real planet. 
The provided detection statistic has a tail-less background distribution with a rate of~$\sim$1~false alarm per search for MES$\sim$7.8. We demonstrate the increase in detection efficiency for MES of~7.5-9 and~$>$4~transits due to the background distribution control. The pipeline was tested to be able to detect most of the faint Confirmed \textit{Kepler} planets.
The pipeline was applied to the entirety of \textit{Kepler} data and detected $\sim50$ candidate events with a high probability of originating from real planets, which will be presented in our future work.}
\end{abstract}


\section*{Abbreviations and notations}
\begin{description}[noitemsep]
    \item[STD] Standard deviation.
    \item[PDF] Probability density function.
    \item[FAR] False alarm rate.
    \item[ISF] Inverse survival function.
    \item[SNR] Signal-to-noise ratio, defined for a detection statistic $\rho$ as 
    $\left(\text{E}(\rho|\mathcal{H}_1)
    -\text{E}(\rho|\mathcal{H}_0)\right)/
    \sqrt{\text{V}(\rho|\mathcal{H}_0)}$, 
    where $\text{E}$ is the expected value, $\text{V}$ is variance, $\mathcal{H}_0$ is a pure noise hypothesis, 
    $\mathcal{H}_1$ is noise plus planet hypothesis.
    
    % \begin{align*}
    %     \left(\text{E}(\rho|\mathcal{H}_1)
    % -\text{E}(\rho|\mathcal{H}_0)\right)/
    % \sqrt{\text{V}(\rho|\mathcal{H}_0)}.
    % \end{align*}
    \item[PSD] Power spectral density as a function of frequency. The diagonal of the covariance matrix in the Fourier domain.
    \item[Whitening filter] $\sqrt{1/\text{PSD}(f)}$ as function of frequency $f$. Filter applied to data to convert correlated noise to white noise.
    \item[Whitening] Applying whitening filter to light curve or template.
    \item[Template] The transit model used to detect signals. 
    \item[SES] Single-event statistic, or single-transit statistic. Result of matched-filtering of a single-transit template and the data.
    \item[MES] Multiple-event statistic, or multiple-transit statistic. Result of matched-filtering of a periodic template and the data, or the result of folding of SES.    
    \item[Period chunk] Part of the period grid that contains a fixed entropy. It means that the number of periods times the number of first transit time (epoch) options for all the chunks is the same. A typical range of one chunk is less than one day.   
    \item[Background] Distribution of statistic scores obtained from running the search on data containing no planets
    \item[Foreground] Distribution of statistic scores obtained from running the search on data containing planets expected according to currently known occurrence rates.
\end{description}

\section{Introduction}
\label{sec:intro}
\input{sec_introduction}

\section{Statistical formalism}
\label{sec:formalism}
\input{sec_formalism}

\section{Pipeline layout} 
\label{sec:layout}
\input{sec_pipeline_layout}

\section{Description of the pipeline methods}
\label{sec:description_of_methods}
\input{sec_pipeline_methods}

\section{Estimating the statistical significance of triggers}
\label{sec:significance}
\input{sec_significance}

\section{Example of a trigger}
\label{sec:trigger_example}
\input{sec_trigger_example}

% \filbreak
\section{Demonstration of pipeline reliability}
\label{sec:reliability_of_pipeline}
\input{sec_reliability_of_pipeline}

% \filbreak
\section{Demonstration of detection efficiency}
\label{sec:detection_efficiency}
\input{sec_detection_efficiency}

% \filbreak
\section{Test on \textit{Kepler} confirmed KOI}
\label{sec:confirmed_koi}
\input{sec_confirmed_koi}

\section{Conclusions}
\label{sec:conclusions}
In this work, we presented an independent search pipeline for the \textit{Kepler} data and demonstrated its performance. The goal of this project is to achieve better reliability of detecting low-SNR long-period planets. Our pipeline used a detection score named IMES which is a matched-filtering statistic with a measured noise PSD, corrected for noise non-Gaussianity and accounting for transiting planets occurrence prior.

In Section~\ref{sec:reliability_of_pipeline} (Figure~\ref{fig:search_triggers_and_scrambled}), we demonstrated that the IMES score ensures the pipeline's reliability by operating the pipeline on scrambled \textit{Kepler} light curves and observing \mbox{$<7\cdot10^{-6}$} events for IMES$>60$ (corresponding to SNR or \textit{Kepler} MES about 7.8).

In Section~\ref{sec:detection_efficiency_real_data}, we probed the detection efficiency of the pipeline via injection and recovery of planetary signals in \textit{Kepler} light curves. We have shown an increase in the detection fraction for signals with detection SNR in the range of 7.5-9 due to the clean background distribution of the IMES score. 

We developed a per-target statistical significance evaluation procedure that uses scrambled searches and injection-recovery campaign to report $P_{\textit{planet}}$, which is the probability for a given trigger to be caused by a real planet rather than by the background noise.

In Section~\ref{sec:confirmed_koi}, we verified that the pipeline is capable of detecting the majority of faint \textit{Kepler} Confirmed KOI with $P_{\textit{planet}}>99\%$.

\paragraph{The results of the pipeline operation}
This pipeline was operated on the entirety of \textit{Kepler} data, detecting $\sim50$ high-likelihood candidate events that do not correspond to Confirmed planets. While most of the detected events were known as \textit{Kepler} KOI, this pipeline allows for a more precise characterization of their origin. 

A detailed description of this search and its results, along with validation tests, will be provided in our forthcoming work~\citep[][]{our_paper_2}.

\paragraph{Future applications of the pipeline for population studies}
In the next stage, we will perform a global pixel-level injection campaign to assess the pipeline completeness across \textit{Kepler} data set.
The detected planetary candidates, together with their individually evaluated probabilities to be real planets and the completeness estimates, will be used to re-evaluate the occurrence rate. 

We will aim at improving the currently available precision of the population measurements \citep[e.g.][]{kunimoto_2020_occurrence_fgk, bryson_2021_occurrence_kepler, dattilo_2023_occurrence} by employing planetary candidates catalog with enhanced reliability.

\section{Discussion} 
\label{sec:discussion}
\subsection{Limitations and improvement directions}
\paragraph{Very long periods}
As shown in Section~\ref{sec:snr_recovery_fraction}, the pipeline detection efficiency declines for 3-transit events, making it impossible to detect planets of $\sim500$-day orbital periods. This is the result of the SNR loss caused by the non-Gaussianity correction penalizing deep transits. On the other hand, not applying the correction would imply an enhanced background rate, which would increase the detection threshold and make the detection of faint events impossible.

In order to detect 3-transit planets, the pipeline should be able to distinguish between noise background tail events and real planets based on 3 available faint SES. This is a challenging task: an alignment of 3 seemingly transit-shaped, equally separated low-SNR SES is likely to appear from the non-Gaussian noise background. 

It is possible that such events can be detected in a subset of targets exhibiting no noise non-Gaussianity, which would enable conducting a limited search without a non-Gaussianity correction.

An additional challenge is encountered when estimating the statistical significance of few-transit events, as the scrambling method reaches the entropy limit of a single light curve (Appendix~\ref{ap:rates_look_elsewhere}). An accurate estimation of the background for such events would require involving more data in the scrambled search.

\paragraph{One planet per target}
Currently, the pipeline operates in the regime of finding one best trigger per target. It is possible to conduct the search iteratively, potentially revealing more candidate events per target. For this, the previously found maximal candidate is masked, and the search is performed again.

Multi-planetary systems are known to be easier for candidate validation~\citep[][]{rowe_2014, valizadegan_2023}.
They provide a higher astrophysical prior of planet occurrence due to the orbital plane alignment. In our search, it would enable lowering the detection threshold and potentially detecting more small candidates. 

Furthermore, TTV measured for an existing planet in the system may serve as a validation for a newly claimed candidate~\citep[][]{jontof_hutter_2021}.

We note, however, that the presence of multiple low-SNR transiting planets would worsen the SNR loss due to the non-Gaussianity correction. Multiple transits would create an effect of a stronger SES distribution tail, thus enhancing the value of the correction.

\paragraph{Strict periodicity}
In addition, our pipeline searches only for strictly periodic events. If a planet has significant TTV, its periodic MES SNR gets "smeared" so that its maximal value may not be sufficient for detection~\citep[][]{leleu_2021_ttv}. In addition, if a transit is deep, the vetting module would vet off-center points of a transit, leading to masking out the potentially detectable smeared periodic signal.

Algorithms are being developed to search for quasi-periodicity~\citep[][]{carter_2013_qats, leleu_2021_ttv}.
In our pipeline, replacing the periodicity search module with a folding algorithm allowing for TTV may potentially enable the discovery of more low-SNR candidates.

It should be mentioned that the inability of a strictly periodic search to detect signals with significant TTV may lead to a biased estimate of the planetary occurrence rates, since we currently do not know how prevalent systems with such TTV are.

\paragraph{Other ways to face non-Gaussianity}
Before applying the non-Gaussianity correction, we performed the SES shape quality vetting, which masked out part of the non-transit-like SES (Figure~\ref{fig:ses_distribution}). If this pre-filtering process was more powerful, it would be possible to clean the SES background distribution and potentially avoid the need for the non-Gaussianity correction. It is possible to enable a soft-threshold vetting, which would carry the information about the SES shape quality to the periodicity search, helping to filter out non-planetary phenomena at the MES stage.

Alternatively, it would be useful to have a better understanding of the non-Gaussianity sources. For example, using ancillary information from the detector could enable the masking of some segments exhibiting instrumental effects. Having models for contaminant phenomena would allow using a more powerful binary hypothesis SES vetting instead of a $\chi^2$ one. Using the \textit{Gaia} mission data to construct a better point spread function in the target pixel files could allow for better-quality light curves.

\paragraph{PSD measurement loss}
The next significant source of SNR loss after the non-Gaussianity correction is the PSD measurement limitations. One \textit{Kepler} quarter does not contain enough data to obtain a high-resolution and high-precision PSD estimate. Using some ancillary information could allow for a better-quality PSD measurement, increasing the detected planetary SNR.

\subsection{Rolling band artifacts}
\label{sec:rolling_bands}
In some of the \textit{Kepler} channels, high-frequency temperature-sensitive amplifier oscillations were detected \citep[][Section 4.2.1]{kepler_data_processing_handbook}. 
When the oscillation frequency is a harmonic of the serial clocking frequency, it can create a shift of the mean bias level of the image that appears as a horizontal band on the CCD. The oscillation frequency is temperature-dependent, leading to a slow drift of the band with time across the detector, known as Rolling Band \citep[][Section 4.2.1]{kepler_data_processing_handbook}.

This effect appears in some of the \textit{Kepler} channels and is a major concern for the long-period planet search because it can mimic transit signals \citep[][Section 11.3.2]{kepler_data_processing_handbook}. The darkened band may appear transit-like, and due to the quarterly roll of the spacecraft, targets may undergo rolling band crossing repetitively when they fall on the corresponding CCD channel. This may appear as planetary signals with a period of $\sim$372~days or a multiple of it.

Even though the effect is not strictly periodic, the quasi-periodicity leads to an increased probability of false alarms in the corresponding period range. The time scrambling method (Section~\ref{sec:scrambling}) measures the background of completely aperiodic contaminants, which may align randomly and produce a planet-like signal. This method does not account for the case when the contaminant can be quasi-periodic, since this periodicity will be destroyed in the scrambling. Therefore, we separate the two issues, leaving the scrambling only as a test against a shift-invariant noise background.

Accounting for the quasi-periodic rolling band contaminants will be addressed in detail in our next work~\citep[][]{our_paper_2} that will deal with post-search vetting of the threshold-crossing events. The approach includes two factors: paying attention to specific periods and using \textit{Kepler} rolling band flags.

Periods having a right risk of rolling band contamination ($\sim$372~days and its multiples) can be completely excluded from the search. In this case, no planet at this orbital period can be detected. As a less conservative approach, the probability of contamination can be estimated as a function of period and used to calibrate the $P_{\text{planet}}$ score. 

The \textit{Kepler} pipeline used the inverted search in order to assess the potentially periodic noise structure~\citep[][]{thompson_2018}. However, the inverted noise has different properties since positive and negative effects in the \textit{Kepler} light curves are not symmetric. Therefore, this approach does not directly represent the real search background. 

However, the only parameter of interest for the rolling band artifacts is the relative probability of event occurrence at the suspect periods and other periods. Therefore, it is possible to use both inverted or non-inverted searches to estimate this relative probability. For this, one needs to assume that the non-rolling-band false alarm rate is a smooth function of orbital period. Then, the occurrence of threshold crossing events should change slowly between, for example, periods of 350 and 400 days. In reality, the threshold crossing event distribution will have a peak at  $\sim$372~days due to the rolling bands. By assessing the baseline rate from the smooth component and comparing it to the peak, one can determine how much more likely a contaminant will occur in the peak area. This coefficient can be used in the $P_{\text{planet}}$ evaluation~(Equation~\ref{eq:p_planet_score}) to amplify the background trigger probability for these periods.

The other approach includes using \textit{Kepler} ancillary information to flag likely non-planet triggers. First, the Dynablack module of the \textit{Kepler} pipeline~\citep[Section~4.1 of][]{kepler_data_processing_handbook} analyzed the full-frame images to identify rolling band artifacts and evaluate their severity level~\citep[Section 4.3.4 of][]{kepler_data_processing_handbook}. Similarly to what was done for the \textit{Kepler} candidates validation, it is possible to exclude the events if a significant fraction of their transits coincide with rolling band flags~\citep[][Section 11.5.1.6]{kepler_data_processing_handbook}.

Second, \textit{Kepler} Target Pixel Files~\citep[][]{kepler_data_characteristics_handbook} can be used for testing against contaminants that have distinct signatures on the pixel maps. Unlike real transits, rolling bands and some other contaminant signals will not be centered at the target star. This enables distinguishing them from the real transits even if their photometry looks the same.

\subsection{Occurrence rate estimation prospects}
The eventual major goal of this project is to re-estimate the occurrence rates for small long-period planets. In this section, we outline the relevance of this pipeline for the occurrence rate estimation and the future steps required for it.

\paragraph{The role of contaminants omitted in this work}
The $P_{\text{planet}}$ score is meant to be a target-level reliability metric. However, it evaluates the triggers only with respect to the noise background of aperiodic nature. This means that periodic transit-like signals of non-planetary origin can score high in $P_{\text{planet}}$. These include astrophysical signals such as various signatures of binary stars, or possibly instrumental signals, such as rolling band artifacts~(Section~\ref{sec:rolling_bands}). In order to evaluate the contamination probability of this nature, a different metric should be introduced. This metric is conceptually similar to the definition of $P_{\text{planet}}$~(Equation~\ref{eq:p_planet_score_definition}), but should contain the new contaminant probability in the denominator. As was mentioned in Section~\ref{sec:challenges}, this pipeline only focuses on distinguishing periodic transit-like signals from the non-Gaussian correlated noise, whereas other diagnostics will be considered separately. The final score used in the occurrence rate calculation should include the contaminants of all natures. This means that it has to include the issues addressed by the classic \textit{Kepler} approach to completeness and reliability presented in~\citep[][]{thompson_2018}.

\paragraph{Global detection metrics}
The $P_{\text{planet}}$ score is a target-specific reliability metric. For the occurrence rate calculations, it is important to take into account the global pipeline completeness over the target list and the statistics of the targets themselves. Every target has different noise properties and physical parameters, hence different planet detectability. For example, the fact that a specific number of candidates can be detected may relate not to the actual number of planets, but to the number of quiet enough targets allowing for such detections. In order to properly account for it, a global injection-recovery campaign should be conducted, scanning detection efficiency over the target list across the period range of interest. Additionally, if one wants to infer the rate for the real astrophysical population of stars, the \textit{Kepler} selection function needs to be taken into account.

\paragraph{The approach to occurrence rate inference}
Detection of planet candidates can be modeled as a Poisson process, where for a class of targets, there exists an underlying rate of observing planetary signals. From this approach, it is possible to infer the astrophysical occurrence rate based on the number of detections, their significance, and the search efficiency \citep[][]{roulet_2020_p_astro_squared}. We intend to follow this approach after all the preliminary steps are completed. A value similar to $P_{\text{planet}}$ is used in \citep[][]{roulet_2020_p_astro_squared} to integrate both confident and marginal events when estimating the astrophysical population. This is a crucial point in this project since in the regime of long-period and small planets, there is a lack of confident detections. Most of the previously known confirmed high-MES planets can be incorporated in this framework as detections with $P_{\text{planet}}=1$. The focus of this project is to account for the information contributed to the occurrence rate inference by the signals with $P_{\text{planet}}<1$.

This meets the previous works on estimating planetary occurrence rates in the effect of reliability on the estimation power. As pointed by~\citep[][]{bryson_2020_occurrence_probabilistic}, accounting properly for reliability near the detection limit results in a significant change in the resulting occurrence rates. Their work also recommends further improvements of the reliability estimation, particularly decreasing the reliability uncertainty and using non-uniform reliability metrics, which are a property of the detection rather than the catalog.

In addition, the formalism of~\citep[][]{roulet_2020_p_astro_squared} does not require the usage of score cuts discussed in ~\citep[][]{bryson_2020_occurrence_probabilistic}. In this formalism, the score plays the role of a weighting factor defining the contribution of every detection to the overall occurrence rate. Hence, having multiple candidates with low $P_{\text{planet}}$ can add a significant information about the occurrence rate, despite the low certainty about each particular candidate.

Thus, the main impact of this pipeline is to be able to use the information from the events that are currently identified as KOI but cannot be used in the occurrence rate estimate due to the inability to classify them as reliable detections.

\subsection{Follow-up possibilities}
\textit{Kepler} mission provides a unique sensitivity, making it hard to follow up the low-SNR \textit{Kepler} candidates with other instruments. 
However, recent developments show that it may be possible to observe some of the \textit{Kepler} candidates using ground-based photometry~\citep[][]{stefansson_2017_ground_follow_up} and radial velocity~\citep[][]{shahaf_2023_rv}. In addition, future space missions, such as \textit{PLATO}~\citep[][]{margin_2018_plato} or \textit{Earth~2.0}~\citep[][]{ge_2024_earth_2.0} may observe the \textit{Kepler} field and provide additional follow-up photometry to validate \textit{Kepler} candidates.

\section*{Acknowledgements}
%This research was partially supported by the Israeli Council for Higher Education (CHE) via the Weizmann Data Science Research Center. 
This research was supported by the Israeli Council for Higher Education (CHE) via the Weizmann Data Science Research Center, and by a research grant from the Estate of Harry Schutzman.
B. Z. is supported by the Israel Science Foundation, NSF-BSF and by a research grant from the Willner Family Leadership Institute for the Weizmann
Institute of Science.

This research has made use of the NASA Exoplanet Archive~\citep[][]{akeson_2013_nasa_archive}, which is operated by the California Institute of Technology, under contract with the National Aeronautics and Space Administration under the Exoplanet Exploration Program.

We acknowledge the use of the \texttt{Python} modules  \texttt{Numpy}~\citep[][]{numpy}, \texttt{Scipy}~\citep[][]{scipy}, \texttt{Matplotlib}~\citep[][]{matplotlib}, \texttt{Jupyter}~\citep[][]{jupyter}, \texttt{Astropy}~\citep[][]{astropy}.

We thank the anonymous reviewer for carefully reading this manuscript and providing insightful comments and remarks.

We thank Jon Jenkins and Steve Bryson for their interest in this work and valuable discussion. We thank Nathan Hara for fruitful conversations. We thank Tejaswi Venumadhav for his insightful comments. 
We thank Sahar Shahaf for his assistance and advice. 
O.I. expresses sincere gratitude to Matias Zaldarriaga for his invaluable help in this project and gratefully acknowledges the hospitality of the IAS during the completion of a part of
this work.

\appendix

\section{Shape of templates and whitened templates} 
\label{ap:templates_shape_losses}
\input{ap_templates_shape_losses}

\section{Math of Multiple-transit statistic}
\label{ap:mes_math}
\input{ap_mes_math}

\section{PSD estimation and its limitations} 
\label{ap:psd_estimation}
\input{ap_psd_estimation}

\section{Amplitude consistency veto} \label{ap:amplitude_consistency_veto}
\input{ap_veto_parts}

\section{Transit depths veto} 
\label{ap:atransit_depth_veto}
\input{ap_transit_depth}

\section{Template bank construction}
\label{ap:template_bank}
\input{ap_template_bank}

\section{Injection parameters}
\label{ap:injection_parameters}
\input{ap_injection_parameters}

\section{Non-Gaussianity correction}
\label{ap:non_gausianity}
\input{ap_non_gaussianity}

\section{SNR loss due to SES non-Gaussianity correction}
\label{ap:loss_score_correction}
\input{ap_loss_score_correction}

\section{Pipeline test on confirmed planets}
\label{ap:confirmed_mes}
\input{ap_confirmed_mes}

\section{$P_{\mathrm{planet}}$ score details}
\label{ap:p_planet}
\input{ap_p_planet}

\section{Event rates, Look-Elsewhere effect}
\label{ap:rates_look_elsewhere}
\input{ap_fp_rates_estimation}

\section{Basic equations reminder}
\label{ap:basic_math}
\input{ap_basic_math}

\section{Robust Gaussianization transformation}
\label{ap:gaussianization}
\input{ap_gaussianization}

\bibliography{kepler_methods_bib.bib}{}
\bibliographystyle{aasjournal}

\end{document}

%% file: sec_introduction.tex
One of the main goals of the \textit{Kepler Mission} \citep[][]{borucki_2010_kepler_mission} was to detect Earth-size
planets in the habitable zone of solar-like stars by means of transit photometry \citep[][]{koch_2010}. The mission achieved unprecedented success in discovering thousands of planets, but few of them were small long-period planets~\citep[][]{lissauer_2024}. The low number of detections makes it difficult to estimate their occurrence rate and study their population
\citep[see e.g ][]{zhu_dong_2021}. These planets are expected to be abundant in nature, but they are hard to detect due to observational biases and low signal-to-noise ratio. They have a low probability to transit due to large separations from the host star, a small number of transits per mission timeline due to long periods, and low signal-to-noise due to small radii. 

\paragraph{Why small planets should be detectable}
Nevertheless, these planets were expected to be detectable in the \textit{Kepler} data. 
An Earth-Sun system in the edge-on orientation would provide a signal of depth 84~ppm, performing 3~transits in the mission operation time with a transit duration of~13~hours. For such a system, observed with the cadence~$\sim$0.49~hours, the allowed photometric uncertainty to be detected with the alleged detection threshold of $7.1\sigma$~\citep[][]{jenkins_2002} is 
\begin{equation}
    \epsilon \sim \frac{84 \text{ ppm} \cdot \sqrt{13/0.49}}{\sqrt{7.1/3}}
    \approx  281 \text{ ppm}.
\end{equation}
The \textit{Kepler} photometric precision for quiet stars can reach 20~ppm in 6.5~hours, which should be enough to detect such an Earth-Sun analog transit \citep[][]{koch_2010, christiansen_2012_cdpp, kepler_data_characteristics_handbook}. For shorter periods, the planet will be able to transit more times, therefore the depth of individual transits can theoretically be even smaller. 

As was argued \citep[][]{burke_2019, mullally_2018_should_not_be_confirmed, thompson_2018}, the main problem is not the insufficient signal-to-noise ratio (SNR) for the true planets, but the excess of false positives.

\paragraph{Current detection challenge}
The lack of detected planets can be explained by the analysis of the \textit{Kepler} catalog  \citep[][]{thompson_2018} reporting low reliability~(37\%) and completeness~(73.5\%) for long-period small-size planets. The estimate in~\citep[][]{mullally_2018_should_not_be_confirmed} suggests that the reliability can be as small as 16\% and points to the difficulty of validating such candidates. 

According to the \textit{Kepler} team analysis~\citep[][]{burke_2019, burke_2015, thompson_2018}, the main complication preventing from reaching better reliability and completeness is presented by the systematic false alarms. The noise, which has a correlated power spectrum and contains non-Gaussian features, may mimic transit signals and produce false positives. For example, if a planet gets \textit{Kepler} Multiple-Event Statistic MES=8, it is hard to ensure that it is a real planet because there are many background events that also got MES=8~\citep[][]{burke_2019}.

These noise properties are known issues addressed in the original \textit{Kepler} pipeline~\citep[][]{kepler_data_processing_handbook_search_chapter, jenkins_2010_pipeline}, but not fully resolved yet. There is ongoing research \citep[see e.g.][]{robnik_seljak_2020_stellar_variability, robnik_Seljak_2021_matched_filtering, kunimoto_2020_search} trying to improve the current ways to treat the noise and conduct searches in it.

\paragraph{Scientific goals of this project}
Our goal is to design an independent search pipeline aimed at achieving better reliability for low-SNR signals. This would allow us to better evaluate the statistical significance of planetary candidates. Having a big enough catalog with well-defined statistical properties is helpful for estimating the occurrence rates and studying the population. Even if those planets are too small for the individual follow-up with currently available facilities, they provide information about the population. For example, if a certain region of parameter space contains 100~candidates, each one with~70\% probability of being real, one can assess that there are approximately 70 planets there, even though it is not known which ones are real. So we aim at conducting the search on the \textit{Kepler} data with a new pipeline, characterize the probability of every candidate to be real, and use them to estimate the occurrence rate.

We target our search at small planets with orbital periods of~50-500~days, where the smallness of a planet is defined by its low SNR. Since the transit depth is set by the planet-star radii ratio, low-SNR threshold will allow to be more sensitive to planets of smaller radii, planets orbiting larger stars, or noisier stars. For reference, we designed the pipeline focusing on \textit{Kepler} Multiple-Event Statistic MES$\lesssim$12.

\paragraph{Technical goal of this project}
Statistically speaking, this pipeline aims at achieving a clean background distribution of non-astrophysical false alarms in order to be able to reliably detect low-SNR signals.

We develop a robust and sensitive detection statistic, 
adopting some precision methods used in gravitational wave astronomy 
\citep[][]{venumadhav_2019_pipeline, zackay_2019_two_detectors, zackay_2019_non_gaussian}.
The detection algorithm is constructed independently and does not use existing transit detection modules. This provides an opportunity to compare the results with other pipelines and have independent conclusions about the origin of the candidates.

\paragraph{Structure of the project}
In this technical paper, we describe our pipeline and its methods and illustrate its performance. In our following papers \citep[][]{our_paper_2, our_paper_3}, we will present the catalog of candidates detected by the pipeline and a population estimation using this catalog. The catalog will include $\sim 50$ candidate events that are likely to correspond to real planets. Most of the pipeline events are known \textit{Kepler} KOI that could not be validated before and had a high risk of being systematic false alarms. 

\paragraph{Structure of this paper}
This paper is organized as follows: the rest of this section will outline particularities of planet detection. Section~\ref{sec:formalism} presents the mathematical foundation behind the pipeline. Section~\ref{sec:layout} outlines the layout of the detection pipeline,  and Section~\ref{sec:description_of_methods} introduces individual methods, whose details are described in more detail in the Appendices. 
Section~\ref{sec:significance} described how we estimate the statistical significance of candidates.

Then, we illustrate the pipeline performance.
Section~\ref{sec:trigger_example} provides an example of the pipeline output for one target. 
Section~\ref{sec:reliability_of_pipeline} provides the background events distribution compared to the true planetary events distribution to prove the pipeline's reliability.
Section~\ref{sec:detection_efficiency} investigates the pipeline detection efficiency and its limits.
In particular, Section~\ref{sec:detection_efficiency_real_data} demonstrates the increase in number of detectable low-SNR events due to better control of the background distribution. 

Finally, Section~\ref{sec:conclusions} concludes the work and outlines the global search in the \textit{Kepler} data that will be presented in~\citep[][]{our_paper_2}.
Section~\ref{sec:discussion} discusses the limitations of the pipeline and possible improvement directions.

\paragraph{Bottom line of this work}
The results of this work are summarized in Figures~\ref{fig:search_triggers_and_scrambled}~and~\ref{fig:detection_efficiency_scores}. Figure~\ref{fig:search_triggers_and_scrambled} presents the distribution of the real events from a search over \textit{Kepler} data, and the background events, displaying the pipeline reliability. Figure~\ref{fig:detection_efficiency_scores} shows that background control procedures lead to an increase in detection efficiency for low-SNR signals. For reference, this pipeline allows detecting signals starting from \textit{Kepler} MES$\sim$7.5.

\filbreak
\subsection{Challenges for detection pipelines}
\label{sec:challenges}
This section outlines the main complications in detecting planetary transits in the time series. The same mathematics of detection, described in Section~\ref{sec:detection_intro}, can be implemented differently, resulting in different tradeoffs between precision, computational speed, simplicity etc. Conceptually, the goal is to detect a periodic transit-like signal in a noise background, which is usually assumed to be Gaussian. The main practical considerations are:
\begin{itemize}[noitemsep, topsep=0pt, left=1pt, label=-]
    \item Noise can be correlated, and its covariance matrix is unknown;
    \item The depth and the parameters of the shape of the transit are unknown;
    \item The periodicity and the initial phase of the transits are unknown;
    \item The real data deviates from the model of pure Gaussian noise;
    \item Signals of non-planetary origin can trigger the detection statistic;
    \item It may be non-trivial to set the detection threshold due to the unknown background distribution;
    \item Computing the detection statistic may be computationally costly.
\end{itemize}
The differences between various pipelines come from different ways of addressing these challenges. 

The success of detection depends on how well a real signal can be distinguished from the background distribution of the detection statistic. When the noise and the signal models are exact, there exists an optimal detection statistic (Equation~\ref{eq:stat_snr}) providing the smallest possible false negative rate under a given false positive rate. When approximations are taken, either the SNR of the signal may be lost, or the background can be enhanced, bringing the background and the true signal distributions closer together and making the pipeline sub-optimal. Each of the challenges listed above requires taking some approximations, and below we outline the commonly used approaches.

\paragraph{Unknown PSD}
The noise covariance is due to stellar noise and instrumental noise correlations \citep[][]{kepler_data_characteristics_handbook, gilliland_2011_kepler_noise}. Often, it is neglected in searches for simplicity, and the noise is assumed to be white after subtracting some low-frequency trend. It can be, for example, a moving average, a smoothing by a Savitzky–Golay filter, or a polynomial fit \citep[e.g.][]{trans_fit, lightkurve}. Effectively, these methods are suppressing the power at low frequencies in an uncontrolled way. In many cases, all of them work well because there is a scale separation: the signal is concentrated at high frequencies, and the noise has excessive power at low frequencies. In the case of long-duration transits, there is no clear frequency scale separation between the noise and the transit Fourier spectra (see Appendix~\ref{ap:templates_shape_losses}). 

The mathematically correct approach is to weigh data and model in the Fourier domain by the inverse power spectrum (Equation~\ref{eq:stat_fourier}). Deviations from these weights either lead to signal SNR loss or inflate the background, when low-frequency noise features mimic transits. Since we aim to detect faint signals that are close to the detection threshold, it is important to use the best possible weights for the frequencies.

The true noise power spectrum is not known and cannot be adequately modeled from stellar properties or other external parameters; it is specific for every star and every \textit{Kepler} scientific quarter \citep[][]{kepler_data_characteristics_handbook}. The only way to treat it is to estimate from the data itself. 
The \textit{Kepler} pipeline \citep[][]{jenkins_2010_pipeline} implements a wavelet-based approach to account for variable stellar noise.

Some modern pipelines use Gaussian processes~\citep[e.g.][]{aigrain_2016_gaussian_processes}. Conceptually, fitting for the Gaussian kernel is similar to estimating the PSD. The periodogram method is generally faster and assumes no fixed functional form.

In our pipeline, we try to directly estimate the PSD implementing a procedure based on Welch's method \citep[][]{welch_1967}. 
We note that the periodogram approach was known since \citep[][]{jenkins_2002}, but for various reasons, it was not used. Our PSD estimation procedure is elaborated in Section~\ref{sec:psd_estimation}. In addition, we account for small PSD changes in time by tracking the time-dependent variance of the score.

\paragraph{Unknown transit shape}
The transit model is defined by the planet and star parameters which are also not known in advance. It is common to use box-fitting algorithms \citep[][]{kovacs_2002_bls}, which are also optimized to make the search faster. However, the mismatch between the box shape and the true transit shape leads to a loss in sensitivity. This effect becomes more significant in the case of correlated noise, as elaborated in Appendix~\ref{ap:templates_shape_losses}. There are searches fitting transit-like shapes to the data~\citep[e.g.][]{hippke_2019}, using, however, a different statistic. In \citep[][]{hippke_2019}, authors also provide a thorough description of other methods used in the community. 

In our pipeline, we use the optimal matched filtering statistic with a template bank constructed from transit-like templates. It is described in Section~\ref{sec:template_bank}. In addition, we incorporate transit parameters prior to our search to make the detection statistic more powerful. The mathematics of prior-informed score is presented in Section~\ref{sec:formalism_integral_score}.

\paragraph{Unknown periodicity} The periodicity of planetary transits is a useful feature helping to robustly detect even small planets. A planet may not be seen in a single transit, but many transits stacked together may provide a statistically significant detection. However, the period is unknown, so the search requires probing all possible options. 

One approach is to fold the light curve over some period, bin it, and then calculate the detection statistic~\citep[][]{kovacs_2002_bls}. Another approach, taken by the \textit{Kepler} team, is to first calculate the single-transit detection statistic~\citep[][]{kepler_data_processing_handbook_search_chapter}, and then perform the periodic folding. Mathematically, both approaches can be efficient, but only if the detection statistic is calculated correctly, taking into account the noise properties. For example, folding a correlated noise and then applying a boxcar filter is mathematically inaccurate and will result in a loss of detection power. 

In our pipeline, we choose first to calculate the single-transit detection statistic and then to fold it over all the periods of the search (mathematical details are provided in Appendix~\ref{ap:mes_math}). In this work, we only search for strictly periodic planets. That is, we do not search for transit timing variations (TTVs)~\citep[e.g.][]{holczer_2016_ttv_list, carter_2013_qats}. This limits us from investigating strongly interacting multi-planetary systems. For now, we do not address such systems, focusing on finding individual planetary candidates of long periods that are not experiencing significant TTVs.

Sometimes, the number of period options in the search is so high that the search becomes computationally unfeasible~\citep[][]{shahaf_22_fBLS}. This problem, appearing for short periodicities, can be solved using a dynamical programming approach~\citep[][]{shahaf_22_fBLS}.
In this work, we consider only long periods of 50-500~days, therefore this issue does not appear. However, our pipeline allows replacing the periodicity search module and enlarging the search to shorter periodicities. 

\paragraph{Non-Gaussianity}
The real \textit{Kepler} data does not fully obey the assumed Gaussian noise model~\citep[see e.g.][]{robnik_seljak_2020_stellar_variability, robnik_Seljak_2021_matched_filtering}. This results in the test statistic having a heavy-tail background distribution so that the true planets get indistinguishable from its tail. A signal quality veto \citep[e.g.][]{seader_2013_chi_2} can partially solve this problem, but it is not powerful enough, especially for low-SNR events. 

The non-Gaussianity can be addressed by applying a Gaussianization transformation to the noise~\citep[][]{robnik_seljak_2020_stellar_variability, robnik_Seljak_2021_matched_filtering} or by using a rank-based score~\citep[][]{venumadhav_2019_pipeline}.
Our pipeline uses parametric modeling of the measured single-transit statistic background distribution and applies the non-Gaussianity correction to the final detection score. See Section~\ref{sec:score_correction} for details.

\paragraph{Statistical significance}
The most common ways to report statistical significance of events is using the false alarm rate (FAR) or Bayes factor. Both approaches have their disadvantages, and improved metrics are being proposed \citep[e.g.][]{hara_2022, robnik_2022_mixed_priors}. Below, we outline the main considerations for our problem and introduce our metric of choice to report statistical significance.

\paragraph{Look-Elsewhere Effect}
Any value of the statistic can emerge from the noise if a sufficient number of checks is made. In large searches, the maximal statistic over a very large number of options is selected. In our search, we look for the best period, transit phase, and transit shape, and we investigate more than a hundred thousand stars. This means that we can obtain a seemingly high score by chance, even though planets are not there, a phenomenon called Look-Elsewhere Effect or Multiple Comparison Problem~\citep[][]{bayer_2020, bayer_2021}.

\paragraph{FAR, Bayes factor, and other metrics}
FAR tells how probable it is to get a score higher than a certain value from the background ($p$-value). It would include the real distribution tail shape and the look-elsewhere effect of the search. In a sense, FAR converts the detection statistic to units of probability.

However, FAR alone does not instruct us what its detection threshold should be. An event may be unlikely to originate from the noise distribution but be even less likely to come from a planet. To set a detection threshold, FAR should be compared to the probability of getting a given score from a true planet population.  

The Bayesian approach already contains the comparison to the true planet rate because it includes the physical prior. However, it may be sensitive to prior choices, and it may be hard to incorporate the look-elsewhere effect in it~\citep[][]{robnik_2022_mixed_priors, bayer_2020}.

Another caveat is that not all the parameters are treated in the same way in the transiting planets search (details in Appendix~\ref{ap_sec:p_planet_period_range}). For example, a planet can be reported without its eccentricity or argument of periastron. However, usually, one cannot claim a planet detection without reporting its period. Therefore, there are metrics reporting the probability of planet existence in a certain period interval~\citep[][]{hara_2022}.

In our pipeline, first, we incorporate the physical prior in the detection statistic at the stage of choosing the best event for each target (see Section~\ref{sec:formalism_integral_score} for details).

Then, to estimate the statistical significance of this best trigger, we use a metric informed by prior rate, pipeline efficiency, and the true background rate. It is referred to as $P_{\mathrm{planet}}$ and shows how probable it is for a given trigger to originate from a planet and not from the background noise.
Similar scores are used in gravitational wave searches~(e.g.~\citep[][]{kapadia_2019_p_astro}.
$P_{\mathrm{planet}}$ focuses on the event period range and uses the empirically measured expected rate of planet triggers in the pipeline and the true background distribution of each separate star. 
After the global search is done, $P_{\text{planet}}$ score can be calibrated based on the number of events to correct for the poorly known prior occurrence rate. 
The details of the statistical significance estimation procedure are provided in~Section~\ref{sec:significance}.

\paragraph{Contaminants}
The astrophysical and instrumental false positives may trigger the detection statistic and need to be filtered out. 

The original \textit{Kepler} analysis filtered the contaminants after the search using the automated \textit{Robovetter} \citep[][]{thompson_2018}. A similar vetting strategy was also used in other searches~\citep[e.g.][]{kunimoto_2020_search}. 

However, letting multiple contaminant events pass to the post-processing stage leads to elevated background levels. All high non-planetary scores contribute to the non-Gaussian background tail that we aim to get rid of.

Therefore, in our pipeline, the vetting is conducted at several stages: after the single-transit search, after the periodicity search, and at the post-search stage. We use multiple $\chi^2$-based tests to ensure transit shape and depth consistency. Sections~\ref{sec:veto_ses} and~\ref{sec:veto_mes} describe the vetting procedure.

At the post-processing stage, we will also use ancillary information such as \textit{Kepler} Target Pixels data. It will be referred to in our future work~\citep[][]{our_paper_2}.

\paragraph{Summary of the pipeline techniques}
We now iterate over features of our pipeline addressing the detection challenges listed above.
\begin{itemize}[noitemsep, topsep=0pt, left=1pt, label=-]
    \item The optimal matched-filter detection statistic for correlated noise is used;
    \item The noise power spectrum is measured from the real data for every star;
    \item The template bank is used to match closely the transit shape;
    \item Non-transit-like signals are filtered in multiple-stage vetting;
    \item The non-Gaussianity of the noise distribution is measured and corrected;
    \item A prior-informed detection statistic is calculated;
    \item The statistical significance of every trigger is computed based on empirically measured per-star per-period distributions of expected planetary and background events.
\end{itemize}
Mathematically, the methods are justified in Section~\ref{sec:formalism}, the implementation is described in Section~\ref{sec:description_of_methods}, and the performance is illustrated in Sections~\ref{sec:trigger_example},~\ref{sec:reliability_of_pipeline},~\ref{sec:detection_efficiency}.

%% file: sec_formalism.tex
In this section, we introduce the mathematical formalism that stands behind this pipeline. We describe the statistical model and the detection scores that we use in the pipeline. Some details are omitted here and can be found in the Appendices.

\subsection{Introduction to the detection problem}
\label{sec:detection_intro}
This section briefly introduces the basic statistical setup for detecting a planet in light curve data.
We represent the \textit{Kepler} light curve data as a vector $\mathbf{d}$ where each component $d(t)$ corresponds to the flux at time $t$. For now, we assume that it can be modeled as a correlated Gaussian noise $n\left(t\right)\sim\mathcal{N}\left(0,C\right)$ with covariance matrix $C$ \citep[see ][for the discussion of noise sources]{jenkins_2002}. If the star has a transiting planet, the data will also contain a transit model $\mathbf{h}$. We assume that the mean flux was subtracted from the light curve, so the noise has zero mean, and $\mathbf{h}$ is zero everywhere outside the transits.

The goal of detection is to decide whether a planet signature is present in the dataset. In the framework of binary hypothesis testing, the $\mathcal{H}_{\mathrm{n}}$ hypothesis is that the data is just noise, and the alternative $\mathcal{H}_{\mathrm{p}}$ hypothesis is that the data is noise plus planet:
\begin{equation}
\begin{aligned}
    d\left(t\right)\Big|\mathcal{H}_{\mathrm{n}}&=n\left(t\right),
    \\ d\left(t\right)\Big|\mathcal{H}_{\mathrm{p}}&=A\,h\left(t\right)+n\left(t\right).
\end{aligned}
\label{eq:data_model}
\end{equation}
$A$ is the amplitude  proportional to the planet-to-star radii ratio squared, and $\mathbf{h}$ is normalized to unit norm $\left\Vert \mathbf{h}\right\Vert=1$. Detecting a planet, or detecting a non-zero amplitude, means rejecting the $\mathcal{H}_{\mathrm{n}}$ hypothesis.

The classical optimal test for this problem \citep[][]{neyman_pearson_1933}  is the log-likelihood ratio test. A reminder of the basic math of it can be found in Appendix~\ref{ap:basic_math}. The resulting test statistic is given by the matched-filtering formula,
\begin{align}
    \rho_{\text{SNR}}=\frac{\mathbf{h}^{T}C^{-1}\mathbf{d}}{\sqrt{\mathbf{h}^{T}C^{-1}\mathbf{h}}},
    \label{eq:stat_snr}
\end{align}
where the superscript $T$ denotes transposition and vector-matrix multiplication is performed. This value is a scalar and can be thought of as an inner product between the data vector and the model vector with weights set by the inverse covariance matrix. This weighting accounts for summarizing optimally the information contained in the data: it down-weighs more noisy entries and de-correlates correlated entries.

The denominator in Equation~\ref{eq:stat_snr} serves for normalization so that in the absence of transit, the statistic will be distributed as $\rho_{\text{SNR}}|\mathcal{H}_{\mathrm{n}}\sim\mathcal{N}(0,1)$. If the obtained value of $\rho_{\text{SNR}}$ deviates significantly from this distribution and crosses a detection threshold $\eta$, we can reject $\mathcal{H}_{\mathrm{n}}$ and claim a detection.
In the presence of a transit, the expected value of $\rho_{\text{SNR}}$ gives the signal-to-noise ratio (SNR) of the transit (see Equation~\ref{eq:snr_definition} for definition). Such normalization is referred to as statistic in units of SNR.

We note that transit amplitude $A$ is not present in the detection statistic because it is a Uniformly most powerful test~\citep[][Theorem 8.3.17]{castella_statistical_inference}. The amplitude can have any value, it is a measured parameter whose estimator is given by Equation~\ref{eq:amplitude_estimator}.

\subsection{Planetary Transit Statistic}
\label{sec:mes_intro}
If the model $\mathbf{h}$ describes multiple equal transits, it can be decomposed into several equivalent single-transit templates $\mathbf{h}_i$. The statistic combining all transits is called Multiple-Event Statistic (MES), following the \textit{Kepler} team terminology. A statistic in which the model contains only one transit is termed the Single-Event Statistic (SES). For $i$-th transit, SES is defined as
\begin{align}
\rho_{\text{SES, }i}=\mathbf{h}_{i}^{T}
C_{i}^{-1}\mathbf{d}_{i},
    \label{eq:ses_i}
\end{align}
where $\mathbf{h}_i$ is the single-transit model, $C_i$ is the noise covariance matrix for that transit, and $\mathbf{d}i$ is the corresponding data segment.

SES can be calculated for all times of transit $\rho_{\text{SES}}(t)$ using convolution, by shifting the model $\mathbf{h}_{i}$ across the dataset and computing the statistic for each time shift.

Assuming that the noise covariance is effectively zero at large separation, the MES statistic can be rewritten using the SES statistic for the individual transits (derivation provided in Appendix~\ref{ap:mes_math}),

\begin{align}
\rho_{\text{MES}}=\frac{
\sum_i\rho_{\text{SES, }i}
}{\sqrt{
\sum_i\text{Var}\left[
\rho_{\text{SES, }i}
\right]
}},
    \label{eq:mes_formula}
\end{align}
where $\text{Var}\left[\rho_{\text{SES, }i}\right]$ is the SES variance for the $i$-th transit that can be calculated as 
\begin{align}
    \text{Var}\left[\rho_{\text{SES, }i}\right] = \mathbf{h}_{i}^{T}
C_{i}^{-1}\mathbf{h}_{i}.
\label{eq:ses_variance_calculated}
\end{align}
Due to this normalization, $\rho_{\text{MES}}$ has units of SNR and follows a standard normal distribution in the absence of planets.

From Equation~\ref{eq:mes_formula}, it follows that one can pre-compute the SES for all the needed transit times and shapes and then combine them to find MES. 

We note that folding the original correlated noise and then matched-filtering the result with a single-transit model is mathematically incorrect. Each transit should be weighted by its corresponding covariance matrix; otherwise, the score will lose sensitivity.

\paragraph{Periodic detection statistic}
In this work, we only search for strictly periodic signals. We can define the timing of all transits in Equation~\ref{eq:mes_formula} for orbital period $p$ and first transit time $t_0$ (which we also refer to as \textit{phase}, and which is commonly called \textit{epoch}). 
Then, we get the periodically folded statistic for every $p$, $t_0$, and transit shape $\mathbf{h}$,
\begin{align}
\begin{split}
    \rho_{\mathrm{UMES}}\left(p, t_0, \mathbf{h}\right) 
    &= \frac
    {\sum_n \rho_{\mathrm{SES}}
    \left(t_0+np, \mathbf{h}\right)}
    {\sqrt{\sum_n \text{Var}\left[\rho_{\mathrm{SES}}\right]\left(t_0+np, \mathbf{h}\right)}}, 
    \\ &n\in \{0,1,...,(T-t_0)/p+1\},
    \label{eq:umes_definition}
\end{split}
\end{align}
where $n$ indexes all transits within the data length $T$. The variances in the denominator are taken at the corresponding times because they can be time-dependent (see Section~\ref{sec:matched_filter}). 
The subscript UMES stands for \textit{Uncorrected Multiple-Event statistic}, as it does not include additional corrections explained later.

\subsection{Non-gaussian noise}
In the case when the noise in Equation~\ref{eq:data_model} has a non-Gaussian tail, the matched-filtering statistic from Equation~\ref{eq:stat_snr} is no longer optimal for detection. 

The mathematically correct strategy would be tailoring a new statistical model for the actual noise distribution and deriving the corresponding maximum likelihood statistic. 
However, in practice, using the Gaussian matched-filtering formula is technically simpler and computationally efficient. In Appendix~\ref{ap:non_gausianity}, we derive a way to represent the test statistic as the matched-filtering score plus correction for non-Gaussianity. We measure the distribution of the SES score, which should be Gaussian if the noise is Gaussian, but in reality, has a heavy tail (see Section~\ref{sec:score_correction}). For every SES, we calculate the following non-Gaussianity correction term:
\begin{align}
    \xi(\rho_{\mathrm{SES}}) = 2\log
    \frac{\mathcal{L}_{\mathrm{G}}
    \left(
    \rho_{\mathrm{SES}}
    |\mathcal{H}_{\mathrm{n}}\right)}{\mathcal{L}_{\mathrm{NG}}
    \left(
    \rho_{\mathrm{SES}}
    |\mathcal{H}_{\mathrm{n}}\right)},
    \label{eq:ses_score_correction}
\end{align}
where $\mathcal{L}_{\mathrm{G}}
\left(\rho_{\mathrm{SES}}\right)$ is the value of the Gaussian likelihood for the SES score $\rho_{\mathrm{SES}}$, and $\mathcal{L}_{\mathrm{NG}}
\left(\rho_{\mathrm{SES}}\right)$ is the same for the empirical non-Gaussian distribution. This correction evaluates the likelihood of a given score arising from a planet in a Gaussian noise rather than from a non-Gaussian noise. When $\rho_{\mathrm{SES}}$ is high and the distribution $\mathcal{L}_{\mathrm{NG}}$ has a heavy tail, it is more likely that this $\rho_{\mathrm{SES}}$ originates from this tail rather than a planetary transit. In such cases, the correction~(Equation \ref{eq:ses_score_correction}) has a large negative value. Conversely, when the SES score is low or when $\mathcal{L}_{\mathrm{NG}}$ is close to Gaussian, the correction is near zero.

\paragraph{Corrected MES score}
The correction (Equation~\ref{eq:ses_score_correction}) is incorporated into the UMES detection score (Equation~\ref{eq:umes_definition}), as derived in Appendix~\ref{ap:non_gausianity}.
It gives rise to the new detection score, \textit{Corrected MES} (CMES),
\begin{align}
\begin{split}
    \rho^2_{\mathrm{CMES}}
    \left(p, t_0, \mathbf{h}\right) 
    =& \rho^2_{\mathrm{UMES}}
    \left(p, t_0, \mathbf{h}\right) 
    +\sum_n \xi 
    \left(t_0+np, \mathbf{h}\right), 
    \\ &n\in \{0,1,...,(T-t_0)/p+1\},
    \label{eq:cmes_definition}
\end{split}
\end{align}
where the correction 
\begin{align}
    \xi (t_0+np, \mathbf{h})
    =\xi \left(\rho_{\text{SES}}
    (t_0+np, \mathbf{h})
    \right)
\end{align}
also undergoes periodic summation.

As will be shown in Section~\ref{sec:detection_efficiency_real_data}, the distribution CMES does not produce the non-Gaussian tail present in the original UMES distribution.

\subsection{Integral statistic score}
\label{sec:formalism_integral_score}
The goal of the search is to determine whether there is any planet transiting the target star. That is to say, planetary parameters are unknown, and we detect whether there is a planet with any parameters in a data set. Among all transit parameters, we emphasize the period $p$ and the first transit time $t_0$, which govern the timing of the transits and which are well-measurable. We denote the remaining parameters (e.g, planet inclination, eccentricity etc) as $\boldsymbol{\theta}$. We assume they only influence the transit shape. As discussed in Section~\ref{sec:template_bank}, the period's influence on the transit shape is degenerate with other parameters, allowing us to treat them separately.

The likelihood of the data containing a planet with any parameters is expressed as:
\begin{align}
    \mathcal{L}=\int dp\int dt_{0}\int d\boldsymbol{\theta}\,\pi\left(\boldsymbol{\theta},p,t_{0}\right)
    \mathcal{L}\left(
    \mathbf{d}|
    \mathbf{h}\left(\boldsymbol{\theta}\right),p,t_{0}\right),
\end{align}
where $\mathcal{L}\left(\mathbf{h}\left(\boldsymbol{\theta}\right),p,t_{0}\right)$ is the probability density function for period $p$, first transit time $t_0$, and transit shape characterized by a single-transit model $\mathbf{h}\left(\boldsymbol{\theta}\right)$. The factor
$\pi\left(\boldsymbol{\theta},p,t_{0}\right)$ is the prior probability of encountering a transiting planet with these parameters. This prior is taken for a given star and can generally depend on the star's properties, such as effective temperature, radius, and others. 

The parameters $\boldsymbol{\theta}$ exhibit a significant degeneracy, as detailed in Section~\ref{sec:template_bank}. That is, many parameter sets correspond to very similar transit shapes. It is enough to choose several dozen templates to approximate any expected transit shape. Such a set is called \textit{template bank} and will be discussed in Section~\ref{sec:template_bank}.

We can approximate the likelihood density for a specific shape $\mathbf{h}\left(\boldsymbol{\theta}\right)$ using the closest template from the bank,
\begin{align}
\mathcal{L}
\left(\mathbf{d}|
\mathbf{h}\left(\boldsymbol{\theta}\right),p,t_{0}\right)
\approx
\mathcal{L}\left(\mathbf{d}|\mathbf{h}_{k},p,t_{0}\right).
\end{align}
Factorizing the prior probability, we get
\begin{align}
    \mathcal{L}\approx\int dp\,\pi\left(p\right)\int dt_{0}\pi\left(t_{0}|p\right)\sum_{k}\pi
    \left(\boldsymbol{h}_{k}\right)
    \mathcal{L}
    \left(\mathbf{d}|\mathbf{h}_{k},p,t_{0}\right),
\end{align}
where the sum over $k$ covers all templates in the template bank. We defined the template prior probability as
\begin{align}
\pi\left(\boldsymbol{h}_{k}|p\right)=
\int_{\boldsymbol{\theta}\in\boldsymbol{\Theta}_{k}}\pi\left(\boldsymbol{\theta}|p\right)d\boldsymbol{\theta}.
\end{align}
It is obtained for every template $\mathbf{h}_k$ by integrating over all the parameters $\boldsymbol{\theta}\in\boldsymbol{\Theta}_{k}$ for which the template $\mathbf{h}_k$ is the closest template. The priors of all the templates sum up to unity. 

The first transit time prior is assumed to be uniform, $\pi\left(t_{0}|p\right)=1/p$. The prior probability to observe a specific period depends on its physical occurrence rate $\eta(p)$ and the probability to transit with this period $\pi_{\mathrm{tr}}(p)$, 
$\pi\left(p\right)\propto \eta\left(p\right)\pi_{\mathrm{tr}}\left(p\right)$. 

The likelihood density $\mathcal{L}\left(\mathbf{h}_{k},p,t_{1}\right)$, up to a normalization constant $\alpha$, can be expressed through the detection statistic (Equation~\ref{eq:cmes_definition}),
\begin{align}
    \mathcal{L}&\approx\alpha\int dp\,\pi_{\mathrm{tr}}\left(p\right)\eta\left(p\right)\int dt_{0}\frac{1}{p}
    \\
    &\times\sum_{k}e^{\rho_{\mathrm{CMES}}^{2}\left(\mathbf{h}_{k},p,t_{0}\right)+\log\pi\left(\mathbf{h}_{k}\right)}.
    \label{eq:likelihood_through_mes}
\end{align}

\paragraph{Marginalized statistic}
We define the template-marginalized detection statistic (MMES) as
\begin{align}
    \rho_{\mathrm{MMES}}^{2}\left(p,t_{0}\right)=\log\left(\sum_{k}e^{\rho_{\mathrm{CMES}}^{2}\left(\mathbf{h}_{k},p,t_{0}\right)+\log\pi\left(\mathbf{h}_{k}\right)}\right).
    \label{eq:mmes_definition}
\end{align}
It represents the likelihood of a planet with any transit shape at a given period and phase.

\paragraph{Integral score}
The resulting MMES score has two dimensions: period and first transit time. If the data contains a planet signature, this score will peak at ($p_{\mathrm{peak}}, t_{0,\mathrm{peak}}$). Since the score appears in the exponent, we assume that the integral in Equation~\ref{eq:likelihood_through_mes} is dominated by values around the peak. 
The actual search is performed on a grid of periods and phases, therefore we approximate the integral by a sum over grid cells around the peak with measure $\Delta p\,\Delta t_{0}$.

In order to fix the normalization constant $\alpha$ in~(\ref{eq:likelihood_through_mes}), we normalize the score with respect to a reference period $p_{\mathrm{ref}}$, which is taken to be the minimal search period. In this way, we keep the units of the original statistic score (approximately units of SNR$^2$) but apply a penalty for priors. A trigger that has a lower prior probability of being real will obtain a penalty and, therefore, will be less likely to contribute to the false positive rate.  At the reference period, the punishment vanishes, and the score is just defined by the MMES statistic. Higher periods typically obtain negative corrections due to their smaller probability to transit.

The resulting score, termed the \textit{Integral MES} (IMES) is given by
\begin{align}
\begin{split}
    \rho_{\mathrm{IMES}}^{2}&=
    \frac{\eta\left(p_{\mathrm{peak}}\right)
    \pi_{\mathrm{tr}}\left(p_{\mathrm{peak}}\right)}
    {\eta\left(p_{\mathrm{ref}}\right)
    \pi_{\mathrm{tr}}\left(p_{\mathrm{ref}}\right)}
    \frac{p_{\mathrm{ref}}}{p_{\mathrm{peak}}}
    \frac{\Delta p\,\Delta t_{0}}
    {\Delta p_{\mathrm{ref}}\,\Delta t_{0,\mathrm{ref}}}    
    \times\\&\times
    \sum_{\mathrm{peak}}
    \rho_{\mathrm{MMES}}^{2}
    \left(p_{\mathrm{peak}},t_{0,\mathrm{peak}}\right).
    \label{eq:imes_definition}
\end{split}
\end{align}
This is the final detection score of the pipeline, where 
\begin{itemize}[noitemsep, label=-, topsep=0pt, left=0pt]
    \item $p_{\mathrm{peak}}$ is the period of the peak MMES;
    \item $p_{\mathrm{ref}}$ is the reference period for score normalization;
    \item $\eta(p_{\mathrm{peak}})$ is the planet occurrence rate density for the peak period and the target star;
    \item $\eta(p_{\mathrm{ref}})$ is the planet occurrence rate density for the reference period and the target star;
    \item $\pi_{\mathrm{tr}}\left(p_{\mathrm{peak}}\right)$ is the probability to transit for a period $p_{\mathrm{peak}}$ and the target star;
    \item $\Delta p\,\Delta t_{0}$ is the measure which is the product of the grid step sizes or a grid cell area, around the peak;
    \item $\Delta p_{\mathrm{ref}}\,\Delta t_{0,\mathrm{ref}}$ is the grid cell area around the reference period. If the grid is not uniform, this area is period-dependent;
    \item $\sum_{\mathrm{peak}}$ indicates the summation over all the periods and phases in the grid cells adjacent to the peak.
\end{itemize}

The IMES score~(\ref{eq:imes_definition}) ensures that the selection of the best trigger is informed by priors, physical rates, and parameter dependencies of the look-elsewhere effect. It incorporates the non-Gaussianity correction inside $\rho_{\mathrm{MMES}}^{2}$, mitigating the impact of noise distribution's heavy tail.
Therefore, when we obtain a high maximal IMES, we know that it is unlikely to be caused by the non-Gaussianity or an enhanced look-elsewhere effect in a physically implausible area. These advantages distinguish the IMES score from the naive UMES score. Comparisons of the score performances are provided in Section~\ref{sec:detection_efficiency_real_data}. 

\paragraph{Summary of detection scores}

\begin{itemize}[noitemsep, label=-, topsep=0pt, left=0pt]
    \item $\rho_{\mathrm{SES}}
    \left(t, \mathbf{h}_k\right)$ (Equation~\ref{eq:ses_i}): Single-transit statistic for a transit at time $t$ and a single-transit template $\mathbf{h}_k$.
    \item $\rho_{\mathrm{UMES}}
    \left(p, t_0, \mathbf{h}_k\right)$ (Equation~\ref{eq:umes_definition}): Uncorrected MES. Naive matched-filtering detection statistic of the full multi-transit ephemeris, for period $p$, first transit time $t_0$, and template $\mathbf{h}_k$. Has a non-Gaussian tail. Conceptually corresponds to \textit{Kepler} MES.
    \item $\rho_{\mathrm{CMES}}\left(p, t_0, \mathbf{h}_k\right)$ (Equation~\ref{eq:cmes_definition}): Corrected MES. It is UMES with the non-Gaussianity correction~(Equation~\ref{eq:ses_score_correction}). Does not exhibit a non-Gaussian tail in its distribution.
    \item $\rho_{\mathrm{MMES}}\left(p, t_0\right)$ (Equation~\ref{eq:mmes_definition}): Marginalized MES. It is CMES marginalized over the templates in the template bank.
    \item $\rho_{\mathrm{IMES}}\left(p_{\text{peak}}, t_{0,\text{peak}}\right)$ (Equation~\ref{eq:imes_definition}): Integral MES. It is MMES integrated over periods and phases around the peak value. It is the final detection score of the pipeline, informed by priors and physical rates.
\end{itemize}
In Section~\ref{sec:description_of_methods}, the practical implementation of these detection statistics in the pipeline will be detailed.

\subsection{$P_{\mathrm{planet}}$  score}
\label{sec:p_planet_formula_formalism}
In the detection problem formulation~(Equation~\ref{eq:data_model}), we considered the distribution of the data under the planet hypothesis $\mathcal{H}_{\mathrm{p}}$ or the noise hypothesis $\mathcal{H}_{\mathrm{n}}$. We addressed the question: "What is the probability of observing this data given this hypothesis?". Then, we derived a detection statistic that summarizes the data. Our question became: "What is the probability to get this detection score given that there is a planet/there is no planet?". Eventually, however, we want to answer the question "What is the probability that there is a planet, given the obtained score?" To answer this question, we use Bayes' formula and define a score that we call $P_{\textbf{planet}}$:
\begin{align}
    P_{\text{planet}}=\frac{\pi_{\mathrm{p}}
    \text{Pr}\left(\rho^{2}|\mathcal{H}_{\mathrm{p}}\right)}
    {\pi_{\mathrm{p}}
    \text{Pr}\left(\rho^{2}|\mathcal{H}_{\mathrm{p}}\right)
    +\left(1-\pi_{\mathrm{p}}\right)
    \text{Pr}\left(\rho^{2}|H_{\mathrm{n}}\right)}.
    \label{eq:p_planet_score}
\end{align}
Here,
\begin{itemize}[noitemsep, label=-, topsep=0pt, left=0pt]
    \item $\rho^{2}$ is the detection statistic (we use IMES, but any score can be used).
    \item $\pi_{\mathrm{p}}$ is the prior probability to have a transiting planet in this dataset.
    \item $\text{Pr}\left(\rho^{2}|\mathcal{H}_{\mathrm{p}}\right)$ is the probability density of the detection score $\rho^{2}$ assuming there is a planet in the data.
    \item $\text{Pr}\left(\rho^{2}|\mathcal{H}_{\mathrm{n}}\right)$ is the probability density of the detection score $\rho^{2}$ assuming there is no planet in the data.
\end{itemize}
Appendix~\ref{ap:p_planet} provides more details about the components of the $P_{\text{planet}}$ score and the way to compute it.

\paragraph{Other alternative hypotheses}
In this formulation, we considered a binary test: the planetary hypothesis, and the background noise hypothesis. There can be other explanations for the trigger origin, such as an eclipsing binary, a field contaminant, or an instrumental effect. The vetting of such contaminants will be addressed in our future study~\citep[][]{our_paper_2}. Here, we concentrate solely on distinguishing planets from the noise. Specifically, we address noise of aperiodic nature; for discussion of quasi-periodic contaminants, see Section~\ref{sec:rolling_bands}.

\subsection{Note about working in the Fourier domain}
\label{sec:fourier}
Throughout this work, we use the representation of the problem in the Fourier domain. Many times, using this basis for vectors (such as light curves and templates) is often more explicative and computationally efficient. 

The detection statistic (Equation~\ref{eq:stat_snr}) can be written using the Fourier image of all the values, denoted with a hat, e.g. $\hat{\mathbf{d}}$. The components $\hat{d}(f)$ of the vector $\hat{\mathbf{d}}$ are now indexed by frequency $f$. While the statistic $\rho$ is a scalar and remains the same whether expressed in the Fourier or time domain, using the Fourier basis offers several technical advantages, as outlined below.

If the noise has the shift-invariance property,
\begin{align}
    \mathrm{Cov}\left[d(t), d(t+\tau)\right] = \mathrm{Cov}\left[d(0), d(\tau)\right],
\end{align}
then the Fourier image of the covariance matrix $C$ will be a diagonal matrix \citep[][]{gray_matrices}, with its diagonal elements forming a vector known as the power spectral density, PSD, denoted $\mathbf{S}$.

With this, the statistic (Equation~\ref{eq:stat_snr}) written in the Fourier domain will not require matrix multiplication,
\begin{align}
    \rho=\frac{\sum_f\frac{\hat{h}^{\dagger}(f)\hat{d}(f)}{S(f)}}
    {\sqrt{\sum_f\frac{\hat{h}^{\dagger}(f)\hat{h}(f)}{S(f)}}},
    \label{eq:stat_fourier_0}
\end{align}
where dagger denotes conjugate transpose. 

This form of the statistic is also intuitive: each data point in the Fourier domain $\hat{d}(f)$ is weighted inversely to its variance $S(f)$. In other words, points with higher noise (larger variance) contribute less to the statistic, while less noisy points have a greater influence. This efficiently summarizes the information contained in the data and explains why this is the optimal detection statistic.

Since the time of the transit is not known, we need to compute the statistic for all possible times, which is equivalent to cross-correlating the data with the model. By applying the Convolution Theorem, this process can be translated into multiplication in the Fourier domain. Thus, it is enough to omit the summation in Equation~\ref{eq:stat_fourier_0} and perform an inverse Fourier transform to obtain a statistic for all the transit times.  

For ease of notation, we will introduce whitened vectors,
\begin{equation}
    \hat{d}_{w}(f)=\hat{d}(f)/\sqrt{S(f)},
    \label{eq:whitening}
\end{equation}
where $1/\sqrt{S(f)}$ will be referred to as whitening filter.
With this, the Fourier-domain statistic for all the possible transit times is
\begin{align}
    \hat{\rho}(f)=
    \frac{\hat{h}_{w}^{\dagger}(f)\hat{d}_{w}(f)}
    {\sqrt{\sum_f \hat{h}_{w}(f)^{\dagger}\hat{h}_{w}(f)}}.
    \label{eq:stat_fourier}
\end{align}

The whitened noise can be understood as noise de-correlated with its inverse covariance kernel. This notation reduces the problem to detection in uncorrelated noise. It is important to note that then the whitening filter $1/\sqrt{S(f)}$ should be applied to both the data and the model, as any filter applied to the data will also distort the expected transit shape. Appendix~\ref{ap:templates_shape_losses} provides an example illustrating the effect of whitening on the transit model.

Working in the Fourier domain significantly reduces computational complexity. Without this transformation, computing the statistic via matrix multiplication and convolution has a time complexity of $N^3$ for data of length $N$. For the Fourier domain formula (Equation~\ref{eq:stat_fourier}), this complexity reduces to $N$, and performing the Fourier transform back to the time domain costs $N\log N$.

%% file: sec_pipeline_layout.tex
\begin{figure*}[h!t]
    \centering
    \includegraphics[width=\textwidth]{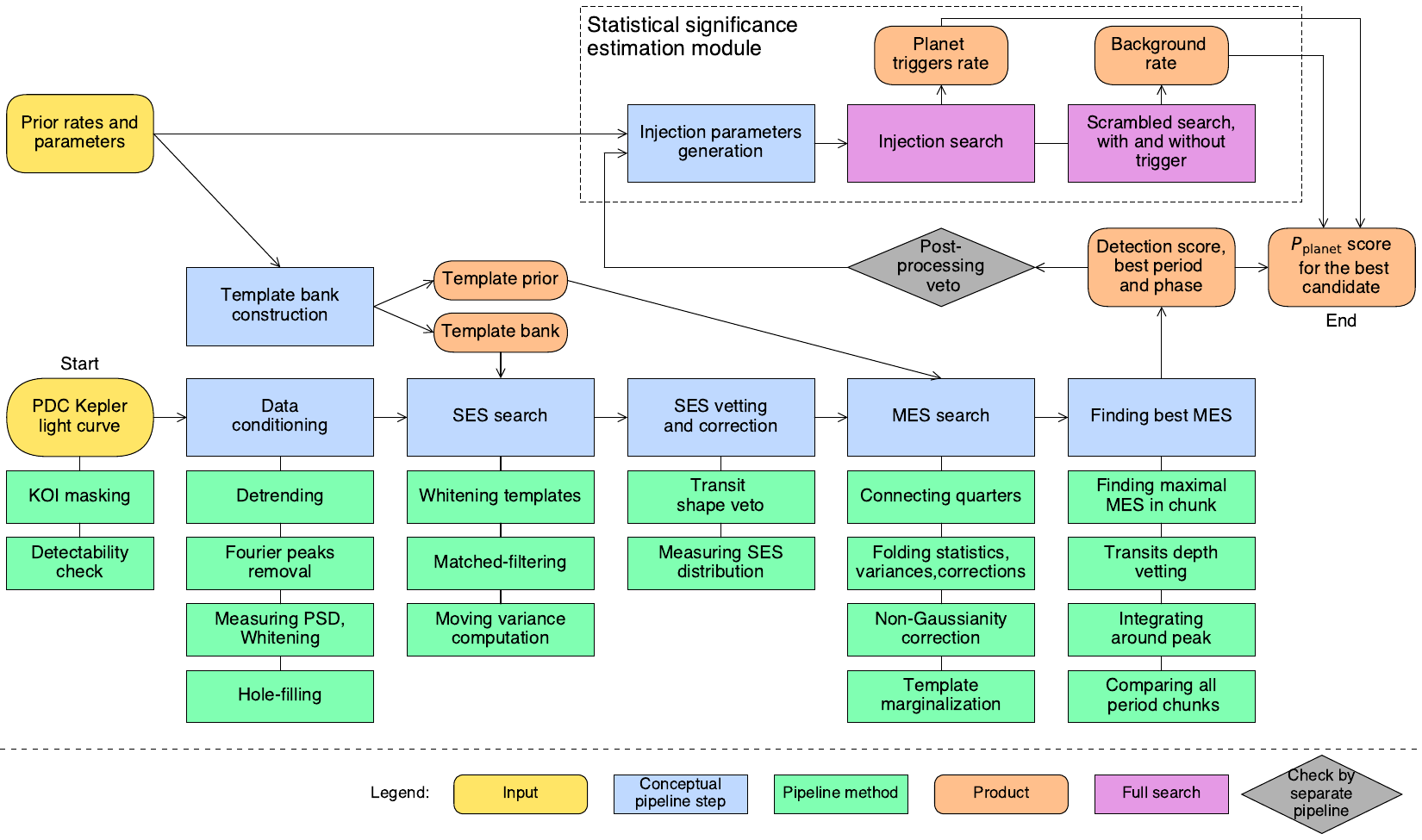}
    \caption{Schematic representation of the pipeline (refer to the legend for color coding). The main flow shows how the light curve is processed to find the optimal IMES detection score. The module framed in the dashed line contains additional searches that are needed to evaluate the statistical significance of a threshold-crossing event.}
    \label{fig:pipeline_scheme}
\end{figure*}

This section presents the pipeline’s workflow (Figure~\ref{fig:pipeline_scheme}) and outlines its key steps. Detailed descriptions of each method are provided in Section~\ref{sec:description_of_methods}.

\begin{figure}[t!]
    \centering
    \includegraphics[width=0.47\textwidth]{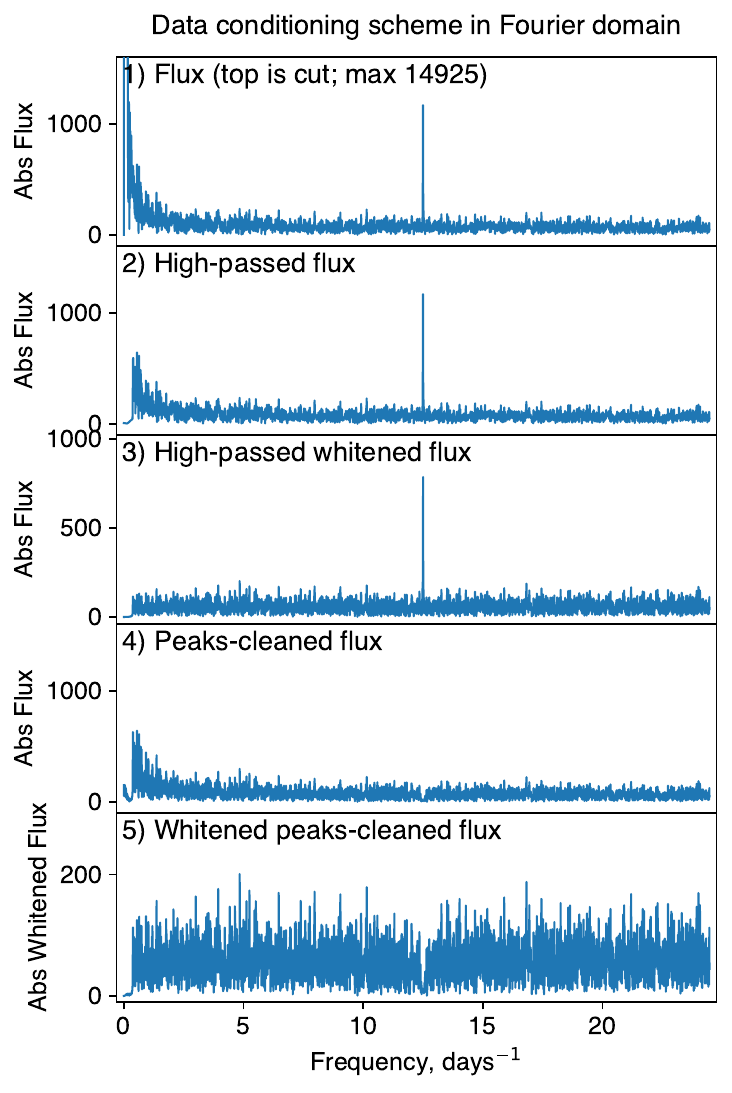}
    \caption{Schematic of data conditioning on a simulated light curve represented in Fourier domain. 1) The original correlated (red-noise) flux, including an additional noise peak. 2) The flux after detrending the low frequencies that cannot be resolved by the measured PSD. 3) The flux whitened after the PSD was measured and the whitening filter was constructed. 4) The flux after the peak detection and removal. 5) The final whitened flux, which will be used for statistic score calculation.}
    \label{fig:conditioning_fourier}
\end{figure}

\begin{figure}[t!]
    \centering
    \includegraphics[width=0.47\textwidth]{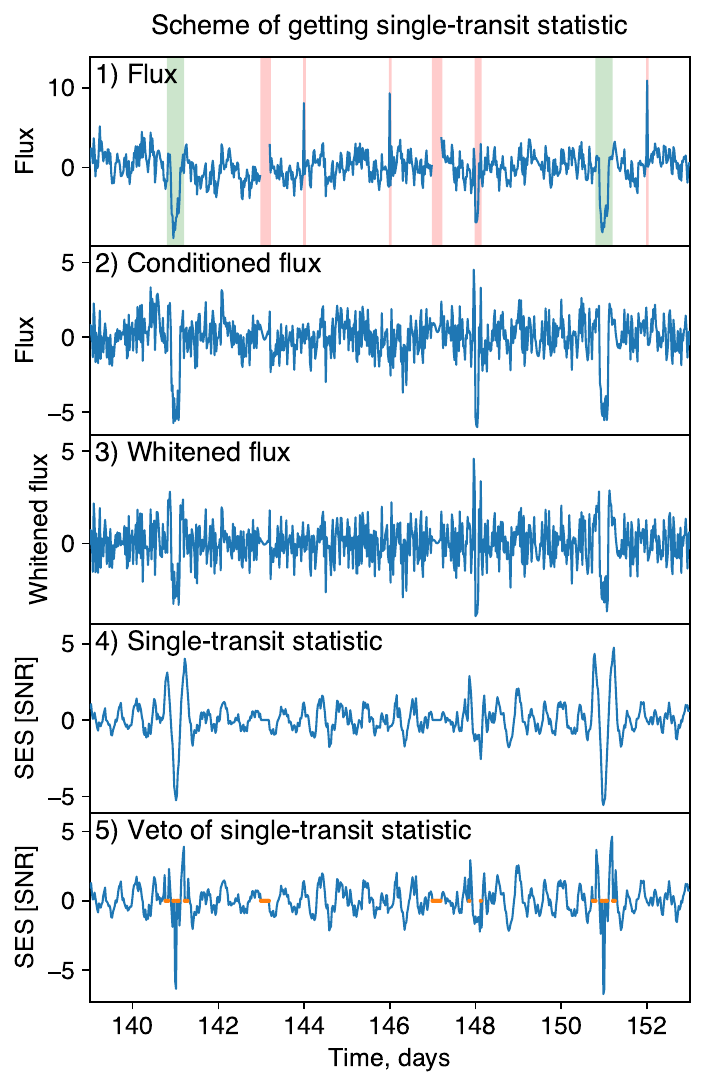}
    \caption{Schematic of processing stages leading to the single-transit statistic shown on simulated data. The final result is then passed to the periodicity search module.
    1) The initial flux is taken, and bad segments are identified (marked in red). The green shading indicates the simulated planet. 2) The flux is conditioned: outliers are cleaned, low-frequency trends and peaks unresolved by PSD measurement are removed, and gaps are filled. 3) The flux is whitened using the measured PSD. 4) The single-transit statistic is obtained from matched filtering with one of the whitened templates. We note that templates are defined to have zero baseline and positive deviation at transits, therefore the statistic is negative at transits. 5) Large non-transit-like scores are identified and masked using the veto. They are included in the mask, and their score is zeroed. The masked points are represented with orange dots and will not have a contribution during the periodicity search. We can see that true transits passed the veto, but the points around them, where the shape is shifted and does not match the template, were masked.}
    \label{fig:scheme_time_domain}
\end{figure}

\begin{enumerate}[left=0pt]
\item \textit{Selecting data}. We check that the quality of a PDC \textit{Kepler} light curve allows for detecting planets of interest. Known planets are masked  (Section~\ref{sec:input_data}).

\item \textit{Constructing the template bank}. We generate a set of single-transit models with different parameters covering the planet parameter space of interest. We calculate the prior occurrence probability for each template. (Section~\ref{sec:template_bank}).

\item \textit{Conditioning data}. We prepare the light curves for statistical score calculation. See Section~\ref{sec:conditioning_and_whitening} and Figure~\ref{fig:conditioning_fourier} for the illustration.
\begin{enumerate}
    \item Masking outliers and bad segments in the data.
    \item Estimating noise PSD using Welch's method (Section~\ref{sec:psd_estimation}).
    \item Piecewise detrending (high-pass filtering) of low frequencies whose PSD cannot be measured (Section~\ref{sec:detrending}).
    \item Masking peaks in the spectrum that are not resolved with the PSD (Section~\ref{sec:notch}).
    \item Whitening the data (Section~\ref{sec:applying_whitening_filter}).
    \item Filling holes (gaps) in the data in such a way that they do not have an impact on the statistic score (Section~\ref{sec:holes}).
    \end{enumerate}
\item \textit{Conditioning the templates}. The templates are whitened using the same filters as the data.

\item \textit{Calculating and vetting the single-event statistic (SES)}. See Section~\ref{sec:ses} and Figure~\ref{fig:scheme_time_domain} for the illustration.
    \begin{enumerate}
    \item \textit{Matched filtering}. We calculate the matched-filtering SES score for all the transit times using the whitened data and the whitened template bank. We measure empirically the moving variance of the score to ensure that its distribution is normalized (Section~\ref{sec:matched_filter}).
    \item \textit{Quality veto of SES}. We apply $\chi^2$-based tests and mask out high scores that do not exhibit a transit-like shape (Section~\ref{sec:veto_ses}).
    \item \textit{Measuring SES distribution}. We measure the SES distribution and calculate the non-Gaussianity correction which will be applied to the final IMES score (Section~\ref{sec:score_correction}).
    \end{enumerate}
\item \textit{Calculating the multiple-event statistic (MES) (Section~\ref{sec:mes_periodicity_search})} 
    \begin{enumerate}
    \item \textit{Preparation}. We connect quarters and define a period grid split into small chunks.
    \item \textit{Period folding}. We co-add SES scores into full planet score for all possible periods and phases for every period grid chunk. We also fold moving variances and non-Gaussianity corrections.
    \item \textit{Correcting non-Gaussianity}.
    We apply the folded non-Gaussianity corrections to the UMES scores obtained from folded SES and their folded variances. As a result, we get CMES.
    \item \textit{Marginalization over templates}. We use the template prior and calculate the MMES score, which is related to the probability of there being any planet with a given period and phase.
    \end{enumerate}
\item \textit{Finding the best event (Section~\ref{sec:best_trigger})}
    \begin{enumerate}
        \item \textit{Finding chunk peak}. We find the best MMES period and phase in every periodicity grid chunk.
        \item \textit{Transit depth veto}. A transit depth consistency check is conducted to ensure that the summed single transits are consistent with being caused by the same planet. If the peak is rejected, a new peak is found, and the procedure is repeated (Section~\ref{sec:veto_mes}).
        \item \textit{Integrating around the peak}. We calculate the IMES score for the peak of every chunk using the prior occurrence rate and the probability to transit for this period. We normalize all the chunks with respect to the reference period (Section~\ref{sec:methods_integral_score}).
        \item {Selecting the best trigger}. We select the best IMES with its period and phase across the chunks.
    \end{enumerate}

\item \textit{Post-processing veto}. We employ folded transit shape veto, contaminant periodicity tests, target pixel tests, centroid tests, and others to reject triggers that are likely to be caused by non-planetary factors. Data external to \textit{Kepler} photometry, such as \textit{Kepler} target pixel files or the properties of stars in the field, are used for the tests. This stage is not included in the search pipeline and is not discussed in this work. It will be addressed in~\citep[][]{our_paper_2}.
\item \textit{Estimating the significance of triggers}. (Section~\ref{sec:significance}). In order to calculate the $P_{\text{planet}}$ score, we evaluate and compare the following probabilities:
\begin{itemize}[noitemsep, label=-, topsep=0pt]
    \item The probability of getting this event from the noise background of this star. (Calculated using scrambled run, Section~\ref{sec:scrambling})
    \item The probability of the pipeline detecting this event from the expected astrophysical population of transiting planets (Calculated using flux-level injection-recovery run, Section~\ref{sec:injections}).
\end{itemize}
\end{enumerate}

%% file: sec_pipeline_methods.tex
\subsection{Input data}
\label{sec:input_data}
As the input to the search pipeline, we use the \textit{Kepler} DR25 \citep[][]{kepler_dr_notes_25} long-cadence pre-search conditioned simple aperture photometry flux (PDCSAP) \citep[][]{pdc_kdph}. The light curves were provided by MAST portal:
\url{doi:10.17909/T9488N}. We pre-filter the targets, selecting only ones that allow for detecting our planets of interest; the selection is described in~\citep[][]{our_paper_2}. 

\textit{Kepler} data are split into 90-day long segments called quarters~\citep[][]{kepler_data_characteristics_handbook}.
In the light curves, we omit the zeroth and the last \textit{Kepler} quarters due to their short length and lower photometric precision.
For each quarter, we subtract the mean flux calculated after outlier clipping. 
Since transit depth is proportional to the flux level (as it masks a relative portion of stellar light), we renormalize the flux of each quarter by its mean, ensuring that the expected transit amplitude is the same in each quarter.

We limit the analysis to targets with \textit{Gaia} data available in the \textit{Gaia}-\textit{Kepler} cross-match table~\citep[][]{bedell_gaia_kepler_fun, gaia_2022_dr3}, from which stellar radii, masses, temperatures, and other relevant parameters are extracted.

\paragraph{Masking known KOI}
If the target has a Confirmed KOI or a Candidate KOI with a \textit{Kepler} MES$>20$ in \textit{NASA Exoplanet Archive}~\citep[][]{akeson_2013_nasa_archive}, we mask them before the search using their transit times and durations reported in \citep[][]{holczer_2016_ttv_list, akeson_2013_nasa_archive}. Masking means that the points corresponding to the transits are excluded from processing.

% \filbreak
\subsection{Template bank}
\label{sec:template_bank}
To calculate the SES, we use a set of single transit models called \textit{templates}. We define the models to have zero baseline and positive deviation at transits. It means that the resulting statistic will be negative at transits. 

\begin{figure}
    \centering
    \includegraphics[width=0.48\textwidth]{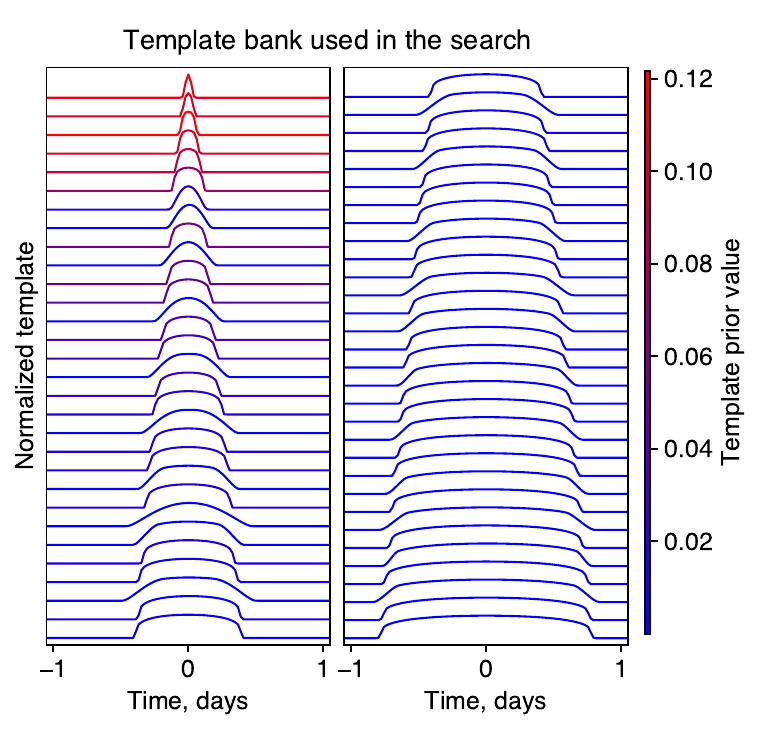}
    \caption{The template bank of 58 templates used by the pipeline. The colors represent the template prior probability. The templates are normalized by their maximum and centered for demonstration purposes. During the SES calculation, these templates will be whitened and cross-correlated with the whitened data.}
    \label{fig:template_bank}
\end{figure}

\paragraph{Coverage of the bank}
When the template does not exactly match the shape of the signal, the signal will be detected with less SNR (see Appendix~\ref{ap:templates_shape_losses} for details). If this SNR is too low, the signal will be missed. We aim to cover the parameter space of interest, meaning that every transit should have a sufficiently similar template in the bank. We construct the bank in such a way that at least 99\% of planetary signals in the parameter space of interest get at least 95\% of their SNR from the closest template in the bank. The details of template generation are explained in Appendix~\ref{ap:template_bank}.
The resulting bank generated using the random placement method~\citep[][]{messenger_2009_random_placement_banks} consists of 58 templates and is shown in Figure~\ref{fig:template_bank}. The templates mainly differ by their duration and by how smooth their shape is. 

\paragraph{Smooth template shape}
As explained in Appendix~\ref{ap:templates_shape_losses}, when working with correlated noise, there is a significant difference between the box-shaped templates and the smooth ones. Our bank is designed using smooth physically modeled templates produced with \textit{Batman} transit modeling package~\citep[][]{kreidberg_2015_batman}. We used a set of random limb darkening coefficients to make sure the bank provides good coverage for different stars. 

\paragraph{Parameter space}. The parameter space of interest is constrained both by physical considerations and technical limitations. 
Physically, we focus on FGK stars, so the search is targeted at stars of radii $0.5R_\cdot<R<3R_\cdot$ and masses $0.3M_\cdot<M<2.5M_\cdot$. We also restrict the search to stars with typical power spectral densities (PSDs) characteristic of FGK stars. While the majority of transits for stars outside this range will still be detected, full coverage for such stars is not guaranteed. 

\paragraph{Transit duration limitations}
Technically, covering durations shorter than \textit{Kepler} exposure time (29.4 minutes, \citep[][]{kepler_data_characteristics_handbook}) is problematic as such transits are under-sampled. Conversely, very long durations are difficult to detect because their spectral support lies primarily in low-frequency regions, where noise power is higher and which are not resolved by the PSD measurement (Section~\ref{sec:detrending}). Therefore, we define the range of interest for transit durations as approximately 1 hour to 1.5 days.

The template duration can be estimated using Kepler's first and second laws, assuming the transit duration is much smaller than the orbital period $p$ of the planet. A detailed review of transit duration estimation is available in \citep[][]{kipping_2010_transit_duration}. The approximate formula is:
\begin{equation}
    \Delta T\approx\frac{p}{\pi}
    \frac{\varrho_c}{a_R}\frac{1}{\sqrt{1-e^2}}
    \sqrt{1-\varrho_c^2 a_R^2\cos^2i},
    \label{eq:transit_duration}
\end{equation}
where $a_R$ is the semi-major axis in units of stellar radii; $\varrho_c$ is the ratio of the star-planet distance at mid-transit to the semi-major axis; $e$ is the orbital eccentricity; $i$ is the inclination.  
These limitations on transit duration introduce constraints on orbital parameters. Qualitatively speaking, rare configurations of very large eccentricities at extreme separations, very large or very short periods, or grazing inclinations cannot be fully covered. Using a Monte-Carlo simulation, we ensure that our bank covers 99\% of possible transit shapes for the stars of interest. The parameter distributions used in these simulations are shown in Figure~\ref{fig:template_bank_parameters}.

\paragraph{Time sampling}. The transit models are generated as continuous functions, and then integrated over the Kepler exposure time of 29.4 minutes \citep[][]{kepler_data_characteristics_handbook} to produce the discrete-time templates. For the shortest-duration templates, sub-cadence shifts can cause significant changes in their sampled shapes. To account for this, we generate multiple sampled instances with varying sub-cadence shifts for such templates. During the periodicity search, we interleave these shifts to replicate how a real short-duration transit at a given periodicity would be sampled. 
For longer-duration templates, transit time shifts are binned by the exposure cadence, as shifts of one cadence produce sufficiently similar templates to recover most of the signal-to-noise ratio (SNR)

\paragraph{Template prior}. We conduct an additional Monte-Carlo simulation with physical priors on stellar geometry to determine the prior probability of triggering for the templates in the template bank. It tells which fraction of transit scenarios will have a given template as the closest template. The details of parameter sampling are described in Appendix~\ref{ap:injection_parameters}. 

This template prior is used when we calculate the template-marginalized statistical score (MMES) (Equation~\ref{eq:mmes_definition}) showing how likely it is for there to be a planet with any transit shape for a given period and phase. The details of this procedure are presented in Section~\ref{sec:mes_periodicity_search}.

\subsection{Data conditioning and whitening}
This section describes the methods that were used to prepare the data for the detection statistic calculation. The processing stages are illustrated in Figure~\ref{fig:conditioning_fourier}, showing the absolute value of the Fourier domain flux and its transformations during the conditioning.
\label{sec:conditioning_and_whitening}
\subsubsection{Power spectral density (PSD) estimation}
\label{sec:psd_estimation}
As shown in Appendix~\ref{ap:psd_estimation}, the maximum-likelihood estimator for a Gaussian noise spectrum is obtained by averaging the Fourier power of this noise. This constitutes the Welch's method \citep[][]{welch_1967}, or modified periodogram. The data is divided into overlapping slices, each multiplied by a window function. The absolute value squared of the Fourier transform of each slice is computed, and the average over slices is taken. This produces a PSD estimate on a coarse frequency grid corresponding to the slice length. The estimated PSD is then interpolated to match the data resolution when constructing the whitening filter.

The slices are taken to be overlapping, and the window function (we used Hann window) zeroes the ends of every slice. This approach ensures that each data point is effectively used only once.

\paragraph{PSD error and its impact}
When the PSD used in the matched-filtering (Equation~\ref{eq:stat_fourier_0}) differs from the true PSD of the noise, a loss in SNR of the detected signal occurs~\citep[][]{zackay_2019_non_gaussian}. This highlights why the white-noise statistic applied to correlated noise is so inefficient: it is equivalent to using a wrong PSD. 

When the PSD is estimated from the data, it will naturally have an error, which will lead to a certain loss in SNR. The error depends on the amount of data available for estimation and on the slice length. Shorter slices result in a coarser frequency grid for the PSD estimate, which may miss details or introduce leakage between frequencies. Large slice length, on the other hand, results in fewer slices, increasing statistical error during averaging. A tradeoff between the effects results in an optimal slice length that is expected to minimize the SNR loss. We selected the slice length of 128 \textit{Kepler} exposure times, approximately 2.6~days. For further discussion on the impact of PSD estimation errors on SNR, see Appendix~\ref{ap:psd_estimation}.

\paragraph{Kepler PSD particularities}
The \textit{Kepler} PSD exhibits a sharp red noise characteristic at frequencies below 1 day$^{-1}$. The best resolution achievable with a Welch slice of 128 bins is $\sim0.4$~day$^{-1}$. This resolution is insufficient to resolve fine details of the PSD, and it does not capture all frequencies below this threshold. The noise power in this region is very high, and the contribution to the useful SNR is less significant than the potential contamination. Therefore, we remove this low-frequency part of the data. The details of how it is done are described in Section~\ref{sec:detrending}, and the impact on the true signal SNR loss is discussed in Appendix~\ref{ap:psd_estimation}. 

The PSD of the \textit{Kepler} noise is quarter-specific, so it needs to be measured separately for each quarter. Given the typical quarter duration of $\sim5000$ samples and a Welch slice length of 128, we obtain $\lesssim$40 slices for averaging, which makes the measurement data-starved. However, the \textit{Kepler} PSD often behaves very predictably at high frequencies where where it flattens out to a white noise tail. This predictable behavior does not require a high frequency resolution and can benefit from a smaller slice size to reduce statistical noise. Therefore, we implement a multi-resolution modification of Welch's method. We use smaller slice sizes at high frequencies and longer sizes at low frequencies or where the spectrum exhibits a detailed structure.

\begin{figure}[ht]
    \centering
    \includegraphics[width=0.47\textwidth]{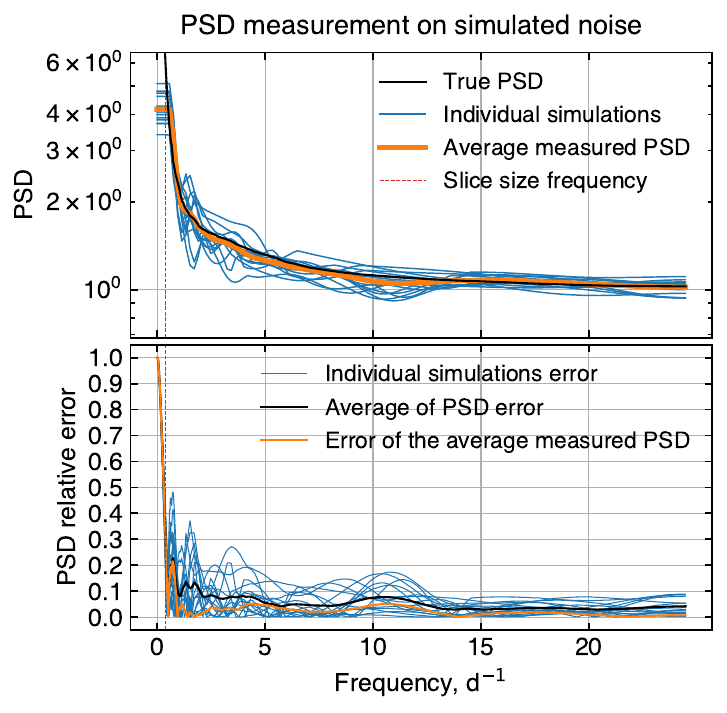}
    \caption{\textit{Top panel}: Example of PSD estimation for several simulated noise samples. The black line is the true PSD used to generate the noise. The blue lines represent individual measurements, while the orange line shows the average PSD from these measurements. The red vertical line marks the resolution limit of frequencies whose power can be measured. Power below this threshold is removed. 
    \textit{Bottom panel}: Relative error of the measurements from the top panel. The blue lines correspond to different noise samples. The black line is the average of the blue lines. The orange line is the error for their average (marked by the orange line in the top panel). }
    \label{fig:psd_measurement_simulation}
\end{figure}

\paragraph{Resulting performance.}
In order to illustrate the work of the method, we generate a representative PSD made by averaging the power spectra of several dozen Kepler stars. Using this averaged PSD, we simulated multiple noise samples, each corresponding to the length of one \textit{Kepler} quarter, and applied our PSD measurement procedure. The results of these measurements, along with a comparison to the true PSD, are shown in Figure~\ref{fig:psd_measurement_simulation}. As can be seen, the average of the measurements converges to the true PSD, demonstrating that the estimator is unbiased. The error is more pronounced at lower frequencies where frequency resolution is a limiting factor. There is a persistent systematic error due to power leakage and interpolation effects. At higher frequencies, the error becomes statistical in nature. The typical error is of the order of 10\%, meaning that the SNR loss resulting from PSD measurement inaccuracy will be of the order of a few percent, as the loss is quadratic with respect to PSD error~\citep[][]{zackay_2019_non_gaussian}.

\paragraph{Overfitting planets}.
The PSD measurement is performed on the same data as the one used for detection. As a result, deep transits can introduce a bias in the spectrum estimation, potentially causing the planetary SNR to be canceled by the overfitted PSD. This issue is particularly pronounced for long transits, which have narrower frequency support, as discussed in Appendix~\ref{ap:psd_estimation}. If the SNR of an individual transit is small, as happens for short periods, then this effect is not very significant. For high-SNR transits, one effective strategy to mitigate overfitting is to reject outliers during the averaging of Welch slices. For example, one can use only the 0.95-percentile of all the slices for PSD estimation, and correct for the selection assuming the exponential distribution.

We note that PSD measurement is not the dominant source of SNR loss. For SNR losses summary and evaluation of the resulting SNR recovery efficiency, the reader is referred to Section~\ref{sec:snr_recovery_fraction}.

\subsubsection{Detrending of low frequencies}
\label{sec:detrending}
As mentioned earlier, the PSD of \textit{Kepler} photometry exhibits a sharp increase in power at low frequencies, with some of these frequencies falling below the PSD measurement resolution limit set by the Welch slice length. For these frequencies, the PSD cannot be adequately estimated. Usually, their contribution to signal SNR is not significant due to high noise power. Therefore, our pipeline applies a high-pass filter (or detrending) to remove power from these inaccessible frequencies.

\begin{figure}[h!]
    \centering
    \includegraphics[width=0.47\textwidth]{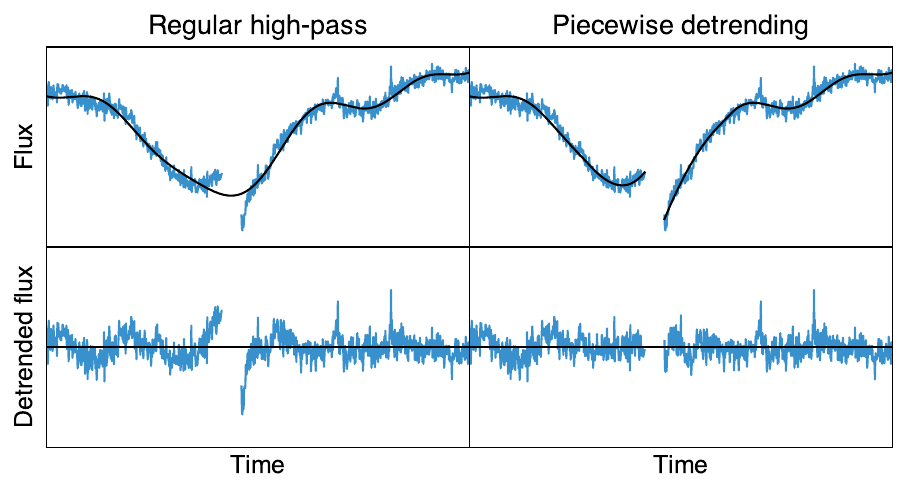}
    \caption{Illustration of regular high-pass filtering compared to the piecewise detrending method shown on one of the \textit{Kepler} targets (KIC003852808). 
    %(target KIC003852808). 
    \textit{Top left}: \textit{Kepler} flux (blue line) and its low-frequency Fourier component (black line).
    \textit{Bottom left}: Flux, after subtracting the low-frequency trend, has a spurious drop which may produce a false positive.
    \textit{Top right}: \textit{Kepler} flux (blue line) and its trend obtained using the piecewise detrending module.
    \textit{Bottom right}: With the piecewise detrending, the break is mitigated.
    }
    \label{fig:detrend_piecewise_example}
\end{figure}

We focus on the outcomes of detrending in the Fourier domain. Methods like subtracting a moving average or fitting polynomials do not control the response in the Fourier domain, meaning they may leave arbitrary amounts of power at low frequencies. Our detrending directly subtracts the low-frequency component from the data.

However, simply taking the Fourier transform and zeroing it at low frequencies can produce artifacts, as illustrated in the left panel of Figure~\ref{fig:detrend_piecewise_example}. Near \textit{Kepler} data gaps, there are sometimes breaks in the light curve behavior, such as due to temperature changes of the detector undergoing relaxation~\citep[][]{pdc_kdph}. These discontinuities can lead to contaminant signals after the low-frequency trend is subtracted, potentially resulting in false positives. 

To mitigate this issue, our pipeline employs a piecewise detrending method. It identifies discontinuously behaving sectors of data split by a gap and subtracts low-frequency component separately from each piece. Two pieces are considered separately if the gap between them exceeds 3 cadences and if they deviate from the common trend by more than 1.5 high-frequency standard deviations. This metric has been selected empirically and uses the \textit{Kepler} data gaps statistics and the fact that at high frequencies, the noise spectrum usually converges to the white noise constant. At the boundaries of the pieces, the pipeline uses polynomial fitting to match the behavior at the open ends. The result is illustrated on the right panel of Figure~\ref{fig:detrend_piecewise_example}.
Figure ~\ref{fig:notch_example}, provides another example made in the Fourier domain where one can see how power at low frequencies was removed.

\subsubsection{Sharp Fourier peaks removal}
\label{sec:notch}
In cases where the noise spectrum contains peaks that are narrower than the resolution limit set by the PSD measurement slice length, these peaks cannot be resolved. It occurs for some fraction of stars and has been referred to as harmonics by the \textit{Kepler} team \citep[][]{kepler_data_processing_handbook_search_chapter}.
To address this, our pipeline identifies these peaks after whitening the data and applies a Notch filter \citep[][]{orphandis_signal_processing} to remove the power associated with them. Rather than simply zeroing out the peak at the identified frequency, we use a filter with a finite width to avoid introducing a broad response function in the time domain.

\begin{figure}[ht]
    \centering
    \includegraphics[width=0.47\textwidth]{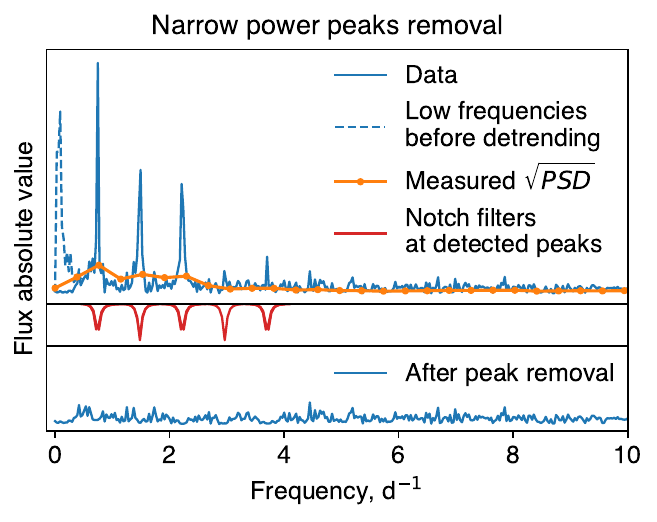}
    \caption{Example of removal of sharp peaks in Fourier domain for one of the \textit{Kepler} targets (KIC000757280).
    \textit{Top panel}: The absolute value of the Fourier transform of \textit{Kepler} data (blue line) alongside the measured PSD which does not resolve the peaks (orange line)
    \textit{Middle panel}:
    Notch filters designed by the pipeline for the identified peaks. 
    \textit{Bottom panel}: The Fourier transform of the data after applying the notch filters, showing the peaks removed. }
    \label{fig:notch_example}
\end{figure}

Figure~\ref{fig:notch_example} provides an example of the peak removal process. The top panel shows the absolute value of the original data in the Fourier domain, along with the measured PSD which is unable to resolve the peaks. The middle panel shows the Notch filters designed by the pipeline to target the identified peaks. These filters multiply the Fourier image of the data, they suppress the peak and converge to unity far away from it. As mentioned, the filters have finite bandwidth in order to control the response length in the time domain. Finally, the bottom panel shows the absolute value of the Fourier transform of the data after the Notch filters have been applied.

We note that the filters are also applied to the templates because the expected shape of transit in the data to which a filter is applied, gets modified.

% \filbreak
\subsubsection{Applying whitening filter}
\label{sec:applying_whitening_filter}
The whitening filter at frequency $f$ is given by $1/\sqrt{S(f)}$, where $S(f)$ is the PSD. The zeroth frequency bin of the whitening filter is always put to zero. 

As required by the matched-filtering formula (Equation~\ref{eq:stat_fourier}), the whitening filter is applied to both the data and all the templates (Equation~\ref{eq:whitening}). It is applied in the Fourier domain, where the PSD was estimated on a coarse frequency grid. In order to bring the data and the filter to the same frequency support, interpolation is necessary. This is done by transforming the whitening filter to the time domain, zero-padding it, windowing it, and then transforming it back to the Fourier domain. Zero padding assumes that there are no correlations in the data beyond the support size of the filter, a condition that was ensured during the detrending stage. The windowing is done using a Tukey window and ensures that the correlation kernel decays smoothly to zero in the time domain, avoiding step artifacts caused by zero-padding. Mathematically, this zero-padding in time domain procedure is equivalent to sinc interpolation. 

An example of whitened noise can be seen in Figures~\ref{fig:conditioning_fourier}~and~\ref{fig:scheme_time_domain}.

\subsubsection{Treating bad pieces of the data}
\label{sec:holes}
\paragraph{Determining the mask}
The \textit{Kepler} data contains segments of missing or low-quality data~\citep[][]{kepler_data_characteristics_handbook}. These excluded segments of the data are consolidated into a mask that is used throughout the search process. Masked-out points do not contribute to the resulting statistical score.

The mask includes \textit{Kepler} PDC light curves gaps (holes) caused by monthly data downlinks, sensitivity drops, cosmic rays, and other impacts~\citep[][]{kepler_data_characteristics_handbook}. Additionally, our pipeline identifies outliers which are points deviating by more than $6\sigma$ from the base level after subtracting a sliding mean. This outlier-detection process is repeated multiple times during different processing stages, as new outliers may emerge after applying filters.

Our pipeline is designed to detect faint planets, and the $6\sigma$ threshold is chosen based on the expected maximum depth of the transits targeted in our search. We assume that deep transits have already been identified as KOI. As discussed in Section~\ref{sec:input_data}, Confirmed KOI and Candidate KOI with \textit{Kepler} MES$>20$ are masked prior to the search. For PSD measurement, we use a stricter mask because losing the planetary signal is not a concern in this step.

\begin{figure}[h]
    \centering
    \includegraphics[width=0.47\textwidth]{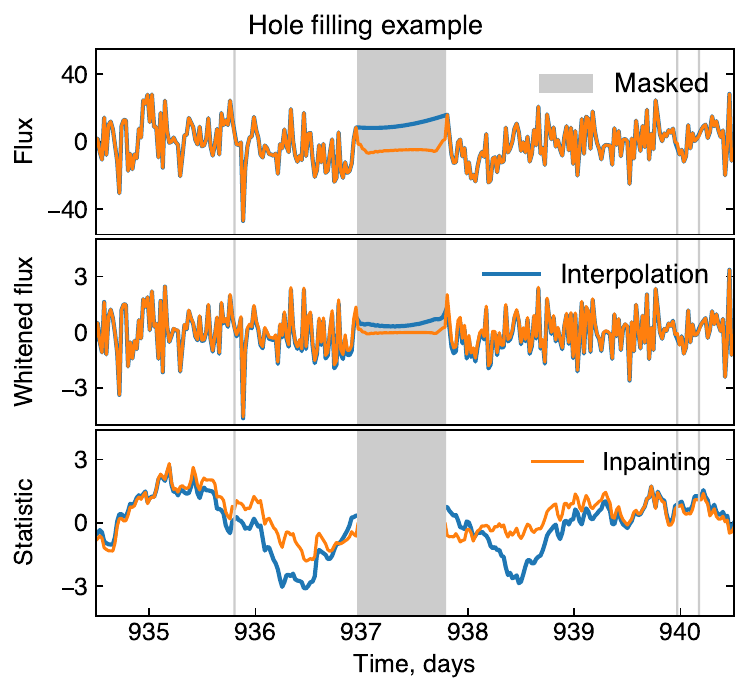}
    \caption{%(target KIC007742408).
    Example of the inpainting filter compared to linear interpolation (shown for target KIC007742408).  
    Gray regions represent masked points. Blue lines depict data with gaps filled using linear interpolation. Orange lines correspond to the gap filling with the inpainting filter. 
    \textit{Top panel}: \textit{Kepler} flux. Interpolation just connects the two points outside the mask. The inpainting filter fills the gap as described in the text, leaving data outside the gaps unchanged. 
    \textit{Middle panel}: Flux after applying the whitening filter. Cross-talk between points inside and outside the gap modifies data outside the gap in the linear interpolation case. 
    \textit{Bottom panel}: Statistic score after matched filtering with a template. The blue line corresponding to linear interpolation exhibits two dips which might mimic transits. The inpainted orange line does not contain these peaks.}
    \label{fig:holes_example}
\end{figure}

\paragraph{Treating the data gaps (holes)}
The distribution of gap lengths is roughly bimodal. Narrow gaps, such as outliers, last for one or two cadences. They can be safely interpolated using basic methods without affecting the analysis.
Wider gaps, such as data downlinks or known transits, may last for a few hours. They must also be filled to maintain data continuity, but improper filling may introduce artifacts that result in false positives. Artifacts can occur when the data is convolved with a kernel (e.g., whitening filter or matched filter), creating cross-talk between the filled region and its neighboring points. This cross-talk can generate spurious signals outside the filled gap, as demonstrated in Figure~\ref{fig:holes_example}

To address this issue, we employ a linear inpainting filter, as proposed by~\citep[][]{zackay_2019_non_gaussian}, which eliminates artifact production. The filter is designed in such a way that the contribution from values within the gap to the test statistic equals zero after the filter is applied. The inpainting process uses the gap mask and the whitening filter as inputs to calculate a linear combination of surrounding data points to fill the gap.

Another way to understand the inpainting filter is that it forces the "blued" data to be zero in the gap. Blued data is data to which the whitening filter is applied twice (or data multiplied by the inverse covariance matrix). It is the vector convolved with a non-whitened template in the matched filtering formula~\ref{eq:stat_snr}. If this vector is zero inside the gaps, convolution with a template results in a zero response from the gap region.

Figure~\ref{fig:holes_example} provides an example of gap filling, comparing linear interpolation and the inpainting filter. As can be seen, linear interpolation leads to contamination around the gap, which can mimic a transit. In contrast, the inpainting filter avoids such contamination.

\subsection{Single-event score (SES)}
\label{sec:ses}
\subsubsection{Matched filtering}
\label{sec:matched_filter}
The matched filtering procedure applies Equation~\ref{eq:ses_i} in the Fourier domain to whitened data $\hat{d}_w(f)$ and whitened templates $\hat{h}_w(f)$, calculating the single-transit statistic (SES),
\begin{align}
    \rho_{\mathrm{SES}}=\mathcal{F}^{-1}\left[\hat{h}^{\dagger}_w(f)\hat{d}_w(f)\right].
    \label{eq:ses_raw_fourier_domain}
\end{align}
This procedure is repeated for all transit times and all whitened templates from the template bank. It is performed per quarter using the data and templates whitened with the whitening filter for that quarter. Since the computation is carried out in the Fourier domain, it effectively performs circular convolution. To avoid cross-talk between the beginning and the end of the quarter, the vectors are padded with zeros at the ends. These padded regions are marked as gaps in the mask and treated accordingly, including gap inpainting (Sec.~\ref{sec:holes}) to avoid their impact on the statistic.

\paragraph{Score variance correction}
The denominator of the MES formula (Equation~\ref{eq:mes_formula}) uses the variances of the SES scores to normalize the resulting MES distribution, converting the MES to units of SNR. These SES variances can be calculated using Equation~\ref{eq:ses_variance_calculated} based on the measured PSD. 
However, due to errors in PSD measurement, the calculated variances deviate from the true SES variances, resulting in incorrect normalization and a subsequent loss of SNR. In addition, the PSD may change slowly over time, exacerbating the error. 
As shown in \citep[][]{zackay_2019_non_gaussian}, this issue leads to an SNR loss that scales linearly with the PSD measurement error ($\propto\epsilon$) rather than the expected quadratic dependence ($\propto\epsilon^2$). It makes a 10\% error in PSD measurement significantly impactful for the SNR loss.

To address this issue, \citep[][]{zackay_2019_non_gaussian} proposed directly measuring the matched-filtering statistic variance from the data rather than calculating it from the PSD. This approach mitigates the impact of the PSD measurement error and can compensate for small PSD changes within one quarter if they are present.  As a result, the distribution of the statistic in units of SNR will have variance one. 

Following this method, we measure the moving variance of the SES and use it for normalization. When computing the moving variance, we apply a window having a gap in the middle. This ensures that if a transit occurs at a given point, it will not bias its own normalization coefficient. After the SES vetting (Section~\ref{sec:veto_ses}), we re-measure the variance to account for possible changes after glitches were removed.

\subsubsection{Quality veto of SES}
\label{sec:veto_ses}
After calculating the matched filtering score for all the templates and transit times, the pipeline applies vetting procedures to all the SES exceeding $2\sigma$. Any triggers failing these veto tests are flagged and registered in the mask. These masked triggers are excluded from contributing to the total score during the subsequent periodicity search, which is performed blindly over the entire parameter space.

The vetting procedures include the following components, described below:
\begin{itemize}[noitemsep, topsep=0pt, left=1pt, label=-]
    \item Excluding the SES outliers larger than 6.5$\sigma$ together with the tails that they produce;
    \item Excluding tails of SES peaks that produce non-transit-like but still possibly significant SES;
    \item Detecting abrupt steps in the light curve and excluding points around them;
    \item Flagging the vicinities of \textit{Kepler} quarter edges and data gaps;
    \item Conducting the transit shape quality veto.
\end{itemize}

\paragraph{SES outlier veto}
This pipeline looks for small planets, and we assume that the reliability and completeness of the existing catalog for MES$>$12 are very good, so that no further search for them is needed. For a typical deep-SES event of MES=10 and 4 transits, the average SES would be 5 (in units of SNR). Based on this, we detect and reject the SES exceeding 6.5 in units of SNR. 

For every template and every considered point in time, we find the most significant SES value within 2 template durations around it, and across templates. If the obtained value exceeds the 6.5$\sigma$ threshold, we consider this point as originating from the tail of the $>6.5\sigma$ SES, and therefore mask it as well. 

We analyze both positive and negative SES outliers because both of them can be associated with excessive SES power around them.

We note that this veto prevents our pipeline from discovering large planets. In order to allow for this, the outlier threshold parameter should be modified.

\paragraph{Tail SES veto}
Since the SES time series are naturally correlated in time due to the convolution with the template, almost any SES peak will have tails of also relatively significant SES. Not always will these less significant SES be vetted by the transit shape quality veto (read below) because it is weaker for low-SNR SES. Since we know that for a real planet, the point of interest will be close to the SES peak, the tail SES are not relevant for the search. The exception is the case of TTV, but they are not the objective of this paper. Therefore, we apply veto to such tail statistics. Around every SES peak, we keep the values that are within 2$\sigma$ of it and mask the remaining SES.

\paragraph{Step veto}
Sometimes, light curves contain steps that were not corrected in the \textit{Kepler} PDC module~\citep[][]{pdc_kdph}. These steps can trigger the transit templates when coinciding with their ingress or egress. In addition, steps and transits look more similar in whitened data.

To detect this contamination, we build a template for a step in the original non-whitened data and calculate the matched-filtering score for it. If the step statistic is too high, it means that the data is better explained by the step than by a transit in correlated noise. In this case, we reject the points adjacent to the detected step.

\paragraph{Veto in the vicinity of data gaps}
It is known that in the vicinity of the \textit{Kepler} quarter edges and the data gaps, possible instrumental glitches can appear~\citep[][]{kepler_data_processing_handbook}. Near the quarter edges, we mask the points that are within one template duration from the edge.

In addition, we account for the enhanced probability of obtaining a glitch near data gaps by increasing there the threshold for the transit shape veto (see below). 

\paragraph{Transit shape veto}
The matched filtering statistic measures the inner product between the data and the template, effectively projecting the data vector onto the direction of the template vector. 
The residuals, representing the orthogonal component, are expected to follow a $\chi^2$ distribution if the data is accurately described by the transit model. Importantly, the vetting should be conducted on the whitened data; otherwise, the residuals will be correlated.

A standard $\chi^2$ test~\citep[e.g.][]{seader_2013_chi_2} is used to ensure that the residuals do not deviate significantly from the $\chi^2$ distribution. However, the power of this test diminishes when the $\chi^2$ distribution has a high number of degrees of freedom, which can be the case for long templates. Whitened templates are even longer than the regular ones, which further increases the number of degrees of freedom.

\paragraph{Piecewise $\chi^2$ veto}
We employ a shape quality test designed to yield a  $\chi^2$ distribution with fewer degrees of freedom. 
Consider a transit detected by a template with an estimated amplitude $A$ (see Equation~\ref{eq:amplitude_estimator} for the amplitude estimator). In this test, the template is split into several segments, and the amplitude of each segment is estimated independently. If the template accurately represents the data, the amplitudes of all segments should be consistent with $A$. Significant deviations from $A$ indicate that the data shape differs from the template.

\begin{figure}[ht]
    \centering
    \includegraphics[width=0.47\textwidth]{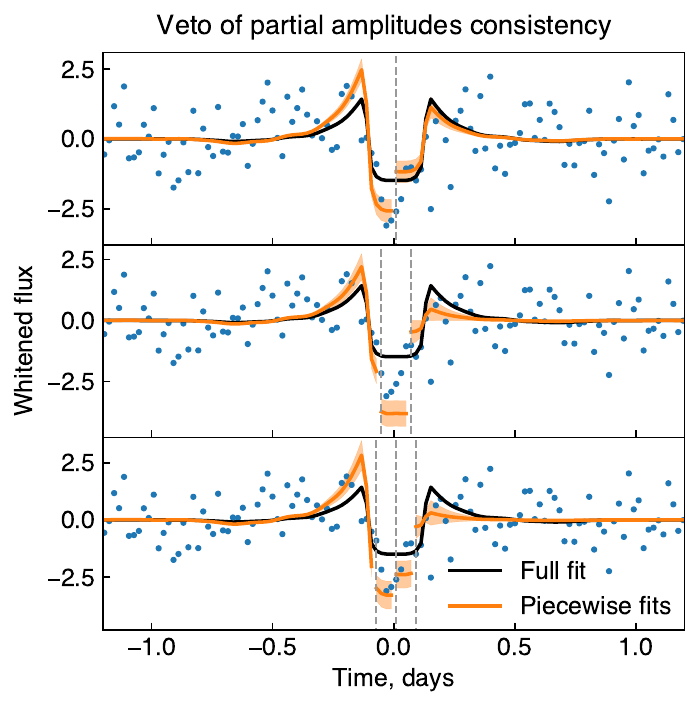}
    \caption{Shape quality piecewise $\chi^2$ veto, illustrated on \textit{Kepelr} target KIC008590354. The blue points represent the whitened flux around a trigger that was identified by the template shown in the black line. Orange lines show the template split into segments, with independent amplitude fits for each. Orange shading represents the 1$\sigma$ confidence interval for these fits. The three panels show the splitting of the template into 2, 3, and 4 parts. In this case, the test identified inconsistency in segment amplitudes, indicating that the template does not adequately describe the data.}
    \label{fig:veto_parts}
\end{figure}

An example is shown in Figure~\ref{fig:veto_parts}, where the template was split into 2, 3, or 4 parts, each contributing equally to the SNR. The test measures the amplitudes for these segments and evaluates a $\chi^2$ score characterizing their consistency. The number of degrees of freedom of the distribution of this score will be related to the number of segments rather than the number of data points, improving the test's sensitivity. The statistical basis and score calculation process are detailed in Appendix~\ref{ap:amplitude_consistency_veto}. In the pipeline, we perform the test both in the time domain and in the Fourier domain, for different numbers of segments.

\paragraph{Dynamic threshold for transit shape veto}
We measure the variance of the vetting score along the data and renormalize the score by it, ensuring it has a proper $\chi^2$ distribution. This is done under the assumption that most of the points in the data are not glitches. Based on this, we calculate the $\chi^2$ p-value for every point of interest.

We reject a point if its p-value falls below the threshold. The threshold is dynamic and is set empirically as a function of the SES value and closeness to a data gap.

We considered a distribution of SES scores of a large target sample with and without injections and calculated the Bayes ratio $\mathcal{L}\left(v|\rho, H_{0}\right)/\mathcal{L}\left(v|\rho, H_{1}\right)$ of the vetting score $v$ for different SES ($\rho$) ranges. We observed that these Bayes factors are SES-dependent, meaning that the probability of getting a glitch for a given vetting score is different for different SES values. Based on it, we build an empirical function for the vetting score p-value threshold: Below SES of 4 (absolute value, in units of SNR), the threshold is $10^{-2}$; above SES of 5, it is $10^{-1.5}$; and in-between, it changes linearly.

We also account for the enhanced probability of getting a glitch near a data gap by setting there a threshold of $0.05$. The vicinity of the gaps for each template is defined as the maximum of 0.75 days or one template width around the gap.

\subsubsection{Non-Gaussianity correction}
\label{sec:score_correction}

Based on the assumed Gaussian noise model (Equation~\ref{eq:data_model}), the SES distribution is also expected to be Gaussian, as it represents a convolution of the data with linear filters. Consequently, the MES, being a weighted sum of multiple SES, should also follow a Gaussian distribution.

However, the Gaussian assumption may not always be correct, producing an SES distribution exhibiting a non-Gaussian tail. During the periodicity search, high SES scores from the tail will participate in multiple combinations with other SES, producing multiple elevated MES. The resulting inflated MES background may make it impossible to detect genuine periodic signals.

\begin{figure}[h!]
    \centering
    \includegraphics[width=0.47\textwidth]{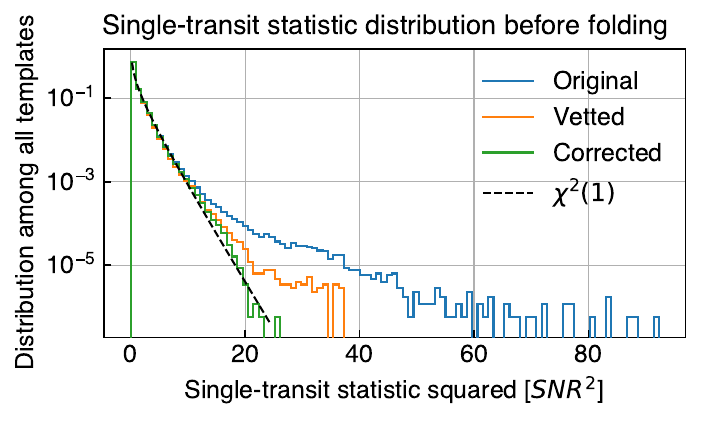}
    \caption{Example of distribution of SES in units of SNR before and after vetting and non-Gaussianity correction for \textit{Kepler} target KIC008590354. 
    Blue line: SES distribution with a non-Gaussian tail caused by the flux non-Gaussianity. Orange line: same distribution after veto. Green line: after the non-Gaussianity correction was applied.}
    \label{fig:ses_distribution}
\end{figure}

An example with an SES distribution having a non-Gaussian tail is shown by a blue line in Figure~\ref{fig:ses_distribution}. While the shape quality veto removes many high SES scores, some remain (orange line). 

To mitigate this issue, we use the non-Gaussianity correction formalism introduced in Section~\ref{sec:formalism}. We measure the SES distribution as described below and calculate from it the correction term (Equation~\ref{eq:ses_score_correction}) for every SES score. The correction will undergo periodicity folding (Section~\ref{sec:mes_periodicity_search}) and will be added the the final MES score. As a result, the distribution of this corrected CMES score will have its tail suppressed. 

In Figure~\ref{fig:ses_distribution}, we added the correction to the SES score for demonstration purposes, illustrating its effect on the SES distribution. We plot the corrected SES score in the green line in Figure~\ref{fig:ses_distribution}.

\paragraph{Finding the non-Gaussian distribution}
The distribution $\mathcal{L}_{NG}$ in Equation~\ref{eq:ses_score_correction} is to be measured empirically from the data. We take SES in units of SNR for all the transit times for every given template and compare their distribution to a Gaussian. If the distribution deviates, we fit it with a mixture model of two Gaussians. One Gaussian has a fixed variance to describe the well-behaved low-SNR part of the distribution, and the other Gaussian has free variance and amplitude fitted to describe the tail. These parameters of the mixture model are determined for every template. Then, they are used in Equation~\ref{eq:ses_score_correction} to calculate the correction for every SES.

\begin{figure}[h!]
    \centering
    \includegraphics[width=0.47\textwidth]{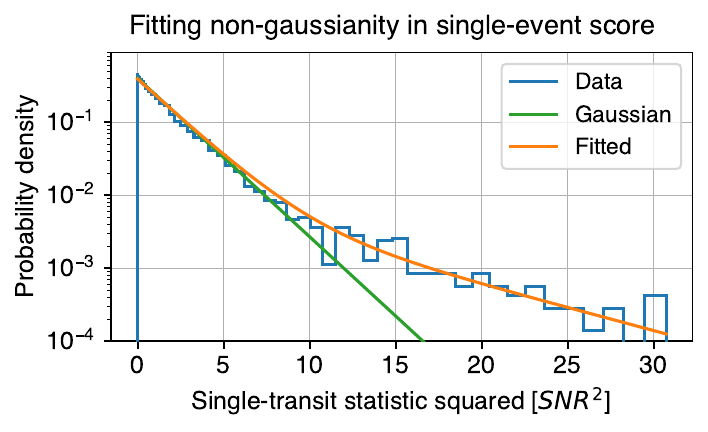}
    \caption{Illustration of SES distribution estimation for one of the Fit of the \textit{Kepler} targets (KIC008553579). \textit{Blue line}: distribution of SES in units of SNR. 
    \textit{Orange line}: fit by the two-Gaussian mixture model. This function will be used to compute the non-Gaussianity correction~(Equation~\ref{eq:ses_score_correction}). The green line shows a Gaussian distribution for reference.
    }
    \label{fig:ses_distribution_fitting}
\end{figure}

The approach of a two-Gaussian mixture has proven effective for typical distributions, as shown in Figure~\ref{fig:ses_distribution_fitting}. 

\paragraph{SNR loss due to non-Gaussianity correction}
The non-Gaussianity correction penalizes not only the noise tail but also the true transits, leading to a very significant loss of SNR. This issue has two main factors:\\
1) The SES distribution measurement is biased by transits. For the distribution measurement, real transits are indistinguishable from noise outliers. They also produce an SES tail and lead to elevated non-Gussianity correction values. \\
2) Transit SES get a non-Gaussianity penalty as any large SES. The correction forces the entire distribution of SES to follow a Gaussian, including real transits.

If the search was done based on SES, this effect would be critical for signal detectability. However, the detection is made by the MES after the periodicity folding. True planets, unlike the noise tail, exhibit periodicity, therefore they combine into significant MES and are detectable despite the correction.

The described losses are minimal for shallow transits, whose SES do not deviate significantly from the noise distribution. For a fixed MES, the depth of the individual SES is defined by the number of transits. Fewer transits require higher individual SES values to achieve the same MES, leading to strong loss due to the non-Gaussianity correction. 
In the limit case of 3 transits, with the target range of this search being MES $\sim10$, the SNR of an individual SES can be as high as 7. The correction in such instances is so strong that detecting three-transit events becomes unlikely. Section~\ref{sec:snr_recovery_fraction} and Appendix~\ref{ap:loss_score_correction} elaborate limitations of the search imposed by this loss.

\paragraph{Significance of the non-Gaussianity correction}
The non-Gaussianity correction is the bottleneck of the pipeline. It significantly enhances its reliability, as shown in Section~\ref{sec:reliability_of_pipeline}, but also leads to a large loss in completeness for few-transit cases, as shown in Section~\ref{sec:snr_recovery_fraction}. Eventually, Section~\ref{sec:detection_efficiency_real_data} shows that the net effect of the correction is positive, leading to increased detection efficiency.

Conceptually, running a search with this correction is equivalent to asking a question: "Does this deviation from Gaussian noise appear more like a non-Gaussian noise or like a periodic planet?". It makes periodicity, rather than transit depth, the main factor that distinguishes planets from noise. For cases with a small number of transits, periodicity is not well-pronounced. For instance, three equally spaced enhanced SES have a high chance of appearing as a noise phenomenon.

\paragraph{Alternative robust Gaussianization}
There is an even stricter way to control for non-Gaussianity, which is to apply a transformation that makes the SES distribution strictly Gaussian. This method was not used in this pipeline, but it can be a good strategy for low-SNR short-period searches. It is described in Appendix~\ref{ap:gaussianization}.

\subsection{Multiple-event statistic (MES)}
\label{sec:mes_periodicity_search}
\paragraph{Summation formulae}

This module conducts the periodicity search, implementing the equations presented in Section~\ref{sec:formalism} to calculate the raw UMES (Equation~\ref{eq:umes_definition}), the non-Gaussianity-corrected CMES (Equation~\ref{eq:cmes_definition}), and the template-marginalized MMES (Equation~\ref{eq:mmes_definition}). 

The periodicity search is performed in a period range of 30-500 days, but only triggers in the period range of 50-500 are selected for post-processing.

\paragraph{Preparation}
Before executing the periodicity search, we connect the quarters for all the templates separately to obtain a continuous time axis. Time samples with no data are marked as gaps (holes) and added to the mask. We connect the SES~(Equation \ref{eq:ses_raw_fourier_domain}), their measured variances, and the SES non-Gaussianity corrections~(Equation \ref{eq:ses_score_correction}).

For the templates requiring sub-cadence shifts~(Section~\ref{sec:template_bank}), we interleave the SES scores obtained from templates generated with different shifts, thereby increasing the time resolution. 

\paragraph{Period grid}
During the periodicity search, the MES is calculated for all the periods and first transit times. The period grid is constructed to satisfy the condition
\begin{align}
    \Delta p = \tau \frac{p}{T},
\end{align}
where $\Delta p$ is the period step size, and $\tau$ is the duty cycle, which is taken to be one \textit{Kepler} cadence. This condition ensures that the total timing error accumulated after $T/p$ repetitions of the transit does not exceed $\tau$. We note that this grid is not uniform; it is denser for small periods. 

The $t_0$ grid has a step size of \textit{Kepler} cadence.

\paragraph{Period chunks}
We divide the period grid into \textit{period chunks} and conduct the periodicity search separately in every chunk. The period range encompassed by one chunk is typically below one day, with ranges of neighboring chunks having a small overlap. Chunks are defined to have equivalent search entropy (effective number of search parameter options), meaning that the product of the number of independent periods and phases is roughly constant across chunks. The chunks have a similar look-elsewhere effect, which allows the look-elsewhere effect of the full search to be established using the number of chunks.

The division into period chunks was introduced primarily for computational reasons but has other practical benefits, for example providing a period range in the statistical significance estimation (Section~\ref{sec:significance}).

\paragraph{Periodicity search}
For each period chunk, the arrays of SES~(Equation \ref{eq:ses_raw_fourier_domain}), measured SES variances, and the SES non-Gaussianity corrections~(Equation \ref{eq:ses_score_correction}) are folded over the period grid. As a result, UMES (Equation~\ref{eq:umes_definition}) is calculated as a function of orbital period, phase, and template index. The non-Gaussianity correction is added to it to produce CMES (Equation~\ref{eq:cmes_definition}). 

\paragraph{Template marginalization}
Next, we perform a marginalization over templates to compute the MMES (Equation~\ref{eq:mmes_definition}). This involves selecting only negative scores corresponding to negative transits and marginalizing over the template index dimension. The resulting MMES score is two-dimensional: period and phase. The marginalization uses the template prior computed via the Monte-Carlo simulation described in Appendix~\ref{ap:template_bank}. Since the prior probability is less than one, the MMES score acquires a negative bias relative to the CMES score. However, as this bias is consistent across all scores, it does not affect the detectability of signals.

\subsection{Finding the best trigger}
\label{sec:best_trigger}
The pipeline performs the described periodicity search for every chunk, identifying and saving the maximum MMES score within each. The maximal trigger must pass the MES veto (Section~\ref{sec:veto_mes}). Then, for the best trigger of every chunk, the IMES score is calculated (Section~\ref{sec:formalism_integral_score}). Finally, the highest IMES score across all chunks is selected and reported as the best target score.

\paragraph{Finding the peak per chunk}
The peak search within each chunk is conducted iteratively. Initially, the highest MMES value is identified and subjected to the MES veto. If the veto is not passed, the peak is masked, and the next highest peak is selected. We make sure that the new best score is not part of the tail around the previously masked peak. Then, we apply the MES veto to the new peak repeating the cycle until the threshold of the lowest MES of interest is reached. 

After iteratively selecting the strongest MES in every period chunk, the pipeline ensures that the best trigger in one chunk is not a leakage from a neighboring chunk peak.

\subsubsection{MES Veto}
\label{sec:veto_mes}
A given MES may arise from the sum of several planetary transits, but it can also result from one or two large SES combined with others consistent with zero. This problem was mitigated by the non-Gaussianity correction (Section~\ref{sec:score_correction}). The correction penalizes the outliers, thus reducing the likelihood of producing large MES through this mechanism. 

For short periods, the individual contributions of the SES in the total MES are relatively small. As a result, achieving a significant MES requires a combination of many consistently periodic SES, which is unlikely to occur in background noise.

For longer periods, only a few transits are summed, making it easier for a few SES outliers to align and generate a strong MES. This effect is mitigated by the non-Gaussianity correction suppressing the large outliers, but cases of inconsistent SES producing a strong MES are still possible. First, some transits may coincide with a gap in the data (hole). For example, an MES might consist of two strong SES and two holes.
To filter out such cases, we consider an MES valid only if it includes at least three non-hole SES and if at least 30\% of all SES in the summation are non-hole.

In addition, we require a valid trigger to pass the transit depths consistency veto. This veto ensures that transit depths are similar, as expected for transits caused by the same planet. The alternative hypothesis tested by the veto is that the signal is dominated by a single large SES, with the others consistent with zero amplitude. The mathematical development and further discussion of this test are provided in Appendix~\ref{ap:atransit_depth_veto}.

\subsubsection{Integral statistic score}
\label{sec:methods_integral_score}

The IMES score is calculated for the best MMES in every chunk using Equation~\ref{eq:imes_definition}. This score incorporates the MMES value summed around the peak with the volume measure, and the prior with a normalization factor.

\paragraph{Prior}
The prior factor is pre-computed on a coarse grid of orbital periods and stellar densities, as detailed in Appendix~\ref{ap:injection_parameters}. We use the physical occurrence rates from~\citep[][]{zhu_dong_2021}, extrapolating them by a constant in regions where data is unavailable. These rates are multiplied by the transit probability, which is also pre-computed as a function of period and stellar density.

The prior assigns relative weights to period chunks based on the expected occurrence of transiting planets. It suppresses triggers that are unlikely to originate from planets and enables the true planetary triggers to be detected. 

As described in Section~\ref{sec:formalism_integral_score}, the IMES score is normalized with respect to a reference period, $p_{\mathrm{ref}}$, set to 30~days. At $p_{\mathrm{ref}}$, the contribution of the normalization factor to IMES vanishes. This normalization makes the prior's influence to be relative, maintaining the IMES score approximately in units of SNR$^2$.

\paragraph{Integration}
For the peak in every chunk, we select an area around the peak where the $\rho_{\mathrm{MMES}}^{2}$ value is at least half of the peak. In this area, we calculate the integral summing the exponentiated MMES with measure $\Delta p \Delta t_0$ taken according to the grid cell size in every chunk.

Since the the $\rho_{\mathrm{MMES}}^{2}$ value appears in the exponent, the integral is usually dominated by the peak itself and the exact integration region is not important. Still, we add overlap to the chunks to make sure that the area surrounding the peak is included. 

\subsubsection{Final products of the pipeline}
After calculating the IMES scores for all chunks, the scores are compared to identify the best trigger. This trigger, representing the most significant detection across all chunks, is reported as the pipeline's final product. 

As a result, for every target, the pipeline run determines:
\begin{itemize}[noitemsep, label=-, topsep=0pt, left=0pt]
    \item Best IMES score~(Equation~\ref{eq:imes_definition});
    \item Best period;
    \item Best first transit time;
    \item Best template found after returning to the MES scores at the best period and first transit time before the marginalization;
    \item Peak MMES for this best integral score;
\end{itemize}

%% file: sec_significance.tex
\subsection{$P_{\text{planet}}$: motivation for the method choice}
Theoretically, the detection statistic~(Equation~\ref{eq:imes_definition}) should exhibit an asymptotic $\chi^2$-like background distribution. The normalization constant of the distribution tail is determined by the look-elsewhere effect of the maximization over the parameter space. If the distribution tail is known, the statistical significance of the trigger should arise directly from the statistic value. 

In practice, however, there are the following caveats:
\begin{itemize}[noitemsep, label=-, topsep=0pt, left=0pt]
    \item Threshold setting. Establishing a detection threshold for the statistic depends on the expected rate of planetary signals. Ultimately, the question is: "Given a trigger, how likely is it to correspond to a planet?" This likelihood depends on the prior probability of a planet being present.
    \item Distribution deviations. The observed distribution of the statistic in real non-Gaussian data may differ from theoretical expectations.
    \item Target-specific effects. Real-data behavior and the look-elsewhere effect are often target-specific, making it impractical to define a universal threshold for all targets.
\end{itemize}
We address these issues in the following way. 

First, we choose to report what we call $P_{\mathrm{planet}}$ score~(Equation~\ref{eq:p_planet_score}). It compares the rate of triggers that are expected in the absence of planets in the data (background) to the rate of triggers that are expected if there are planets from the currently known population (foreground).
The score is to answer the question, "What is the probability that this trigger is caused by a planet, rather than by a background noise?". It is meant to be a target-level estimate of detection reliability.

While we also evaluate the false alarm probability (FAP), it does not directly indicate whether a trigger corresponds to a real planet. For instance, a low FAP trigger may still be unlikely to originate from a planet if the expected planetary trigger rate for that specific target is even lower. 

Second, we use the real data to assess the background and the foreground rates, as explained below.

Third, the analysis is conducted on a per-target basis to account for noise characteristics specific to each target.

The foreground rate is determined via an injection-recovery campaign (Section~\ref{sec:injections}). The background rate is found through a scrambled data search (Section~\ref{sec:scrambling}). Details on how the $P_{\mathrm{planet}}$ score is calculated can be found in Section~\ref{sec:calculating_p_planet}.

\paragraph{Period range of scrambling and injections}
To estimate the background and foreground rates for a given trigger in a quick and efficient way, we confine the scrambled and injection searches to a narrow period range around the trigger period. 
The rationale for this restriction is explained in Appendix~\ref{ap_sec:p_planet_period_range}. Intuitively, expanding the period range increases both the numerator and denominator of Equation~\ref{eq:p_planet_score} proportionally. The numerator grows because the prior planetary occurrence rate scales with the parameter space volume. The denominator increases due to the enhanced look-elsewhere effect from the larger parameter space. 

\paragraph{Full false alarm probability}
The background rate relates to the false alarm density per bin in the statistic value and a narrow range of orbital periods.

If one is interested in calculating a false alarm rate over all the search periods, it can be derived by scaling the narrow-range background rate. Practically, we use the range set by the period grid chunks introduced in Section~\ref{sec:mes_periodicity_search}. Given the uniformity of the effective number of independent parameter sets across chunks, the look-elsewhere effect for the entire multi-chunk search can be assessed from the single-chunk value and the total number of chunks.

To obtain an integral p-value-like false alarm probability, we calculate the integral of the extrapolated background distribution from the trigger score to infinity. Together with the previously described scaling by the period range, we can obtain the full p-value-like false alarm probability for a given trigger in the search.

\subsection{Scrambled search}
\label{sec:scrambling}
To estimate the background distribution, we conduct what we call a scrambled search. Similar techniques are used by the gravitational wave community \citep[][]{venumadhav_2019_pipeline} and by the \textit{Kepler} team to estimate the catalog reliability \citep[][]{thompson_2018}. Unlike the \textit{Kepler} team, we do not use inverted search, as noise properties differ for positive and negative values.

In our pipeline, we take the original data of a given target, split every quarter into a few segments, and apply random shifts to their time axis. Then, we run the search as usual. 

The size of the segments is larger than the noise correlation time scale addressed by the PSD (about 5 days), but smaller than the minimum candidate period (50 days). 

The time shifts are larger than the transit duration, and segments are also reordered. In this way, we preserve the noise features, including its spectrum, holes, glitches, and outliers -- while destroying any true periodicity that could be in the data.

For each target, we conduct $10^5$ scrambled searches. The best score from each iteration is saved, and their distribution is analyzed and compared to the original trigger~(see example in Figure~\ref{fig:background_injections_histogram}).

\paragraph{Bias due to the trigger}
As elaborated in Appendix~\ref{ap_sec:p_planet_background_rate}, background estimation can be biased if a planet is present in the data. While the periodicity of the planet is destroyed during scrambling, the individual SES can still contribute if they are sufficiently strong. On the other hand, masking the trigger could underestimate the true background if the trigger is not planetary. To mitigate these biases, we perform half the scrambled simulations with the trigger masked and the other half without masking. The two rates are then summarized in the $P_{\mathrm{planet}}$ score as described in Appendix~\ref{ap:p_planet}.

\paragraph{Background distribution extrapolation}
Given the large search volume, a statistically significant trigger must be exceptionally rare (the detection statistic has to be large). With $\sim10^5$ target stars and $\sim5\cdot 10^5$ effectively independent parameter sets per star (see Appendix~\ref{ap:rates_look_elsewhere}), millions of repetitions would be needed to observe rare background events in scrambled searches, which is computationally unfeasible. Therefore, it is hard to evaluate the background rate for large triggers from the scrambling alone. Luckily, due to the non-Gaussianity correction, the background distribution tail decay has approximately the $\chi^2$ behavior. This enables us to use extrapolation to assess the alleged background rate for rare events. Technical details of this extrapolation are provided in Appendix~\ref{ap_sec:bg_extrapolation}. An example of an extrapolated background can be found in Figure~\ref{fig:background_injections_histogram}.

\subsection{Injection-recovery campaign}
\label{sec:injections}
The numerator for the $P_{\mathrm{planet}}$ score (Equation~\ref{eq:p_planet_score}) represents the probability of identifying a real planet trigger with score $\rho^2$ in the search. This foreground rate includes the following factors:
\begin{itemize}[noitemsep, label=-, topsep=0pt, left=0pt]
\item The prior rate of planetary occurrence,
\item The probability to transit given the orbital and stellar parameters,
\item The probability of being identified correctly by the pipeline under the star's noise conditions,
\item The probability of achieving the score $\rho^2$.
\end{itemize}
To estimate this foreground rate, we conduct a flux-level injection campaign. We perform $10^4$ injection searches for every trigger, with the injected parameter sets following the expected transiting prior distribution.
For every search, mask the real trigger from the light curve, inject there a planet with needed parameters and random phase, and run our pipeline to determine the resulting detection statistic. Below, we provide further details.

\paragraph{Injection parameters}
The injection parameters for each trigger are generated based on the star's mass, radius, and approximate limb-darkening coefficients. As mentioned in Section~\ref{sec:significance}, we focus on planet scores rate only in the vicinity of the real trigger period. To get the expected number of planets at a given period, we use the prior occurrence rates calculated by~\citep[][]{zhu_dong_2021}. We apply geometric priors for orbital parameters (inclination, periastron argument), and a flat prior for eccentricity.

The prior radius distribution of the injected planets is also sourced from~\citep[][]{zhu_dong_2021}. However, not all radii can result in a detection statistic close to the trigger statistic $\rho^2_{\mathrm{trigger}}$. Scores far away from $\rho^2_{\mathrm{trigger}}$ will not contribute to the distribution density estimation in the vicinity of $\rho^2_{\mathrm{trigger}}$. In addition, very small radii will be non-detectable, and very large radii will considered outliers by the pipeline. In order to make the injection campaign more computationally efficient, we pre-select planet radii based on their expected statistic, estimated using 
an average star's measured power spectral density (PSD) and Equation~\ref{eq:umes_definition}. The pre-selection is conservative, resulting in a broad distribution of detection statistics, 
as shown in Figure~\ref{fig:background_injections_histogram}. To maintain correct units, we re-normalize the planet rates to account for the pre-selection. Further discussion on rates is provided in Appendix~\ref{ap_sec:p_planet_prior_rate}, and the parameter generation is detailed in Appendix~\ref{ap:injection_parameters}.

\paragraph{Probability to transit}
Not all naively generated parameters result in a transiting planet. To select only transiting parameters and calculate the transit probability, we use pre-filtered rejection sampling. For each combination of eccentricity and periastron argument, we identify inclinations that allow for a transit. Then, we generate a transit model \citep[][]{kreidberg_2015_batman} and accept the sample if the model indeed contains a noticeable transit.

Each accepted sample is weighted by the volume of its parameter space element. The integral over all volume elements equals unity, while the integral over accepted samples represents the transit probability.

\paragraph{Producing artificial transits}
First, the true search trigger is masked in order not to interfere with detecting the injections.

For each injection, the first transit phase is selected randomly. The transit model produced with~\citep[][]{kreidberg_2015_batman} is scaled by the mean flux of the quarter and added to the light curve.

\paragraph{Search for injections} 
Each injection undergoes full data processing, accounting for any biases the injection introduces to PSD measurement, non-Gaussianity correction, and other methods, as would occur in a real search.

The periodicity search is limited to one period grid chunk around the injection period, as results from other empty chunks are irrelevant. At the end of each run, we check whether the 
injection timing was successfully recovered and record its score.

\subsection{How the $P_{\mathrm{planet}}$ score is calculated}
\label{sec:calculating_p_planet}
For each trigger crossing a preliminary threshold, we initiate the additional scrambled and injection searches performed on this target's data. These searches provide the background and the foreground scores distribution. From these distributions, we determine rates corresponding to the trigger's IMES and orbital period value.

Equation~\ref{eq:p_planet_score} is a conceptual definition of the $P_{\mathrm{planet}}$ score. In practice, a more precise equation \ref{eq:p_planet_formula} is used. It provides a minor improvement of accuracy, taking into account the scrambled rates with and without masking the trigger (details in Appendix~\ref{ap_sec:p_planet_background_rate}).

\subsection{Errors in the $P_{\mathrm{planet}}$ estimation}
\label{sec:p_planet_errors}
The estimated background rate is a random variable subjected to errors, such as finite data error or statistical extrapolation error (discussed in Appendix~\ref{ap_sec:p_planet_errors}). The injection rate also has an error, mainly due to the uncertainty in the prior planetary occurrence rates. 
The resulting $P_{\mathrm{planet}}$ score is thus an estimate of the "true $P_{\mathrm{planet}}$ given the true background and foreground rates." 

Eventually, the $P_{\mathrm{planet}}$ score has two functions:\\
1) Assigning a weight to the candidate for population study;\\
2) Predicting the likelihood of a planet in potential follow-up observations.

The provided estimate addresses these two functions, however, it can be made more precise by:
\begin{itemize}[noitemsep, label=-, topsep=0pt, left=0pt]
\item Evaluating the distributions of the measured background and foreground rates around their true values and incorporating this uncertainty in the estimate;
\item Using a self-consistent planetary occurrence estimation that accounts for all candidates simultaneously.
\end{itemize}
These methods will be implemented in our forthcoming works
\citep{our_paper_2, our_paper_3}.

%% file: sec_trigger_example.tex
In this section, we provide an example of the pipeline performance on one of the \textit{Kepler} targets.

\begin{figure}[h]
    \centering
    \includegraphics[width=0.47\textwidth]{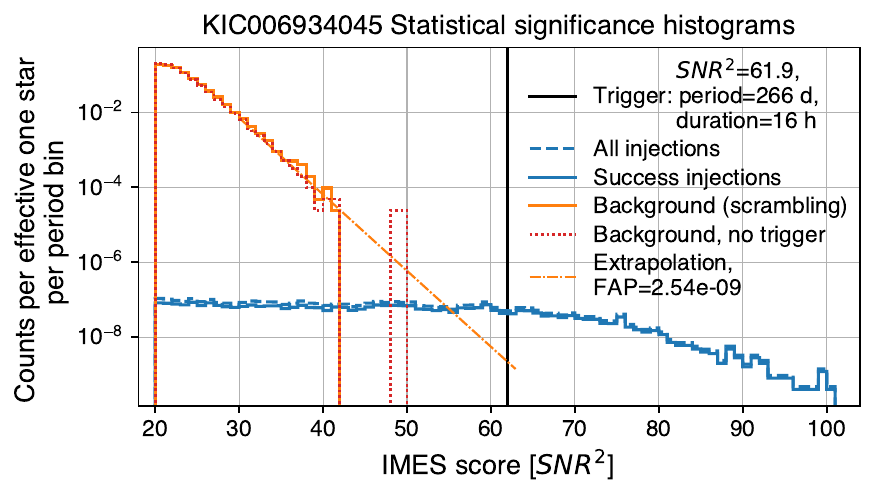}
    \caption{Example of a trigger statistical significance analysis for one \textit{Kepler} target KIC006934045. The horizontal axis represents the IMES statistic, and the vertical axis shows the probability density of obtaining a given IMES per star in a search in the selected period range (0.35 days around the trigger period of approximately 266 days). 
    \textit{Black vertical line}: IMES value of the trigger obtained in the main search.
    \textit{Orange solid line}: Background distribution obtained from $5\cdot 10^4$ scrambled searches.
    \textit{Orange dashed line}: Extrapolation of the background distribution (orange solid line).
    \textit{Red dotted line}: Same as orange solid line, but when the trigger was masked before running the scrambled searches. The extrapolation for this background distribution is not shown in this plot.
    \textit{Blue solid line}: Foreground distribution obtained in the injection-recovery campaign, normalized to the expected transiting planet rate in this period range.
    \textit{Blue dashed line}: Foreground distribution including injections for which the pipeline failed to detect the correct timing.
    }
    \label{fig:background_injections_histogram}
\end{figure}

The light curve was processed by the pipeline, and the best trigger was threshold-crossing. It passed the vetting process and underwent the statistical significance estimation procedure. Figure~\ref{fig:background_injections_histogram} demonstrates the output, showing the background and the foreground IMES score distributions, along with the trigger IMES. 

The background was obtained by running the scrambled searches (Section~\ref{sec:scrambling}) in the period range of width about 0.35 days. The same range was used for the injection-recovery campaign (Section~\ref{sec:injections}). 
The expected background rate at the trigger IMES was determined through extrapolation (Appendix~\ref{ap_sec:bg_extrapolation}), and, along with the injections rate (blue distribution in Figure~\ref{fig:background_injections_histogram} at the trigger MES), was used to calculate the $P_{\mathrm{planet}}$ score (Section~\ref{sec:calculating_p_planet}). The resulting $P_{\mathrm{planet}}$ is approximately 0.94, indicating that the pipeline predicts a 94\% probability of this trigger corresponding to a real planet.

\begin{figure}[h]
    \centering
    \includegraphics[width=0.47\textwidth]{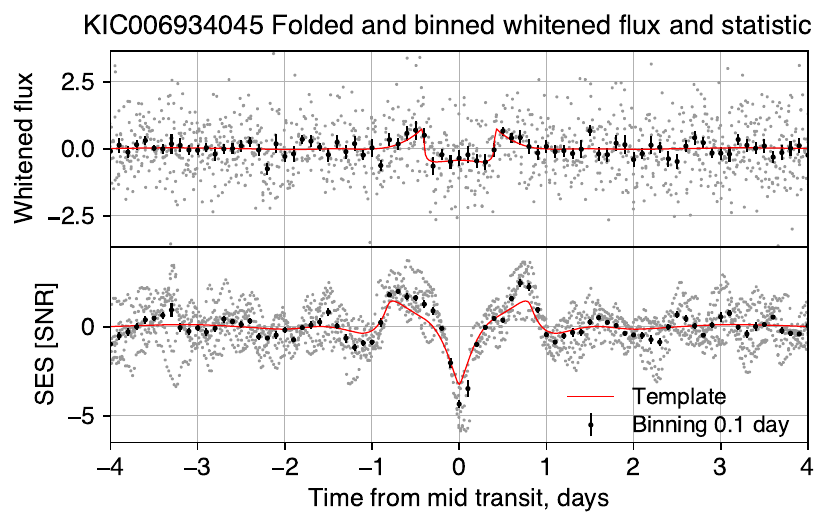}
    \caption{Folded and binned time series for the trigger analyzed in Figure~\ref{fig:background_injections_histogram}. Gray dots represent folded time series, meaning that their time axis was taken modulo the trigger period. Black dots represent the binning of the gray dots. The plots are zoomed-in on the transit; the full plots would span the entire orbital period of about 266 days.
    \textit{Top panel}: Folded whitened flux (Section~\ref{sec:applying_whitening_filter}). The red line shows the best-match whitened template from the template bank, with its amplitude estimated using Equation~\ref{eq:amplitude_estimator}.
    \textit{Bottom panel}: Folded SES time series obtained by cross-correlating the closest whitened template with the whitened flux (Section~\ref{sec:matched_filter}). The red line represents the cross-correlation of this template with itself, showing the expected SES response.
    }
    \label{fig:example_trigger_folds}
\end{figure}

For illustration purposes, the whitened flux and the SES statistic were folded at the trigger period and binned. They are shown in Figure~\ref{fig:example_trigger_folds}, zoomed-in around the transit time. The figure also shows the whitened template that achieved the highest CMES score and its the expected SES response.

%% file: sec_reliability_of_pipeline.tex
In this section, we analyze the quality of the false-alarm control of the pipeline. It is done by comparing the distribution of the triggers obtained by operating the pipeline on the real \textit{Kepler} light curves and on the scrambled light curves.

\begin{figure*}[ht]
    \centering
    \includegraphics[width=0.8\textwidth]{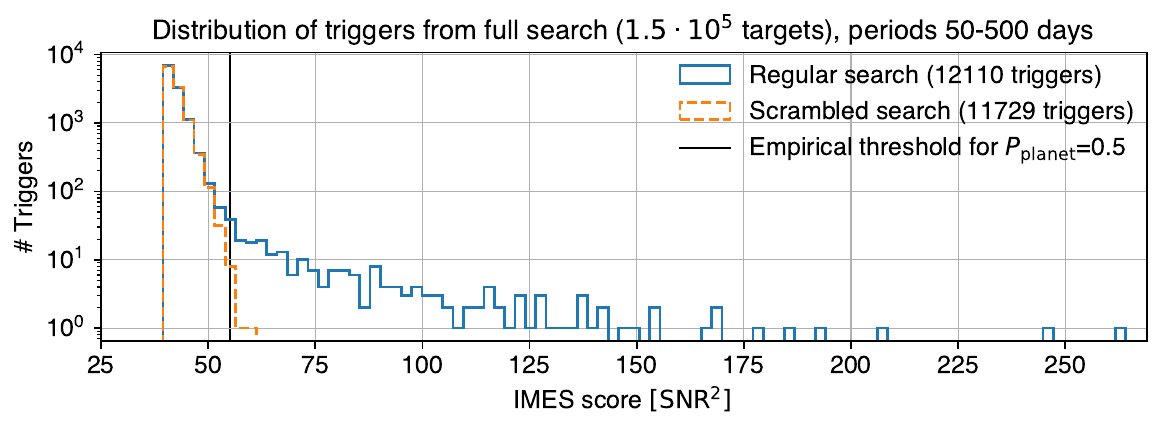}
    \caption{Distributions of IMES detection score for regular search and scrambled search over about $1.5\cdot 10^5$ targets. The vertical axis shows the number of obtained triggers in this IMES bin. 
    \textit{Blue line}: Scores from the regular, non-scrambled search.
    \textit{Orange line}: Scores from the scrambled search.
    \textit{Black vertical line}: Approximate empirical threshold for which the triggers result in $P_{\mathrm{planet}}\approx0.5$.
    }
    \label{fig:search_triggers_and_scrambled}
\end{figure*}

The search covered around $1.5\cdot 10^5$ targets, 
pre-selecting light curves with sufficient data size and noise levels allowing detection of planets of interest. The details of target selection are not crucial here, as the goal is just to compare the regular and the scrambled searches; they will be provided in our forthcoming work~\citep[][]{our_paper_2}. 

The regular and the scrambled searches were performed in the same way, differing only by the time-axis shuffling applied to the data before the scrambled search (Section~\ref{sec:scrambling}). Before both searches, we masked the \textit{Kepler} bright candidates of \textit{Kepler} MES$>$20, listed in the \textit{NASA Exoplanet Archive}~\citep[][]{akeson_2013_nasa_archive}.

The maximal triggers with IMES values exceeding 40 were retained and are displayed in Figure~\ref{fig:search_triggers_and_scrambled}. As can be seen, the regular search resulted in a long-tailed IMES distribution, whereas the scrambled search lacked this feature. 
Assuming scrambled data fairly represents the no-signal case, the long tail in the regular search is attributed to true periodic signals in the data. These could be planetary transits or other periodic phenomena, but they are distinct from the background.

The black vertical line in Figure~\ref{fig:search_triggers_and_scrambled} marks the approximate empirical IMES$=$55 threshold where triggers'  $P_{\mathrm{planet}}$ equals 0.5. This threshold aligns with the divergence between regular and scrambled search histograms, suggesting that scores above this value correspond to real periodic signals. Further details on triggers and $P_{\mathrm{planet}}$ scores will appear in~\citep[][]{our_paper_2}.

This experiment demonstrates the reliability of low-MES triggers in the real search. 
Starting from IMES$\sim$60, triggers are unlikely to be background contaminants, which ensures the reliability of the future catalog at these values.
This value of IMES$\sim$60 is consistent with an approximate estimate of the expected background rate of the search provided in Appendix~\ref{ap:rates_look_elsewhere}.

However, reliability may come at the expense of completeness. While the false alarm tail was suppressed, we must also verify that true planets are not missed due to reduced SNR. In Section~\ref{sec:detection_efficiency_real_data}, we will analyze the resulting detection efficiency and demonstrate that it increases due to the suppressed background distribution.

%% file: sec_detection_efficiency.tex
This section analyzes the pipeline's capacity to detect injected planets in both simulated and real light curves. 

As discussed in Section~\ref{sec:score_correction}, measures to enhance pipeline reliability, particularly the non-Gaussianity correction, reduce the SNR of true signals while effectively controlling the background. This section quantitatively assesses this SNR loss, defines detectability limits, and shows that the correction ultimately improves the pipeline's ability to detect small planets. 

In Section~\ref{sec:reliability_of_pipeline}, we established that the IMES score provides a clean background distribution which is responsible for pipeline reliability. 

In Section~\ref{sec:snr_recovery_fraction}, we will consider constraints on the completeness of the IMES score. We will examine the recovered fraction of injected SNR across periods in simulated light curves and reveal the theoretical limits on detection completeness that it sets.

In Section~\ref{sec:detection_efficiency_real_data}, we will examine the pipeline's detection efficiency for real light curves at a representative orbital period. We will demonstrate that the net effect of the corrections used in the IMES score leads to improved detection efficiency. 

\subsection{Detection efficiency limits due to the SNR loss (simulated data)}
\label{sec:snr_recovery_fraction}
In this section, we investigate the loss in SNR due to the pipeline background control procedures, mostly the non-Gaussianity correction. For this, we simulated the light curves, injected a signal with known SNR, and operated the pipeline on this data. 

\paragraph{Light curve simulation}
For this test, we used simulated light curves and not the real data because it is impossible to inject a signal with known SNR into the real data. The definition of SNR (Equation~\ref{eq:snr_definition}) includes the true PSD of the noise, which is not known for the real data. In our simulation, we use an effective PSD that we obtained by averaging many \textit{Kepler} targets in the Fourier domain. With this PSD, we generated light curves of the length similar to \textit{Kepler} data. 

\begin{figure}[h]
    \centering
    \includegraphics[width=0.47\textwidth]{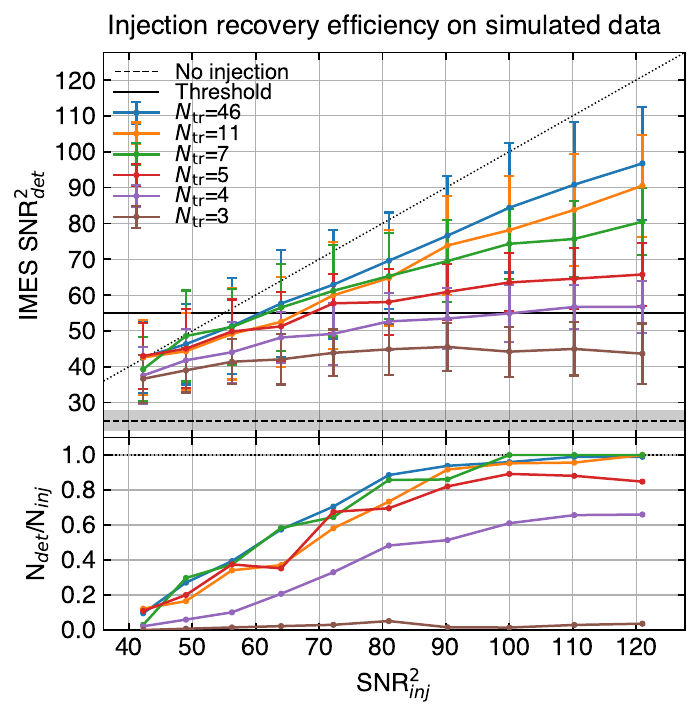}
    \caption{SNR recovery fraction and detection efficiency in simulated light curves.
    \textit{Top panel}: Dependence of the recovered SNR$^2$ on the injected SNR$^2$ for different number of transits. The duration of injected transits was 0.4 days. Each point is an average of multiple simulations, with the error bar showing the 1$\sigma$ dispersion.
    The solid horizontal line shows the empirical detection threshold of IMES corresponding to $P_{\text{planet}}\approx 0.5$.
    The dashed horizontal line shows the scores obtained in this simulation when no planet was injected.
    \textit{Bottom panel}: Relative number of injections that were identified at correct times and crossed the empirical detection threshold. It is plotted as a function of the injected SNR$^2$ and for different numbers of transits.
    }
    \label{fig:fraction_sr_recovered}
\end{figure}

In these light curves, we injected planetary signals of fixed transit duration of 0.4 days and known SNR. We varied the period of the planet to investigate the dependence of the SNR recovery efficiency on the number of transits. A signal with fewer transits has a higher SNR per transit, which, as discussed in Section~\ref{sec:score_correction}, makes it more susceptible to suppression by the non-Gaussianity correction. Figure~\ref{fig:fraction_sr_recovered} presents the dependence of recovered SNR on injected SNR for different numbers of transits.

\paragraph{Discussion of SNR loss}
The detected SNR deviates from the injected SNR for strong signals and for small number of transits. This discrepancy primarily arises from the non-Gaussianity correction, which penalizes large SES scores. 

Simulations performed without the non-Gaussianity correction showed significantly smaller SNR losses and weaker dependence on the number of transits. The result of this simulation can be found in Figure~\ref{fig:fraction_snr_recovered_no_correction} in Appendix~\ref{ap:confirmed_mes}. The remaining SNR decrease summarizes the losses caused by PSD measurement inaccuracy, false-negative vetting, and other factors.

\paragraph{Detection efficiency}
While SNR loss is not inherently problematic, it can hinder detection if it lowers the SNR below the detection threshold. With a horizontal line in Figure~\ref{fig:fraction_sr_recovered}, we show the empirical threshold of IMES=55 corresponding to $P_{\text{planet}}\approx0.5$ (see Section~\ref{sec:reliability_of_pipeline}). We define a detected signal here as a signal whose timing was recovered correctly and which acquired an IMES$\geq$55. 

In the bottom panel of Figure~\ref{fig:fraction_sr_recovered}, we plot the fraction of detected injections as a function of injected SNR, stratified by the number of transits. The very low detection fraction for planets having only 3 transits signifies that currently, this pipeline is not suitable for detecting periods $\sim500$ days. As can be seen in Figure~\ref{fig:fraction_snr_recovered_no_correction} in Appendix~\ref{ap:confirmed_mes}, without the non-Gaussianity correction, the dependence of the detection efficiency on the number of transits largely disappears. 

However, these results are based on simulated correlated Gaussian noise, where the non-Gaussianity correction provides no benefit because there is no real non-Gaussianity to be corrected. For the real non-Gaussian noise, detectability depends on the actual background distribution and can benefit from the non-Gaussianity correction, as will be shown in Section~\ref{sec:detection_efficiency_real_data}. 

We also note that in this section, we discussed the theoretical average losses and detectability limits for the pipeline. Real data may involve additional losses due to factors such as missing data or transit coinciding with holes. Variations in photometric quality, PSD shape, and light curve length can also impact detectability.

\subsection{Demonstration of detection efficiency in real data}
\label{sec:detection_efficiency_real_data}
This section has two goals:
\begin{itemize}[noitemsep, label=-, topsep=0pt, left=0pt]
\item Illustrate the pipeline's detection efficiency in the real  \textit{Kepler} data;
\item Demonstrate that the net impact of the corrections used in the IMES score improves detection efficiency for faint signals.
\end{itemize}

As was shown in Section~\ref{sec:snr_recovery_fraction}, the non-Gaussianity correction leads to SNR loss. From the other side, as shown in Section~\ref{sec:score_correction}, it controls the background distribution tail, helping distinguish the signal from the noise. Here, we demonstrate that the latter effect is dominant for real \textit{Kepler} light curves.

\paragraph{Comparing IMES, CMES, UMES scores}
We remind the definitions of the detection scores introduced in Section~\ref{sec:formalism}:
\begin{itemize}[noitemsep, label=-, topsep=0pt, left=0pt]
\item UMES: matched-filtering MES detection statistic with measured noise PSD, SES vetting, and normalization by measured moving SES variance;
\item CMES: UMES with non-Gaussianity correction;
\item IMES: CMES with template marginalization, prior, MES vetting, and likelihood integration around the peak.
\end{itemize}
All the scores have units of SNR$^2$.
For reference, it is demonstrated in Figure~\ref{fig:imes_umes_kmes_comparison} in Appendix~\ref{ap:confirmed_mes}, that the UMES score is similar to the squared \textit{Kepler} MES score. By comparing the UMES and IMES scores' performance, we assess the net effects of the corrections used in the IMES score on the planet detectability. 

For simulated Gaussian data and fixed detection threshold, UMES score exhibits significantly less SNR loss than IMES, resulting in higher detection efficiency (Figures~\ref{fig:fraction_sr_recovered},~\ref{fig:fraction_snr_recovered_no_correction}). 
Here, we compare UMES and IMES performance in real data. We will show that using UMES results in a background distribution with a strong tail, raising the detection threshold. IMES allows for lowering the detection threshold due to its clean background distribution. This effect is more significant than the SNR loss effect, therefore the IMES score increases the fraction of detected injections.

\paragraph{The idea of the analysis}
To evaluate detection efficiency, we conducted the following test:
\begin{enumerate}[noitemsep, topsep=0pt, left=0pt]
\item We run a background search on scrambled \textit{Kepler} light curves with no planetary signal. From the distribution of UMES and IMES scores in this search, we determine the detection threshold for each score.
\item We run the same search with injected planets and measure the fraction of injections that surpassed the threshold for each score. 
\end{enumerate}

\paragraph{Light curves used in the test}
We selected a subset of \mbox{$\sim1.5\cdot 10^5$} \textit{Kepler} light curves used in Section~\ref{sec:reliability_of_pipeline}. We made sure that the selected targets have at least 12 \textit{Kepler} quarters and do not contain KOI. For every target, we conducted several scrambled searches (Section~\ref{sec:scrambling}) with and without injection.

For demonstration purposes, we choose nominal parameters for search and injections. We use an orbital period window centered at 200 days of width \mbox{$\approx 0.35$} days. We note that the effect of corrections on background and injected SNR tends to increase for longer periods.

The injections were made for a circular orbit and transit on the line of sight. Planetary radii were selected to imitate small planets roughly corresponding to the expected SNR$^2$ in our range of interest between 60 and 80. Recovered scores can vary greatly both because of statistical factors and the roughness of SNR$^2$ prediction. 

\begin{figure*}
    \centering
    \includegraphics[width=0.8\textwidth]{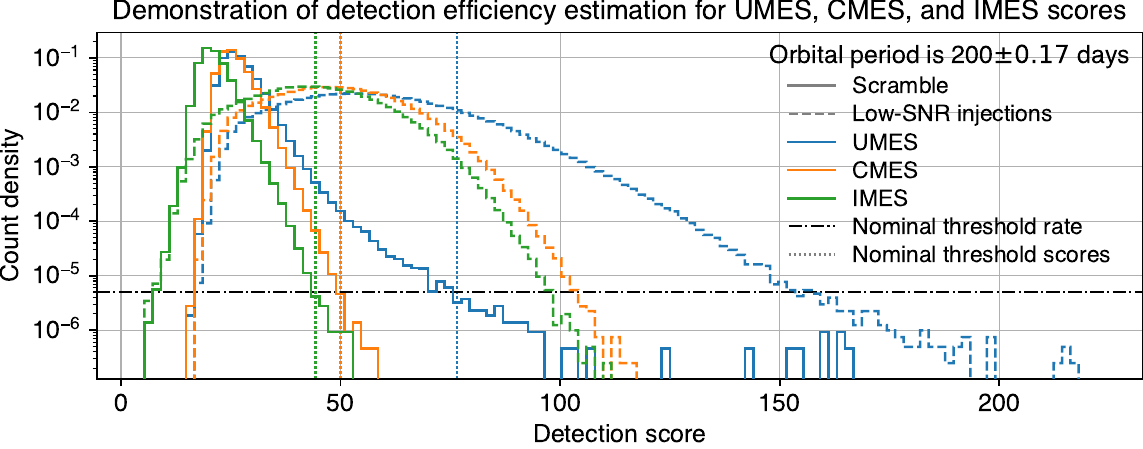}
    \caption{Distribution of different scores from empty scrambled searches and searches with injection. 
    \textit{Solid lines}: UMES, CMES, and IMES scores from empty scrambled searches. These distributions define the corresponding threshold
    \textit{Black dash-dotted line}: Nominal background density distribution defining the detection threshold.
    \textit{Vertical dotted}: Threshold UMES, CMES, and IMES values, at which the corresponding score distributions cross the threshold density.
    \textit{Dashed lines}: UMES, CMES, and IMES scores from injected searches.
    }
    \label{fig:detection_efficiency_histograms}
    \vskip0.2cm
\end{figure*}

\paragraph{Resulting distribution of scores}
Figure~\ref{fig:detection_efficiency_histograms} shows the score distributions for empty and injection searches, normalized to have unit integral. We add also the CMES score to show in isolation the effect of the non-Gaussianity correction. IMES and CMES scores are suppressed with respect to the UMES score in both empty and injection runs. We proceed to determine the net effect of this on the detection efficiency.

\paragraph{Detection threshold determination}
To determine the detection threshold, we identify the detection score that corresponds to a given density of the background distribution. Here, we chose a nominal density of $5\cdot 10^{-6}$ (shown by a dot-dashed line in Figure~\ref{fig:detection_efficiency_histograms}). This value was selected for illustration purposes; in the real search, the required background density would be determined based on the expected planet occurrence rate, as the ratio of the two defines the $P_{\text{planet}}$ score. Typically, this threshold background density would be smaller, as shown, for instance, in Figure~\ref{fig:background_injections_histogram}.

Alternatively, the threshold definition can be established using the false alarm rate (FAR), which considers the integral of the PDF of the background distribution. This would increase the difference between UMES and IMES thresholds, as the integral of the extended UMES tail decays is slower than the PDF itself.

\begin{figure}[h!]
    \centering
    \includegraphics[width=0.47\textwidth]{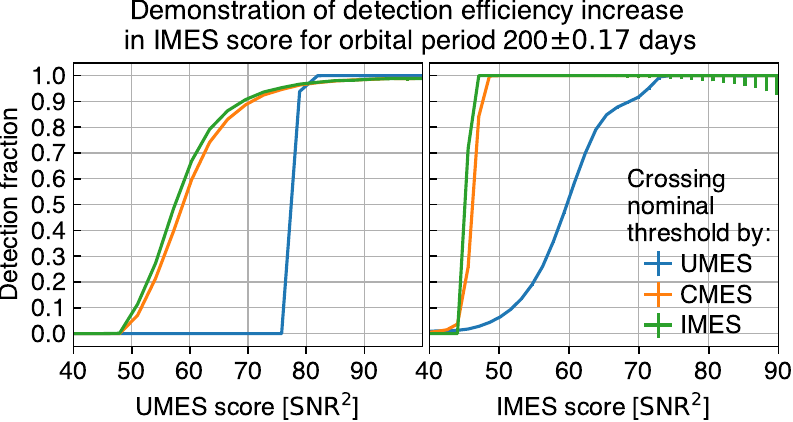}
    \caption{Comparison of detection efficiency of UMES, CMES, and IMES scores. 
    \textit{Vertical axis}: part of all injections that crossed the detection threshold for the corresponding score. The detection thresholds and the injection distributions are shown in Figure~\ref{fig:detection_efficiency_histograms}. The timing of the transits was identified correctly for all the injections that crossed the detection threshold.
    \textit{Left panel}: Detection fraction presented as a function of the UMES score of the injections.
    \textit{Right panel}: Same data as the left panel presented as a function of its IMES score.}
    \label{fig:detection_efficiency_scores}
\end{figure}

\paragraph{Detection efficiency increase}
Once the detection threshold score is set based on the background distribution, we evaluate the fraction of injections surpassing this threshold. Figure~\ref{fig:detection_efficiency_scores} displays this fraction as a function of the UMES and the IMES score obtained by the signals. As can be seen, for a low-SNR trigger, the detection fraction is higher when the detection is made using the IMES compared to the UMES score. 
The net effects of corrections used in the IMES score allow the detection of those triggers of UMES 55 to 80 (roughly SNR from 7.5 to 9) that would otherwise be inaccessible. 

Comparing the CMES and IMES scores, we see that the largest improvement with respect to the UMES score comes from the non-Gaussianity correction. There is a slight improvement in IMES score compared to CMES score, but this is not the reason IMES should be used. In this test, the injections did not follow physical prior distributions, we used fixed parameters. In the search on the real planetary population, we use IMES because it is prior-informed and thus adjusts its detection threshold automatically for the expected planet rate.

We note that Figure~\ref{fig:detection_efficiency_scores} was produced as a demonstration for a particular period window. In order to broadly understand the detection efficiency of this pipeline, see its limitations, and compare it to the previous searches, the same investigation should be repeated for other orbital periods. This will be done in our future work as part of the general pipeline efficiency evaluation needed to estimate the planet occurrence rates.

\begin{figure}[h!]
    \centering
    \includegraphics[width=0.47\textwidth]{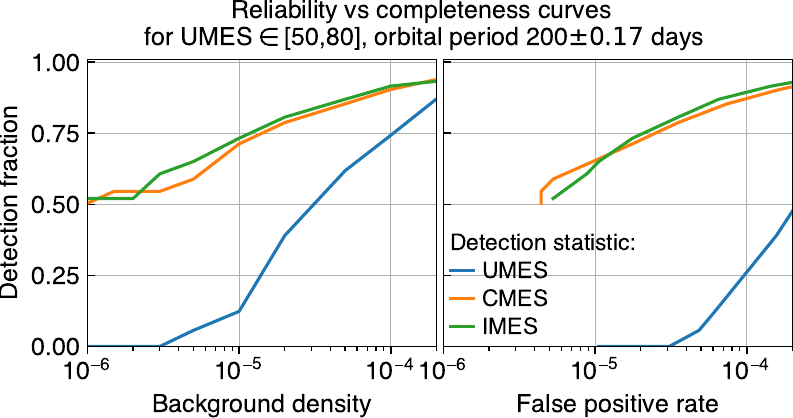}
    \caption{Receiver operating characteristic curves for injections in a range of UMES from 50 to 80.
    \textit{Vertical axis}: fraction of the injections (Figure~\ref{fig:detection_efficiency_histograms}) that crossed the detection threshold. 
    \textit{Left panel}: detection fraction as a function of chosen background rate density (horizontal line in Figure~\ref{fig:detection_efficiency_histograms}).
    \textit{Right panel}: Detection fraction as a function of the measured false positive rate.}
    \label{fig:roc}
\end{figure}

\paragraph{Varying background threshold rate}
We investigated the detection fraction, varying the nominal background density rate that was fixed in Figure~\ref{fig:detection_efficiency_histograms}. We focus on UMES from 50 to 80 and calculate the fraction of detected injections in this range for different threshold background densities, shown in the left panel of Figure~\ref{fig:roc}. The right panel of Figure~\ref{fig:roc} shows the detection fraction as a function of the measured false positive rate, which is the integral of the background distribution beyond the detection threshold.

%% file: sec_confirmed_koi.tex
% Some KOI were rejected by the search because the gaia-kepler crossmatch catalog does not have infor about them
% Two tarhets had 2 relevant KOI each, so far only one per each was searched.

% *** MAKE COMPARISON PLOTS WHICH ARE SHOWING THEIR MES, THEIR SNR (god knows what it is), MY MES, MY P-PLANET. DISCUSS LOSS DUE TO MISSING DATA.
This section checks whether the pipeline detects the Confirmed \textit{Kepler} planets. We selected a subset of KOI marked as Confirmed by the \textit{NASA Exoplanet Archive}~\citep[][]{cumulative_koi_table}.
% \footnote{Cumulative \textit{Kepler} KOI Table DOI 10.26133/NEA4}
% \footnote{https://doi.org/10.26133/nea4}
We only used faint KOI with \textit{Kepler} MES$\leq$ 15 and having \textit{Gaia} data available in the \textit{Gaia}-\textit{Kepler} Cross-match table~\citep[][]{bedell_gaia_kepler_fun}. 

\begin{figure}[h!]
    \centering
    \includegraphics[width=0.47\textwidth]{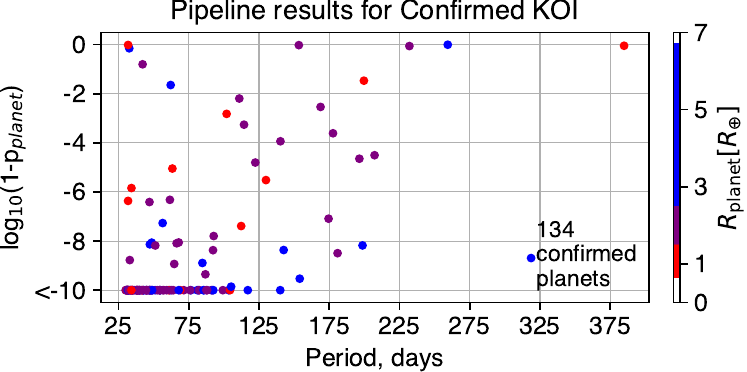}
    \caption{$P_{\text{planet}}$ for confirmed \textit{Kepler} candidates as function of \textit{Kepler} period. Color coding represents planetary radii in units of Earth radius.}
    \label{fig:p_planet_confirmed}
\end{figure}

We operate the pipeline and save IMES and UMES for each target (the results can be seen in Figure~\ref{fig:imes_umes_kmes_comparison} in the Appendix). The IMES scores are systematically lower than \textit{Kepler} MES scores because of the non-Gaussianity correction. The UMES scores are comparable to \textit{Kepler} MES scores, up to statistical noise. However, detectability is defined according to the $P_{\text{planet}}$ score, therefore we run the statistical significance estimation procedure~(Section~\ref{sec:significance}) for each target and report the results in Figure~\ref{fig:p_planet_confirmed}.

As can be seen, most of the processed 134 faint confirmed planets have $P_{\text{planet}}>99\%$. However, there are several cases having $P_{\text{planet}}<50\%$, and they are discussed below. A summary for the targets with $P_{\text{planet}}<99\%$ is provided in Table~\ref{table:unsuccessful_kois}.

The main reason for the low $P_{\text{planet}}$ is the non-Gaussianity correction punishing very deep transits. 
As can be seen in Table~\ref{table:unsuccessful_kois}, the uncorrected UMES score for most of the targets is greater than 60, which usually corresponds to large $P_{\text{planet}}$. 

The non-detection due to the non-Gaussianity correction happens when the number of transits is small ($\leq5$), so that each individual SES is large.  The number of transits can be large either because of a long period, or because of a low number of the observed \textit{Kepler} quarters. Generally, an SES of SNR$\sim4.5$ is already a deep transit, getting a significant non-Gaussianity correction. Most of the considered confirmed planets are in this regime, so they show a significant SNR loss. The SES scores larger than 7 are considered too deep for the mode in which our pipeline operates; therefore, such planets may be masked as outliers. Additionally, transits in \textit{Kepler} quarters 0 and 17 were not processed by our pipeline, so if there are transits there, they would be missed and would not contribute to the total UMES or IMES.

\begin{deluxetable*}{ccccccccc} % Use 'deluxetable*' for two-column width, 'deluxetable' for single column
\tablecaption{Confirmed KOI that got $P_{\text{planet}}<99\%$ \label{tab:example}}
\tablehead{
\colhead{KIC} & \colhead{KOI name} & \colhead{$P_{\text{planet}}$} & \colhead{$\text{SES}_{\text{KOI}}$} & \colhead{$N_{\text{tr, KOI}}$} & \colhead{$P_{\text{KOI}}$ [days]} & \colhead{${\text{MES}^2}_{\text{KOI}}$} & \colhead{UMES} & \colhead{IMES}
}
\startdata
011037818 & Kepler-1638 b & 0.01 & 7.7 & 5 & 259.3 & 119.9 & 68.4 & 44.8 \\ 
006368175 & Kepler-1703 c & 0.03 & 4.1 & 41 & 31.8 & 59.1 & 47.0 & 45.6 \\ 
009896558 & Kepler-937 c & 0.04 & 5.1 & 7 & 153.3 & 100.3 & 66.4 & 47.4 \\ 
008311864 & Kepler-452 b & 0.10 & 4.1 & 4 & 384.8 & 57.8 & 73.7 & 49.3 \\ 
010055126 & Kepler-311 d & 0.12 & 7.8 & 5 & 232.0 & 156.4 & 107.9 & 49.4 \\ 
006221385 & Kepler-1641 c & 0.29 & 7.2 & 5 & 32.7 & 185.8 & 110.1 & 43.7 \\ 
008150320 & Kepler-55 c & 0.84 & 7.3 & 28 & 42.1 & 169.1 & 152.2 & 55.2 \\ 
009205938 & Kepler-1126 c & 0.97 & 6.2 & 7 & 199.7 & 123.0 & 92.9 & 63.9 \\ 
008745553 & Kepler-1633 b & 0.98 & 4.8 & 8 & 62.1 & 92.8 & 90.4 & 63.0 \\   
\enddata
\tablecomments{
$\text{SES}_{\text{KOI}}$ is the \textit{Kepler} maximal single-event statistic;\\
$N_{\text{tr, KOI}}$ is the \textit{Kepler} number of transits;\\
$P_{\text{KOI}}$ is the orbital period of the KOI;\\
$\text{MES}_{\text{KOI}}$ is the \textit{Kepler} multiple-event statistic.\\
All the \textit{Kepler} values originate from the Cumulative KOI table from \textit{NASA Exoplanet Archive}~\citep[][]{cumulative_koi_table} 
}
\label{table:unsuccessful_kois}
\end{deluxetable*}

We note that target \textit{KIC009896558} has only 6 valid transits in Q1-Q16. Additionally, its transits exhibit a significant SES variation, so that the deepest transits get an enhanced non-Gaussianity correction. The same issue of varying SES appears for target \textit{KIC011037818}.

Two targets experience significant transit timing variations (TTV): \textit{KIC006368175}, \textit{KIC008150320}. Due to vetting~(Section~\ref{sec:veto_ses}), in-transit epochs that are significantly offset from the transit center can be masked as non-transit-like and not contribute to the MES score. Therefore, in order to include events with significant TTV in the strictly periodic search, the vetting should be modified accordingly.

% K03503.02 Kepler-1703 c \textit{KIC006368175}
In addition, as reported by the \textit{NASA Archive}~\citep[][]{cumulative_koi_table}, the confirmation of the \textit{KIC006368175} candidate was done by \citep[][]{jontof_hutter_2021} using the TTV method. The \textit{Kepler} MES is as low as 7.69, and the disposition score is not provided for this candidate. It may be that the MES alone is not sufficient to make a significant detection. 

The target \textit{KIC008150320} is a 5-planet system, and we masked the remaining 4 planets while searching for the candidate of interest. Since the masked candidates have short periods, many data points were affected by the mask, and it could decrease the score of \textit{Kepler-55 c} that we targeted in our search.

% It is kepler mes 7.69, disposition score not provided. has large ttvs [Jontof-Hutter 2021]. I do not see ttvs there, but i do not see anything at all. river diagram is blurred, perhaps it is ttvs
% https://iopscience.iop.org/article/10.3847/1538-3881/abd93f/meta#ajabd93ff15

% K07016.01 Kepler 452-b
Finally, the candidate for \textit{KIC008311864} (Kepler 452-b) was first announced as a confirmed planet, but then re-evaluated \citep[][]{mullally_2018_should_not_be_confirmed, burke_2019}, and is still controversial.

%% file: ap_templates_shape_losses.tex
This Appendix discussed two questions:\\
1) The influence of the template shape used for detection on transit detectability;\\
2) How this influence changes when the noise is correlated.

The expected value of the detection statistic in Equation~\ref{eq:stat_snr} equals the SNR of the transit, assuming that the detection is made with a template describing precisely the signal shape. Consider the following case: we use template $\mathbf{h}_{(1)}$ for detection in data that contains a transit of a different shape $\mathbf{h}_{(2)}$. In this case, the detected SNR, according to Equation~\ref{eq:stat_snr}, is
\begin{align}
    \text{SNR}_{\text{detected}}
    =\frac{\mathbf{h}_{(1)}^{T}C^{-1}
    \mathbf{h}_{(2)}}
    {\sqrt{\mathbf{h}_{(1)}^{T}C^{-1}\mathbf{h}_{(1)}}}
    =\frac{\mathbf{h}_{(1),w}^{T}
    \mathbf{h}_{(2),w}}
    {\sqrt{\mathbf{h}_{(1),w}^{T}\mathbf{h}_{(1),w}}}.
    \label{eq:snr_templates}
\end{align}
Here, we used the definition for the whitened templates (subscript $w$), which are templates to which the whitening filter 
was applied (Section~\ref{sec:fourier}).

If the correct template $\mathbf{h}_{(2)}$ was used, then we would detect the full SNR of the transit. With the incorrect template, we detect only part of it, which is 
\begin{align}
    \frac{\text{SNR}_{\text{detected}}}
    {\text{SNR}_{\text{full}}}
    =\frac{\mathbf{h}_{(1),w}^{T}
    \mathbf{h}_{(2),w}}
    {\sqrt{\mathbf{h}_{(1),w}^{T}\mathbf{h}_{(1),w}}
    \sqrt{\mathbf{h}_{(2),w}^{T}\mathbf{h}_{(2),w}}
    }
    \equiv \cos \left(\mathbf{h}_{(1)}, \mathbf{h}_{(2)} \right).
    \label{eq:snr_loss_template_cos_angle}
\end{align}
Here, we introduced the definition of the cosine angle between the two templates. It is a metric of template closeness that shows how similar their shapes are and indicates the part of SNR that can be recovered by applying one template to the transit shaped by the other template. When the two templates are identical, the metric equals unity.

We emphasize that the definition of the metric includes the inverse covariance matrix of the noise. Two transit models can be close in white noise but have a significant angle between them when the noise is correlated.

\begin{figure}[h]
    \centering
    \includegraphics[width=0.94\textwidth]{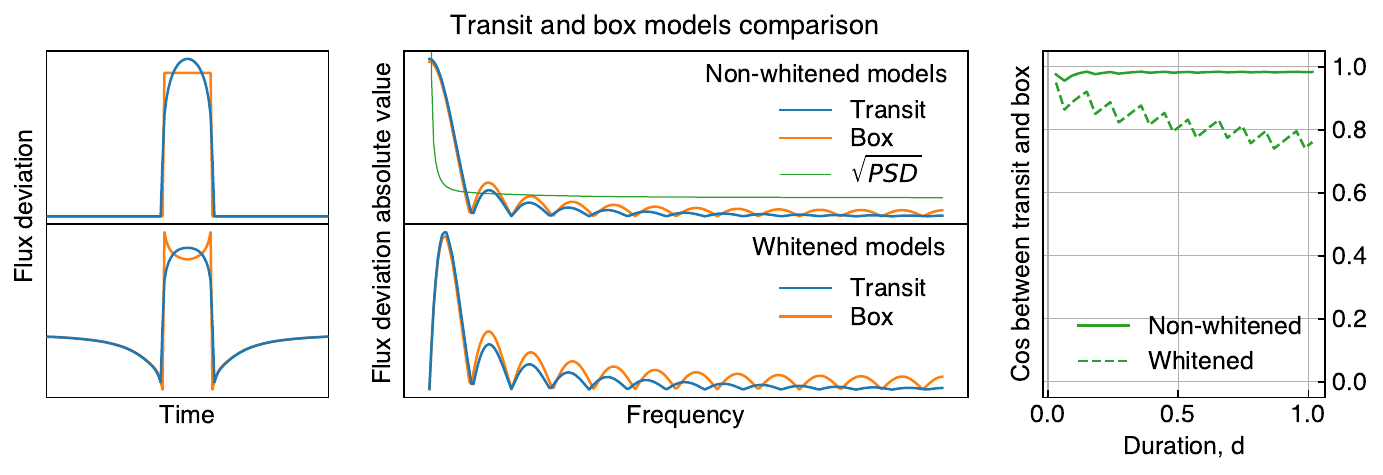}
    \caption{Illustration of a box-shaped and a smooth templates match in the case of white noise (upper panel) and "red" correlated noise (lower panel)
    \textit{Left panel}: Examples of box template and smooth template in the time domain. The top plot shows the non-whitened template which would be used if the noise is white. The bottom plot shows the whitened templates which have to be used if the noise is correlated.
    \textit{Middle panel}: The two templates in the Fourier domain. The green line shows the fiducial PSD that was used for whitening in this example.
    \textit{Right panel}: The cosine angle closeness metric between the templates as a function of their duration. The solid line corresponds to the match of non-whitened templates, or to the case when the noise is not correlated. The dashed lines stand for the whitened templates which have to be used if the noise is correlated.}
    \label{fig:whitened_transit_box}
\end{figure}

From Equation~\ref{eq:snr_loss_template_cos_angle}, we conclude that if a mismatching template is used in the search, loss in SNR may occur, resulting in a missed detection. The size of this effect depends on the covariance matrix of the noise. Specifically, the "red" noise power spectrum typical for the \textit{Kepler} light curves increases the mismatch between templates. To understand this, we consider the Fourier image of a template, shown in the middle panel of Figure~\ref{fig:whitened_transit_box}. In the Fourier domain, most of the template power is concentrated at low frequencies, in the main lobe defined by the transit duration. The high-frequency tail carries information about the shape details. When the noise is white, the whitening filter is flat, and most of the power comes from the main lobe. Therefore, if two templates have the same duration making their Fourier main lobes similar, their cosine angle will be close to unity. 

In the case of a "red" power spectrum, the whitening filter suppresses the lowest frequencies, reducing the amount of power in the main lobe. It enhances the contribution of the tail making the shape difference influence the mismatch.

This mechanism explains why the popular method of using box-shaped templates can lead to a loss in SNR when the noise is correlated. Figure~\ref{fig:whitened_transit_box} provides an example of a physical smooth transit shape and a box template of the same duration. They are very similar in the main lobe, but the box-shaped template has a heavier high-frequency tail due to its sharp edges. As a result, when the main lobe is suppressed by the whitening filter, the match between the two shapes decreases. 

The right panel of Figure~\ref{fig:whitened_transit_box} compares the cosine angle between these two templates as a function of their duration for the case of white noise (no whitening is needed) and correlated noise (templates have to be whitened). As can be seen, the mismatch is insignificant when no whitening is needed, but can lead to noticeable loss in SNR when whitening is required. The effect becomes more prominent for long durations when the power of the template gets more focused in the low frequencies.

In order to avoid extra losses in SNR, smooth templates should be used instead of box-shaped ones. The exact shape of the smooth template is not crucial, as long as it does not have a heavy tail at long frequencies arising from the sharp edges.  

For reference, if the cosine angle between two templates is 0.8, then for the true SNR$^2=70$ (which is well-detectable in the search), the pipeline would detect only SNR$^2=44$, which is below the detection threshold. 

%% file: ap_mes_math.tex
In this appendix, we show how the multiple-transit statistic (MES) can be expressed in terms of the single-transit statistic (SES) of the individual transits.

Assume that the data vector from Equation \ref{eq:data_model} can be written as a concatenation of two parts, and each of them contains one transit,
\begin{align}
    \mathbf{d}=
    \left(\mathbf{d}_{1}\;\mathbf{d}_{2}\right)
    =\left(\mathbf{n}_{1}\;\mathbf{n}_{2}\right)
    +A\left(\mathbf{h}_{1}\;\mathbf{h}_{2}\right),
    \label{eq:concatenation}
\end{align}
where $\mathbf{d}$, $\mathbf{n}$, and $\mathbf{h}$ are the data, noise and template vectors, and $A$ is the amplitude coefficient.

Assuming that the correlation length of the noise is small compared to the duration of the two parts of the data, we can neglect the correlations between the two parts and write the data covariance matrix in a block-diagonal form. Then, the matched filtering score (numerator of Equation~\ref{eq:stat_snr}) can be split into two terms,
\begin{equation}
\begin{split}
    \left(\mathbf{h}_{1},\mathbf{h}_{2}\right)^{T}C^{-1}\left(\mathbf{d}_{1},
    \mathbf{d}_{2}\right)&=\left(\begin{array}{cc}
\mathbf{h}_{1}^{T} & \mathbf{h}_{1}^{T}\end{array}\right)\left(\begin{array}{cc}
C_{1}^{-1} & 0\\
0 & C_{2}^{-1}
\end{array}\right)\left(\begin{array}{c}
\mathbf{d}_{1}\\
\mathbf{d}_{2}
\end{array}\right)
\\&=\mathbf{h}_{1}^{T}C_{1}^{-1}\mathbf{d}_{1}+\mathbf{h}_{2}^{T}C_{2}^{-1}\mathbf{d}_{2}.
\label{eq:matched_filter_two_parts}
\end{split}
\end{equation}
Applying this to the numerator and the denominator of the test statistic (Equation \ref{eq:stat_snr}) gives that the multiple-transit statistic (MES) can be expressed in terms of the single-transit statistic (SES),
\begin{align}
\begin{split}
    \rho_{\mathrm{MES}}=
    \frac{\mathbf{h}_{1}^{T}C_{1}^{-1}\mathbf{d}_{1}
    +\mathbf{h}_{2}^{T}C_{2}^{-1}\mathbf{d}_{2}}
    {\sqrt{\mathbf{h}_{1}^{T}C_{1}^{-1}\mathbf{h}_{1}
    +\mathbf{h}_{2}^{T}C_{2}^{-1}\mathbf{h}_{2}}}
    =
    \frac{\rho_{\mathrm{SES},1}
    +\rho_{\mathrm{SES},2}}
    {\text{Var}[\rho_{\mathrm{SES},1}]
    +\text{Var}[\rho_{\mathrm{SES},2}]}.
    \label{eq:snr_concatenated}
\end{split}
\end{align}
This equation provides the mathematically correct way to sum SES, adding separately the numerators and the denominators. The resulting MES has units of SNR. Generalizing Equation~\ref{eq:snr_concatenated} for multiple transits, one gets Equation~\ref{eq:mes_formula}, the formula for getting multiple-transit statistic from single-transit statistics.

We note that the vector $\mathbf{h}$ had a unit norm, whereas vectors $\mathbf{h}_1$ and $\mathbf{h}_2$ do not.
In the general case, the individual transits may have different amplitudes, and then norms of $\mathbf{h}_{1,2}$ will be different. 
However, since we re-normalized the data quarters by their mean flux, the expected transit depth is the same for all the transits. Then, the template norm cancels out in Equation~\ref{eq:snr_concatenated}, and the templates $\mathbf{h}_{1,2}$ can be assumed to be normalized.

%% file: ap_psd_estimation.tex
This section provides details about the power spectral density (PSD) estimator and discusses the errors of the PSD measurement.
\paragraph{PSD estimator}
Consider a correlated Gaussian noise $\mathbf{n}$ having PSD $\mathbf{S}$. The absolute value squared of its Fourier image $|\hat{\mathbf{n}}|^2$ will follow the exponential distribution

\begin{equation}
    |\hat{n}(f)|^{2}\sim\frac{1}{S(f)}\exp\left(-\frac{|\hat{n}(f)|^{2}}{S(f)}\right). 
    \label{eq:power_exp_distr}
\end{equation}
Assume we are provided $k$ independent realizations of such data, 
$\left\{
|\hat{n}(f)|_j
\right\}_{j=0}^k$. Their likelihood function yields
\begin{equation}
    \log\left[\mathcal{L}\left(\left\{ \left|\hat{n}(f)\right|_{j}^{2}\right\} _{j=0}^{k} \Big| S(f)\right)\right]=\log\left[\prod_{j}^{k}\frac{1}{S(f)}e^{-\frac{\left|\hat{n}(f)\right|_{j}^{2}}{S(f)}}\right]=\log\left[\frac{1}{S(f)^{k}}e^{-\frac{\sum_{j}^{k}\left|\hat{n}(f)\right|_{j}^{2}}{S(f)}}\right]=-k\log S(f)-\frac{k\left\langle \left|\hat{n}(f)\right|^{2}\right\rangle}{S(f)},
\end{equation}
where $\left\langle .\right\rangle$ denotes the average of $k$ samples. From here, the maximum-likelihood estimator of $S(f)$ can be obtained:
\begin{equation}
    \frac{d}{dS(f)}\log\left[\mathcal{L}\right]
    =0,
\end{equation}
which yields
\begin{equation}
    S_{\text{MLE}}(f)=\left\langle \left|\hat{n}(f)\right|^{2}\right\rangle,
    \label{eq:psd_mle_est}
\end{equation}
meaning that the PSD estimator is the average of the noise realizations at the corresponding frequency.  

However, in the real search, we are provided with only one sample of data. To create several samples to average over, data is sliced into pieces, ideally longer than the correlation length of the noise. Averaging over multiple slices reduces the statistical error of the estimator. However, it also lowers the frequency resolution defined by the length of the slice. These two effects result in PSD estimation errors that will be discussed below, together with SNR losses associated with them.

\paragraph{Statistical error of PSD estimation}
The standard deviation of the distribution in Equation \ref{eq:power_exp_distr} is equal to its expected value $S_i$. When averaging over $k$ measurements, the standard deviation decreases as $S_i/\sqrt{k}$. If one quarter of \textit{Kepler} has length $N$, and the length of the slice used in the PSD estimator is $n$, then $k\sim N/n$, and the standard deviation of the PSD estimator \mbox{is 
$\sim S_i \sqrt{n/N}$}. 

\paragraph{Rounding error of PSD estimation}
The resolution of the PSD estimation is also defined by the slice length $n$. The frequency resolution of the resulting PSD estimator is $\Delta f=1/(n \Delta t)$, where $\Delta t$ is the data sampling time step. When applying the PSD for whitening, all the data in the frequency bin $\Delta f$ will be multiplied by the same coarse-grained value of the estimated PSD. The average squared PSD error in the frequency bin $f_i$ is
\begin{equation}
    \frac{1}{\Delta f}\int_{-\frac{\Delta f}{2}}^{\frac{\Delta f}{2}}df\left(S\left(f_{i}+f\right)-S\left(f_{i}\right)\right)^{2}
    \approx\frac{1}{\Delta f}\left(\frac{dS}{df}\Big|_{f_{i}}\right)^{2}\int_{-\frac{\Delta f}{2}}^{\frac{\Delta f}{2}}dff^{2}=\left(\frac{dS}{df}\Big|_{f_{i}}\right)^{2}\frac{\left(\Delta f\right)^{2}}{12}.
\end{equation}
As a result, the PSD estimation error will behave \mbox{like 
$\propto \frac{dS}{df} \frac{1}{n\Delta t} $}.

All the power in the lowest frequencies $f<\Delta f$ will be lost because the coarse resolution of the estimator does not allow measuring PSD there. In order to prevent contamination from these frequencies, they undergo detrending (Sec.~\ref{sec:detrending}). The bandwidth of the signal lost due to detrending will decrease with growing $n$ \mbox{as $\frac{1}{n\Delta t}$}.

\paragraph{SNR loss due to PSD estimation errors} 
All the errors in PSD measurement result in a loss in SNR. Assume the PSD measurement error is $\epsilon(f)$, so that the measured PSD yields
\begin{equation}
    S(f)=S_{\text{true}}(f) (1-\epsilon(f)).
    \label{eq:psd_error}
\end{equation}
As calculated in \citep[][]{zackay_2019_non_gaussian}, the difference $\Delta \text{SNR}$ between the detected and the true SNR to the leading order is given by

\begin{equation}
    \frac{\Delta \text{SNR}}{\text{SNR}_{\text{true}}}
    \approx
    -\sum_{f}I\left(f\right)\epsilon^{2}\left(f\right)
    +\left(\sum_{f}I\left(f\right)\epsilon\left(f\right)\right)^{2},
    \label{eq:snr_loss_psd}
\end{equation}
where
\begin{equation}
    I\left(f\right)=
    \frac{|\hat{h}_w(f)|^2}
    {\sum_f |\hat{h}_w(f)|^2},
    \label{eq:renorn_whitened_template}
\end{equation}
with $\hat{h}_w$ being the whitened template. We note that the loss in SNR is quadratic in PSD error only when the score variance correction is used (Section~\ref{sec:matched_filter}). 

Substituting different PSD measurement errors in this equation, one can assess the values of the SNR loss due to these errors.

\paragraph{Summary of loss factors}
Some of the PSD errors grow with the length of the estimator slice $n$, while others decay. As obtained above,

Frequency cut: $f_{\text{min}} \propto 1/n$; %$f_{\text{min}}=\frac{1}{n\Delta t}$

Statistical error: $\propto \sqrt{n}$; %$\propto \sqrt{\frac{n}{N}}$

\begin{figure}[h!]
    \centering
    \includegraphics[width=0.94\textwidth]{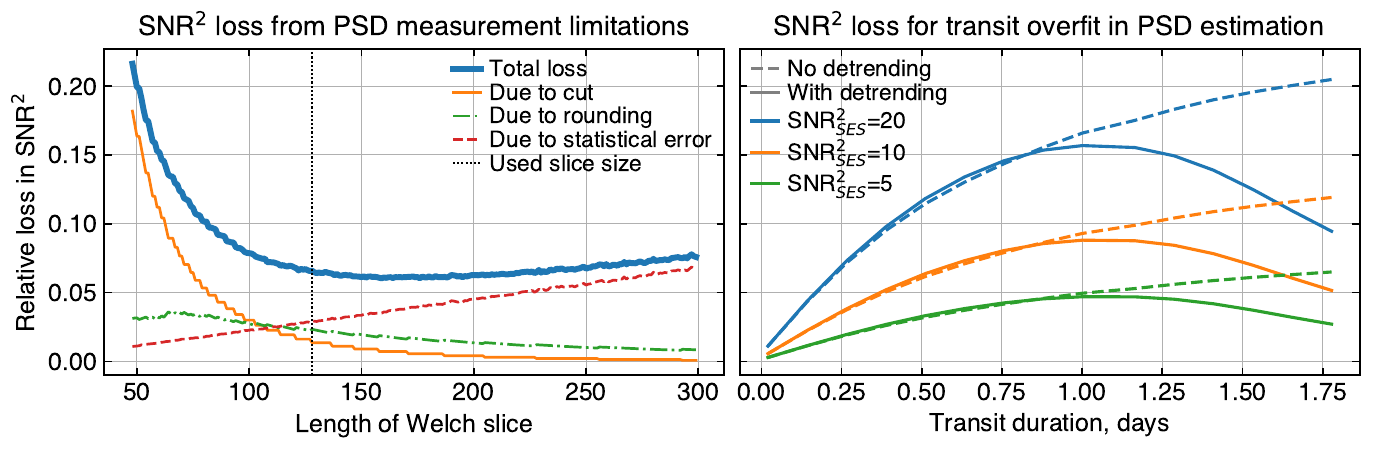}
    \caption{\textit{Left panel}: Estimation (performed on simulated noise) of SNR losses due to the PSD measurement errors as a function of the PSD estimator slice length. Different lines represent losses due to different sources of PSD measurement error discussed in the text. The thick blue line shows their sum, which is the total SNR loss.
    \textit{Right panel}: Loss in $\text{SNR}^2$ caused by planetary transit biasing the PSD measurement. Different colors denote cases of different SNR of the overfitted transit. Dashed lines correspond to the case when a preliminary detrending (high-pass filtering) was performed. For long transit durations, detrending already removes a significant part of their SNR concentrated in low frequencies, therefore the residual loss from the PSD overfitting becomes less significant.}
    \label{fig:psd_error_losses_nperseg}
\end{figure}

Rounding error: $\propto 1/n $. %$\propto\frac{1}{\sqrt{12}} \frac{dS}{df} \frac{1}{n\Delta t} $
\\
It is expected that there should be an optimal $n$ minimizing the total SNR loss resulting from the three factors.
To find this length, we explore the SNR loss numerically using an effective PSD obtained by averaging Fourier amplitudes of many \textit{Kepler} light curves. Figure~\ref{fig:psd_error_losses_nperseg} presents the SNR losses estimated using Equation \ref{eq:snr_loss_psd} as a function of the PSD estimator slice length $n$. As can be seen, the total loss has a minimum at $n\sim150$. In the pipeline, the value of $n=128$ was selected for convenience, since it is the closest power of 2.

We note that the chosen $n$ should be larger than the longest template duration. If the template is too long, most of its support in the Fourier domain will fall below the PSD resolution limit, and most of its power will be lost.

\paragraph{Overfitting the planets}
If the star hosts a planet, some data slices used in the PSD estimation might contain transits. Sufficiently deep transits can bias the PSD measurement. As a result, part of the detected signal may get canceled by the whitening filter. The significance of this effect is assessed below.

For long orbital periods, one \textit{Kepler} quarter can contain at most one transit. If it happens, one of the $k=N/n$ slices will contain the signal, whose power will be suppressed by a factor of $1/k$ after averaging. The resulting measured power, contaminated by the transit contribution $A\hat{h}(f)$ (consult Equation~\ref{eq:data_model} for definitions) yields
\begin{equation}
    S_{p}(f)=S_{
    \text{}}(f)+\frac{1}{k}A^{2}\left|\hat{h}(f)\right|^{2}
    =S_{
    \text{}}(f)\left(1
    +\frac{\text{SNR}^2_{\mathrm{SES}}(f)}{k}\right),
\end{equation}
where $S_p$ is the PSD estimate contaminated by the planet, and $S$ is the estimate that would be made in the absence of the planet. For a typical multi-transit $\text{SNR}^2$ of 60, the single-transit $\text{SNR}^2_{\mathrm{SES}}$ can vary between about 2 and 20, depending on the number of transits. This power is spread among several frequency bins $\text{SNR}^2_{\mathrm{SES}}(f_i)$, depending on the transit duration. Varying the overfitted transit duration, we calculate the expected $\text{SNR}^2$ of the signal (Equation~\ref{eq:snr_definition}) with the true PSD and the biased PSD, plotting the result in Figure~\ref{fig:psd_error_losses_nperseg}. We also do it for the case when a preliminary detrending (Section~\ref{sec:detrending}) was made, removing all power from the lowest frequencies. 

This SNR loss can get significant for long transits with very long periods. It can be mitigated by using outlier clipping during PSD estimation. However, the PSD error is not the dominant source of SNR loss, as discussed in Section~\ref{sec:snr_recovery_fraction}. The 3-transit events cannot be detected by the pipeline primarily due to the noise non-Gaussianity correction, making the PSD error effect secondary.

%% file: ap_veto_parts.tex
In this appendix, we derive a statistical test ensuring that the template describes well the shape of the data. It is done by splitting the template into several segments and comparing the independently measured transit amplitudes for these segments. If the amplitudes are the same, then the template describes well the shape of the data. The possibility to select the number of parts allows for control of the number of degrees of freedom in the distribution of the test statistic.

Consider the data vector $\mathbf{d}$ and a tentative template $\mathbf{h}$. We split the template into $N$ non-overlapping segments and create $N$ sub-templates $\mathbf{h}_i$ that equal to zero everywhere except for the corresponding segment, resulting in $\sum_{i}\mathbf{h}_{i=1}^N=\mathbf{h}$. Segments are designed to have the same expected part of SNR$^2$ of the overall signal. In the Fourier domain, splitting is done differently, as will be clarified at the end of this Appendix.

We formulate a binary hypothesis test for this problem. In the "good" case $\mathcal{H}_0$, data is well described by the template $\mathbf{h}$ so that all the sub-templates amplitudes are the same. In the alternative case, all the sub-templates $\mathbf{h}_i$ have different amplitudes:
\begin{align}
\begin{split}
    &\mathcal{H}_{0}:\,\mathbf{d}	
    =A\mathbf{h}+\mathbf{n},
    \\&\mathcal{H}_{1}:\,\mathbf{d}	=\sum_{i}A_{i}\mathbf{h}_{i}+\mathbf{n},
\end{split}
\end{align}
where the noise $\mathbf{n}$ is assumed to be Gaussian. Writing the log-likelihood ratio test~(one may consult Appendix~\ref{ap:basic_math} for a reminder) results in the test statistic
\begin{align}
    \mathcal{T}&=\sum_{i}A_{i}\left\langle \mathbf{d},\mathbf{h}_{i}\right\rangle -A\left\langle \mathbf{d},\mathbf{h}\right\rangle.
\end{align}
Here, the angular brackets denote the inner product with weights of noise inverse covariance matrix $C$, \mbox{$\left\langle \mathbf{d},\mathbf{h}\right\rangle = \mathbf{d} C^{-1} \mathbf{h}$.} If the noise is standard Gaussian, then \mbox{$\left\langle \mathbf{d},\mathbf{h}\right\rangle = \mathbf{d} \mathbf{h}$.}

However, the amplitudes $A_i$ and $A$ are not known, so their maximum-likelihood estimates should be used. To find $A_i$ corresponding to the extremum of the likelihood, we solve
\begin{align}
    \frac{\partial}{\partial A_{i}}\left|\mathbf{d}-\sum_{i}A_{i}\mathbf{h}_{i}\right|^{2}=0,
\end{align}
which yields the solution
\begin{align}
    \hat{A}_{i}=\sum_{j}\left(H^{-1}\right)_{ij}\left\langle \mathbf{d},\mathbf{h}_{j}\right\rangle,
\end{align}
where $H$ is a matrix with $H_{ij}=\left\langle \mathbf{h}_{i},\mathbf{h}_{j}\right\rangle $.

The maximum-likelihood estimate for $A$ is
\begin{align}
    \hat{A}=\frac{\left\langle \mathbf{d},\mathbf{h}\right\rangle }{\left\langle \mathbf{h},\mathbf{h}\right\rangle }.
\end{align}
Thus, the resulting test statistic yields
\begin{align}
    \mathcal{T}&=\sum_{i,j}\left\langle \mathbf{d},\mathbf{h}_{i}\right\rangle \left(H^{-1}\right)_{ij}\left\langle \mathbf{d},\mathbf{h}_{j}\right\rangle -\frac{\left\langle \mathbf{d},\mathbf{h}\right\rangle ^{2}}{\left\langle \mathbf{h},\mathbf{h}\right\rangle }.
    \label{eq:veto_parts_stat_general}
\end{align}
If the template parts are orthogonal, $\left\langle \mathbf{h}_{i},\mathbf{h}_{j}\right\rangle \propto\delta_{ij}$, which is the case for the implemented splitting, then
\begin{align}
    \mathcal{T}&=\sum_{i}\frac{\left\langle \mathbf{d},\mathbf{h}_{i}\right\rangle ^{2}}{\left\langle \mathbf{h}_{i},\mathbf{h}_{i}\right\rangle }-\frac{\left\langle \mathbf{d},\mathbf{h}\right\rangle ^{2}}{\left\langle \mathbf{h},\mathbf{h}\right\rangle }.
    \label{eq:veto_parts_stat_orthogonal}
\end{align}
This result can be written as
\begin{align}
    \mathcal{T}&=\sum_{i}\left\langle \mathbf{d},\mathbf{h}_{i}\right\rangle ^{2}\frac{1}{\left\langle \mathbf{h}_{i},\mathbf{h}_{i}\right\rangle }-\sum_{i,j}\frac{\left\langle \mathbf{d},\mathbf{h}_{i}\right\rangle \left\langle \mathbf{d},\mathbf{h}_{j}\right\rangle }{\left\langle \mathbf{h},\mathbf{h}\right\rangle },
\end{align}
which is a quadratic form for a vector $\{\mathbf{d},\mathbf{h}_{i} \}_{i=1}^N$. Under $\mathcal{H}_0$, it is a vector of normal random variables. As mentioned above, the sub-templates are designed to have the same expected SNR$^2$, meaning that we can set
\begin{align}
    \left\langle \mathbf{h},\mathbf{h}\right\rangle=1,\;\;\;
    \left\langle \mathbf{h}_i,\mathbf{h}_i\right\rangle=1/N.
\end{align}
Therefore, the associated matrix of the quadratic form takes shape $1/N\cdot I - U$, where $I$ is the unit matrix, and $U$ is a matrix full of ones. 
This quadratic form can be diagonalized, having $N-1$ degenerate eigenvalues and one zero eigenvalue. Diagonalizing the form, we see that the detection statistic is a sum of $N-1$ $\chi^2(1)$-distributed random variables, thus
\begin{align}
    \mathcal{T}|\mathcal{H}_{0}\sim\chi^{2}\left(N-1\right).
\end{align}
This provides a test which has $N-1$ degrees of freedom in the background distribution, where $N$ can be chosen arbitrarily.

\paragraph{Correctness of inner product} We note that the correct noise covariance matrix should be used in the inner product computation. Since the test is performed on whitened vectors, the covariance matrix should be unity. However, since the noise was detrended (Section\ref{sec:detrending}), it will lack power at the lowest detrended frequencies. In addition, the whitening is not exact due to the PSD measurement error. This issue was addressed in Section~\ref{sec:matched_filter} by normalizing the score with its empirically calculated variance. We employ this solution here, normalizing terms of  Equation~\ref{eq:veto_parts_stat_orthogonal} with their measured variances before subtracting them.

\paragraph{Splitting the template in the Fourier domain} The test is written for general vectors and can be applied both in the time domain and in the Fourier domain.
In the time domain, we divided the template into non-overlapping segments resulting in orthogonal sub-templates providing a simplified form of the statistic (Equation~\ref{eq:veto_parts_stat_orthogonal}).
The same division in the Fourier domain is problematic since the sharp boundaries of the sub-templates result in them having infinite support in the time domain. Therefore, in the Fourier domain, we split the template into overlapping segments with smooth boundaries. Then, we use Equation~\ref{eq:veto_parts_stat_general} for the test statistic.

%% file: ap_transit_depth.tex
This appendix provides details about MES veto of transit depth consistency. 

The goal of this veto is to check whether the individual SES amplitudes are consistent with being caused by the same transiting planet. This goal could be achieved by implementing an equivalent of the transit shape test (Section~\ref{ap:amplitude_consistency_veto}) with template segments replaced by individual transits. The “good shape,” in this case, is when all the transits have the same depth. 

However, this test is less convenient for the considered problem. If the number of transits is large (for example, 10), the test will have a large number of degrees of freedom. In Section~\ref{ap:amplitude_consistency_veto}, the number of degrees of freedom was reduced by combining individual points into segments. This approach was convenient because individual points of one transit behave smoothly, enabling a common amplitude fit. For separate transits, amplitudes can be completely independent, making it not helpful to combine different transits and fit together. 

If the number of transits is small (for example, 3), the large number of degrees of freedom is not an issue. However, $\chi^2$ testing is still less powerful than binary hypothesis testing. For a small number of transits, the scenario that we want to test for is when the entire MES is dominated by one transit, and the others are consistent with zero. Since the MES that we are investigating are small, this is possible even after correcting for non-Gaussianity.

Therefore, we design the following test. We select the strongest transit and test between the two hypotheses: 
\begin{align}
    \mathcal{H}_{0}:&\; \text{The SNR from the remaining transits is consistent with this amplitude;}
    \\
    \mathcal{H}_{1}:&\; \text{The SNR of the remaining transits is consistent with zero.}
\end{align}

As follows from Equations~\ref{eq:ses_i} and~\ref{eq:ses_variance_calculated}, the SES score has a distribution
\begin{align}
    \rho_{\text{SES},i}\sim\mathcal{N}\left(A\eta_{i},\,\sigma=\sqrt{\eta_{i}}\right),
\end{align}
where 
\begin{align}
    \eta_{i}\equiv\mathbf{h}_{i}^{T}C_{i}^{-1}\mathbf{h}_{i}.
\end{align}
Say we select transit number 1 and estimate its amplitude (Equation~\ref{eq:amplitude_estimator}) as
\begin{align}
    \hat{A}_{1}=\frac{\rho_{\text{SES,1}}}{\eta_{1}}.
\end{align}
The SNR of the remaining transits yields
\begin{align}
    \text{SNR}_{2...N}=\frac{\sum_{i=2}^{N}\rho_{\text{SES},i}}{\sqrt{\sum_{i=2}^{N}\eta_{i}}}.
\end{align}
Under the two considered hypotheses, this value behaves as
\begin{align}
    H_{0}&:\text{SNR}_{2...N}=A_{1}\sqrt{\sum_{i=2}^{N}\eta_{i}}+\mathcal{N}\left(0,1\right),
    \\H_{1}&:\text{SNR}_{2...N}=\mathcal{N}\left(0,1\right).
\end{align}
A centered and normalized log-likelihood ratio test statistic for this model reads
\begin{align}
    \mathcal{T}=\frac{\frac{\rho_{\text{SES,1}}}{\eta_{1}}-\frac{\sum_{i=2}^{N}\rho_{\text{SES},i}}{\sum_{i=2}^{N}\eta_{i}}}{\sqrt{\frac{1}{\eta_{1}}+\frac{1}{\sum_{i=2}^{N}\eta_{i}}}}.
\end{align}
This test statistic will have a standard normal distribution if the true amplitudes of all the transits are the same. 

\paragraph{Look-elsewhere effect} The test was formulated focusing on transit number 1, but in reality, the deepest transit is selected. It is equivalent to conducting the test for all the transits and selecting the most significant value. This results in a look-elsewhere effect, requiring a proper adjustment of the threshold p-value.

%% file: ap_template_bank.tex
In this appendix, we describe the construction of the template bank used for the search. The template bank was created once and applied to all targets.

\begin{figure}[h!]
    \centering
    \includegraphics[width=0.57\textwidth]{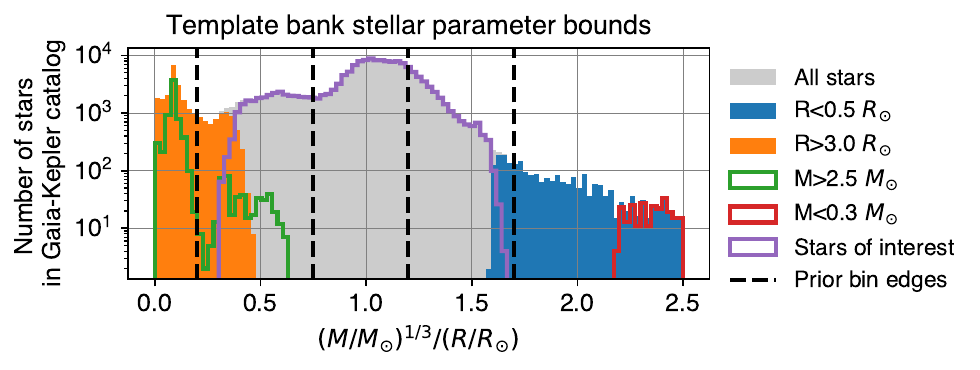}
    \caption{Distribution of the scaled density parameter of \textit{Kepler} stars from the \textit{Gaia-Kepler} cross-match catalog \citep[][]{bedell_gaia_kepler_fun}. 
    Objects excluded from the search, filtered based on stellar mass and radius are shown in different colors, with the purple contour indicating the stars ultimately selected. Vertical black dashed lines mark the density bin edges used for prior calculations.}
    \label{fig:star_selection_bank}
\end{figure}

\paragraph{Stellar parameters}
The stellar parameters of interest for the search were introduced in Section~\ref{sec:template_bank}. To obtain these parameters, we utilized the \textit{Gaia-Kepler} cross-match catalog \citep[][]{bedell_gaia_kepler_fun}, based on the \textit{Gaia} DR3 data \citep[][]{gaia_2016_mission, gaia_2022_dr3}.
Figure~\ref{fig:star_selection_bank} illustrates the distribution of the scaled stellar density parameter for the selected targets, alongside the limits imposed by the search criteria.

The geometry of a planet projection crossing the stellar disk can be described in terms of the ratio $a/R$ between the star-planet distance and the star radius. The transit duration is also defined by this ratio and by the orbital period, as shown in Equation~\ref{eq:transit_duration}. Using Kepler's third law of planetary motion, the ratio $a/R$ can be expressed as
\begin{equation}
    \frac{a}{R} = \left(\frac{G p^2}{4\pi^2}\right)^{\frac{1}{3}}
    \frac{M^{\frac{1}{3}}}{R},
    \label{eq:kepler_3rd_law}
\end{equation}
where $M$ is the stellar mass, $p$ is the orbital period, and $G$ is the gravitational constant. As evident from this equation, the quantity defining $a/R$ is the scaled stellar density parameter, $M^{1/3}/R$, meaning that the dependence on stellar mass and stellar radius is exercised only through this parameter. For main sequence stars, the range of scaled densities is relatively narrow, as shown in Figure~\ref{fig:star_selection_bank}, due to the mass-radius relation of these stars. 

We note, however, that limb darkening coefficients, which also influence the transit shape, depend on additional stellar properties, including temperature, metallicity, and surface gravity.

\paragraph{Planetary parameters}
A transit is defined by five orbital parameters: the orbital period, semi-major axis, eccentricity, inclination, and the argument of periastron. Additionally, the ratio of the planetary radius to the stellar radius is needed. However, this 6-dimensional parameterization is redundant since different parameter combinations can produce nearly identical transit shapes. This enables us to choose a small subset of 58 transit models covering most of the transits in the parameter space of interest.

We note the difference between the orbital period value used to generate the templates, and the orbital period defining transit periodicity in the search. In the periodicity search, the period is the actual measure of when the transits appear. In the template bank generation, the period is one of the latent degenerate parameters that influences the transit duration. The search is not informed about any physical parameters defining the templates. The detection statistic is calculated for all the search periods paired with all the templates.

\paragraph{Similarity metric}
The similarity of templates is quantified using the template overlap metric, the cosine angle between the two template vectors (introduced in Appendix~\ref{ap:templates_shape_losses}). The metric takes values from 0 for orthogonal templates to 1 for identical ones. It characterizes the loss in SNR due to template shape mismatch. This metric depends on the whitening filter, which is generally star-specific. For the template bank construction, we use an averaged whitening filter, approximating the expected template's similarity.  

\begin{figure}[h!]
    \centering
    \includegraphics[width=0.96\textwidth]{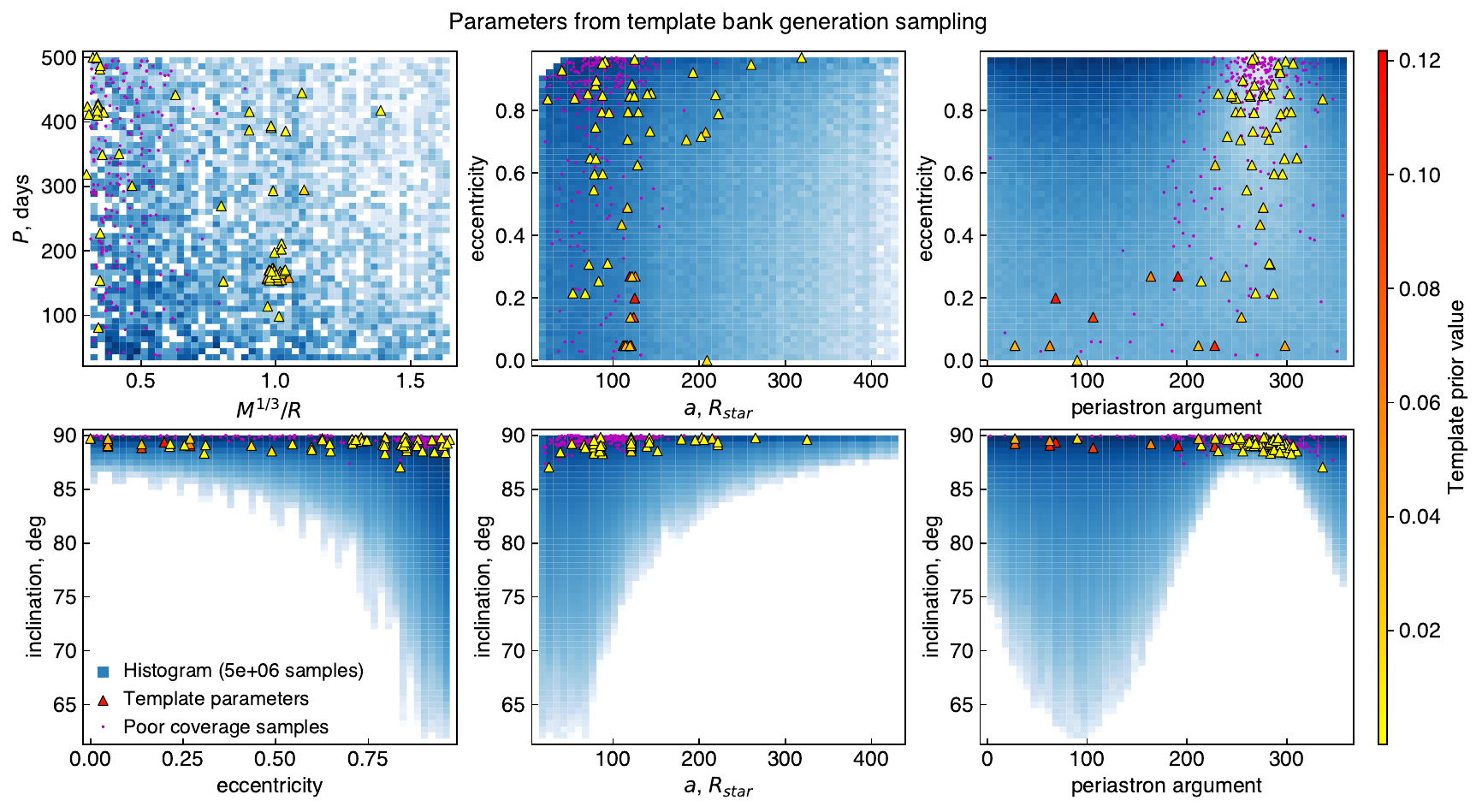}
    \caption{\textit{Blue shading}: histograms for the distributions of the stellar and planetary parameters used in the Monte-Carlo simulation for template bank and prior generation. \textit{Triangles}: parameters of the templates in the template bank, color-coded by the template prior values. \textit{Magenta dots}: parameters of transits which got poor coverage by the bank ($<0.97$ match).}
    \label{fig:template_bank_parameters}
\end{figure}

\paragraph{Template selection}
To construct a descriptive subset of templates, we performed a Monte-Carlo simulation as described in Appendix~\ref{ap:injection_parameters}. In each iteration, the simulation samples parameters, generates a transit shape, and calculates its cosine angle (closeness metric) with all existing templates in the bank.  If no template in the bank achieves a cosine value of at least 0.97, the new template is added to the bank. This procedure resulted in 58 templates, with their parameters shown in Figure~\ref{fig:template_bank_parameters}.

\begin{figure}[h!]
    \centering
    \includegraphics[width=0.47\textwidth]{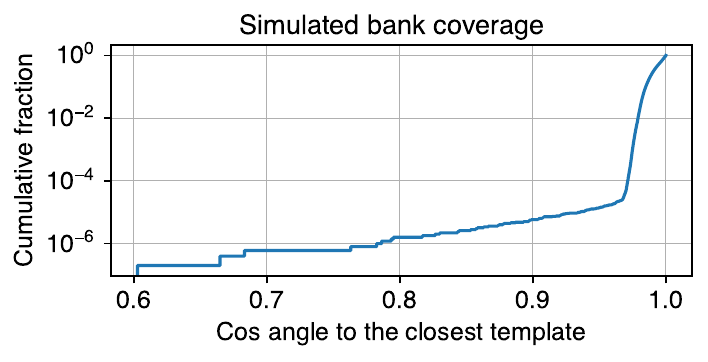}
    \caption{The distribution of cosine angles (closeness metric, see Appendix~\ref{ap:templates_shape_losses}) between generated transit models and their best-matching template from the bank, assuming a fiducial averaged whitening filter. It was obtained using a Monte Carlo simulation covering the parameter space of interest.}
    \label{fig:template_bank_cosines}
\end{figure}

To evaluate the template bank coverage, we re-ran for the fixed template bank. For each generated transit, the maximum cosine angle with the templates was recorded. The resulting distribution is shown in Figure~\ref{fig:template_bank_cosines}. As can be seen, 99\% of all simulations yielded a match with a cosine angle of at least 0.97 with one of the templates.

Figure~\ref{fig:template_bank_parameters} presents the histogram for parameter sets sampled during the simulation. It highlights cases with poor matches (cosine $<0.97$), which typically correspond to highly eccentric orbits near apoapsis. Such orbits lead to exceptionally long transit durations, which are not included in the template bank.

\paragraph{Template prior}
The same simulation also tracked how frequently each template was the best match for simulated transit models. This frequency reflects the likelihood that a real transiting planet would produce a signal resembling a given template, thereby triggering it in the search. These likelihoods are stored as template priors and used in the calculation of the marginalized detection statistic (Equation~\ref{eq:mmes_definition}). The template priors are represented via color-coding in Figure~\ref{fig:template_bank_parameters}.

%% file: ap_injection_parameters.tex
This appendix describes how the injection parameters for the injection-recovery search are generated. A similar procedure was used for the template bank generation.

\paragraph{Orbital parameters and shapes of priors}
Orbital periods are sampled from the specified range, which, in the case of the injection-recovery search, is one period chunk around the period of interest. Since the chunk size is small, no prior shape is applied.

The semi-major axis is calculated based on the period and the known stellar mass and radius using Equation~\ref{eq:kepler_3rd_law}.

The limb darkening coefficients are taken from the table\footnote{Table~~J/A+A/510/A21/table2.dat.gz~~at~~http://cdsarc.cds.unistra.fr} \cite{sing_2010_limb_darkening} for the corresponding stellar parameters.

The eccentricity prior is uniform between 0 and 0.97, with higher values excluded due to numerical issues. If the stellar radius provides a stricter eccentricity limitation, this limitation is adopted.

Priors for the periastron argument and orbital inclination are geometric, reflecting a uniform distribution of the orbital angular momentum vector directions on the sphere.

The first transit time is assigned a uniform prior.

\paragraph{Modeling the transit}
Single-transit models were simulated using the \textit{batman} package \citep{kreidberg_2015_batman}. Time sampling was set to match the \textit{Kepler} cadence. The planetary radius was varied by scaling the same transit model. Transit duration was estimated as the time encompassing 90\% of the model norm.

\paragraph{Sampling}
Each star received a total of $10^4$ sets of injection parameters, sampled using rejection sampling with pre-filtering. 
For every period, eccentricity, and periastron argument, the inclination range consistent with a planetary passing across the stellar disk was calculated, and samples were only drawn from this range. For each valid inclination, a flux model was generated, and its norm was verified to ensure a transit occurred.

An effective whitening filter was applied to roughly estimate the expected SNR of each transit as a function of planetary radius. From this, radii corresponding to the SNR range of interest were identified. Radii outside this range were excluded from the injection-recovery run as they result in detection scores beyond the operation range of the pipeline. Since the goal of the injection-recovery search is to find the foreground rate for a trigger, we run the search only for radii that can contribute to score distribution around the trigger location (see Figure~\ref{fig:background_injections_histogram} for illustration).

% Too small radii will not be detected, whereas the large radii will be masked as outliers, therefore modeling them is irrelevant.

We emphasize that it is impossible to make injections with known SNR because determining the true SNR requires knowing the true PSD of the data. The detection score is obtained empirically from running the pipeline on planets with controlled radii defining the relative depth of the transit. 

Each sample was assigned a weight corresponding to its measure in the parameter space. The sum of all weights was normalized to provide the total prior occurrence rate of planets. Accepted samples correspond to the rate of planets within the relevant ranges of periods and radii that are transiting. This rate was used to normalize the integral of the injection distribution shown in Figure~\ref{fig:background_injections_histogram}.

The total prior occurrence rate for planets within this period range was estimated using the table provided by \cite{zhu_dong_2021}. For parameters beyond the range covered in \cite{zhu_dong_2021}, occurrence rates were extrapolated by a constant.

\paragraph{Resulting parameter distributions}

\begin{figure}[h!]
    \centering
    \includegraphics[width=0.84\textwidth]{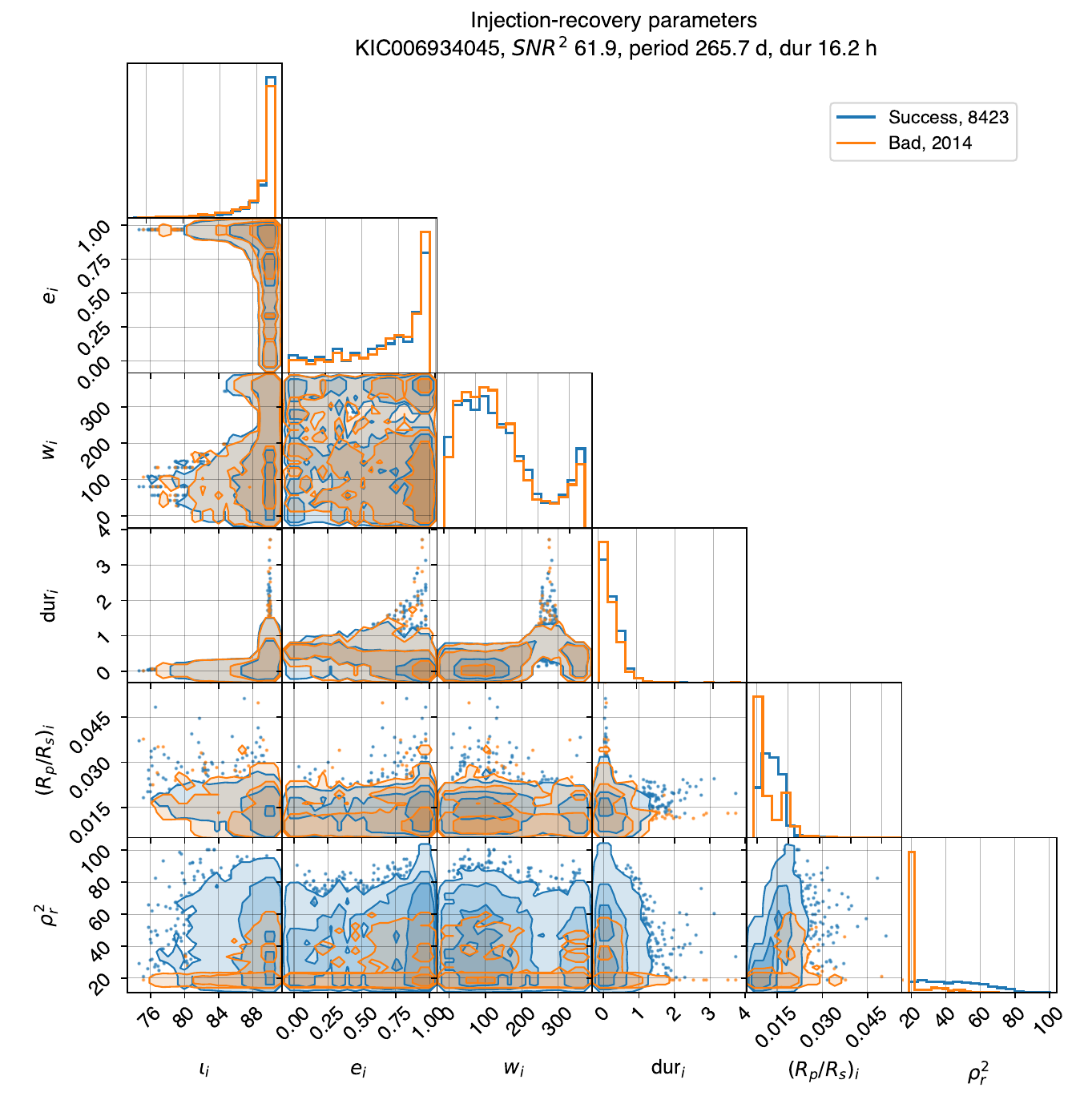}
    \caption{Injection parameters generated for one of the triggers using the scheme described in the text. Notation: $\iota_i$ is the injected inclination, $e_i$ is the injected eccentricity, $w_i$ is the injected periastron argument, $dur_i$ is the injected transit duration, $(R_p/R_s)_i$ is the injected planet-to-star radius ratio, $\rho^2_r$ is the recovered detection score.
    Blue distributions represent injections whose timing was successfully identified by the pipeline, while orange represents those missed. The 1D histograms are normalized to units of density. The total number of successful and failed injections is shown in the legend.}
    \label{fig:injection_parameters_corner}
\end{figure}

Figure~\ref{fig:injection_parameters_corner} illustrates an example of the resulting injected orbital parameter distributions. These distributions are shaped by the transit probability. For instance, a periastron angle of $w=90^\circ$ aligns the periastron with the line of sight, maximizing the probability of transit due to the planet's closest approach to the star, which allows a broad range of inclinations.

In contrast, at $w=270^\circ$, the planet is the farthest from the star during transit, reducing the probability of transit and limiting high eccentricities and inclinations. Furthermore, transits at $w=270^\circ$ are generally longer because the planet's low orbital speed is lower at apoastron. 

Figure~\ref{fig:injection_parameters_corner} distinguishes between successfully detected injections and those missed by the pipeline. By successful detection here we mean correct timing identification. Typically, detections occur above a sharp SNR threshold, dictated by the highest noise peak in the given data sample. Below this SNR, the pipeline will always trigger at this noise peak. 

As can be seen, the main difference between the successful and the missed detections is in the planetary radius which scales the SNR. For the other parameters, there is no significant bias of detection. 

%% file: ap_non_gaussianity.tex
This appendix provides a derivation for the non-Gaussianity correction (Equation~\ref{eq:ses_score_correction}).

We consider a model similar to Equation~\ref{eq:data_model}, but where the noise distribution is non-Gaussian. Specifically, the noise converges to a Gaussian distribution at low values but exhibits heavy tails. Our goal is to test whether the data $\mathbf{d}$ consists solely of noise or contains transits at specific times. The multiple-transit log-likelihood test statistic for this case would be
\begin{align}
    \rho^2_{\mathrm{MES}} = 2\log\frac{\mathcal{L}_{NG}\left(\mathbf{d}|\mathcal{H}_{1}\right)}{\mathcal{L}_{NG}\left(\mathbf{d}|\mathcal{H}_{0}\right)},
\end{align}
where $\mathcal{L}_{NG}$ denotes the likelihood for the non-Gaussian noise. Consult Appendix~\ref{ap:basic_math} if the definitions need to be clarified.

The cases $\mathcal{H}_{0}$ (pure noise) and $\mathcal{H}_{1}$ (transits present) differ only during the transits. Otherwise, they give the same values, therefore the contribution of all the data points outside the alleged transits cancels out:
\begin{align}
    \rho^2_{\mathrm{MES}} =2\log\frac{\mathcal{L}_{NG}\left(\mathbf{d}_{\mathrm{tr}}|\mathcal{H}_{1}\right)}{\mathcal{L}_{NG}\left(\mathbf{d}_{\mathrm{tr}}|\mathcal{H}_{0}\right)},
\end{align}
where $\mathbf{d}_{\mathrm{tr}}$ represents the data associated with the alleged transits.

Given the assumption of additive noise, the likelihood with transits depends on the residuals: $\mathcal{L}(\mathbf{d})=\mathcal{L}(\mathbf{d}-A\mathbf{h})$, where $A$ is the best-fit amplitude. It means that the test would automatically minimize the residuals, pushing them into the Gaussian regime of the noise distribution. Therefore, we can approximate the test statistic as
\begin{align}
    \rho^2_{\mathrm{MES}} \approx2\log\frac{\mathcal{L}_{G}\left(\mathbf{d}_{\mathrm{tr}}|\mathcal{H}_{1}\right)}{\mathcal{L}_{NG}\left(\mathbf{d}_{\mathrm{tr}}|\mathcal{H}_{0}\right)}.
\end{align}
In other words, the test evaluates whether the trigger is more likely to be induced by a planet or by noise non-Gaussianity.

Now consider matched-filtering the data with single-transit templates. Assuming that the resulting score $\rho_{\mathrm{SES}}$ is a sufficient statistic~\citep[][]{van_trees_book, castella_statistical_inference}, the test statistic will be a function of this score:
\begin{align}
    \rho^2_{\mathrm{MES}} =2\log\frac{\mathcal{L}_{G}\left(
    \{\rho_{\mathrm{SES}}\}_{\mathrm{tr}}|\mathcal{H}_{1}\right)}{\mathcal{L}_{NG}\left(
    \{\rho_{\mathrm{SES}}\}_{\mathrm{tr}}|\mathcal{H}_{0}\right)},
\end{align}
where ${\rho_{\mathrm{SES}}}_{\mathrm{tr}}$ denotes the collection of matched-filtering scores for the alleged transits.

Through an equivalence transformation, this expression can be rewritten as:
\begin{align}
    \rho^2_{\mathrm{MES}} 
    = 2\log\frac{\mathcal{L}_{G}\left(
    \{\rho_{\mathrm{SES}}\}_{\mathrm{tr}}|\mathcal{H}_{1}\right)}{\mathcal{L}_{G}\left(
    \{\rho_{\mathrm{SES}}\}_{\mathrm{tr}}|\mathcal{H}_{0}\right)}
    + 2\log\frac{\mathcal{L}_{G}\left(
    \{\rho_{\mathrm{SES}}\}_{\mathrm{tr}}|\mathcal{H}_{0}\right)}{\mathcal{L}_{NG}\left(
    \{\rho_{\mathrm{SES}}\}_{\mathrm{tr}}|\mathcal{H}_{0}\right)}
    =\rho^2_{\mathrm{MES,G}} + 2\log\frac{\mathcal{L}_{G}\left(
    \{\rho_{\mathrm{SES}}\}_{\mathrm{tr}}|\mathcal{H}_{0}\right)}{\mathcal{L}_{NG}\left(
    \{\rho_{\mathrm{SES}}\}_{\mathrm{tr}}|\mathcal{H}_{0}\right)},
\end{align}
where $\rho^2_{\mathrm{MES,G}}$ denotes the familiar statistic for the Gaussian model. The second term represents a correction accounting for the non-Gaussianity of the noise distribution. The total correction can be expressed as a sum of individual corrections of all transits indexed by $i$,

\begin{align}
    \rho^2_{\mathrm{MES}} 
    =\rho^2_{\mathrm{MES,G}} 
    + \sum_i 2\log\frac{\mathcal{L}_{G}\left(
    \rho_{\mathrm{SES, i}}|\mathcal{H}_{0}\right)}{\mathcal{L}_{NG}\left(
    \rho_{\mathrm{SES, i}}|\mathcal{H}_{0}\right)}.
    \label{eq:non_gaussianity_corrected_mes}
\end{align}

As can be seen, the corrections for individual transits can be computed as differences between the Gaussian and the non-Gaussian log-likelihoods for a given $\rho_{\mathrm{SES}}$. The single-transit corrections can be folded over the corresponding period and added to the pre-computed Gaussian MES score.

%% file: ap_loss_score_correction.tex
In this appendix, we present the results of the same simulation as in Section~\ref{sec:snr_recovery_fraction}, but performed without applying the non-Gaussianity correction. Since the simulations use Gaussian-noise light curves, the correction is unnecessary and only leads to a loss of SNR for the true signal.
This experiment has two goals:\\
1) Compare the SNR loss with and without the correction, helping to isolate the component of the loss attributable to the correction itself; \\
2) Investigate the remaining SNR loss caused by other factors, such as low PSD resolution, PSD measurement error, false negative rejection by veto, and template mismatch error.

\begin{figure}[h!]
    \centering
    \includegraphics[width=0.48\textwidth]{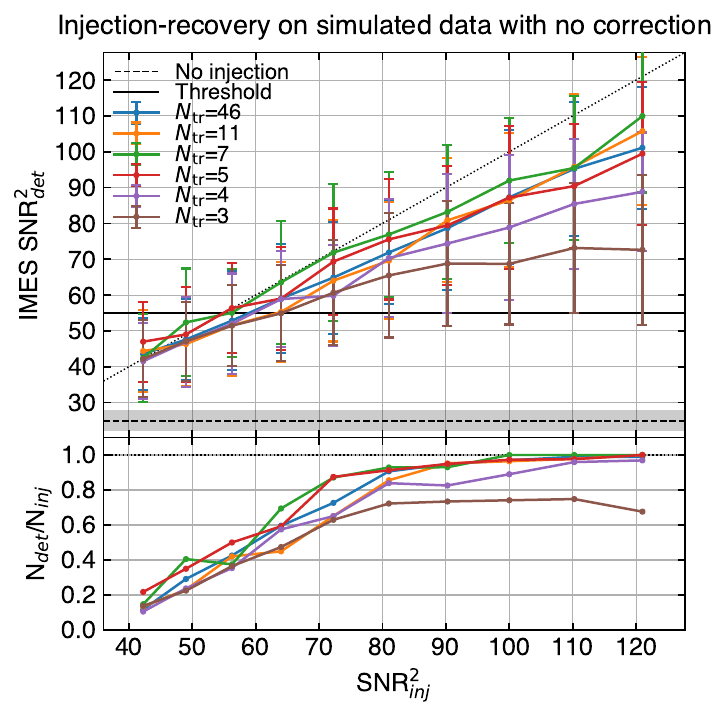}
    \caption{Same as Figure~\ref{fig:fraction_sr_recovered} but for the UMES score (i.e., without non-Gaussianity correction, template marginalization and peak integration).
    \textit{Top panel}: Recovered UMES as a function of the injected SNR$^2$ for various numbers of transits. The deviation from the identity shows the SNR loss not associated with the non-Gaussianity correction.
    \textit{Bottom panel}: Fraction of successfully detected injections crossing the fiducial detection threshold of 55.}
    \label{fig:fraction_snr_recovered_no_correction}
\end{figure}

Figure~\ref{fig:fraction_snr_recovered_no_correction} replicates the analysis of Figure~\ref{fig:fraction_sr_recovered}, but using the UMES score instead of the IMES score. As evident from the figure, the recovered UMES values are significantly closer to the injected SNR$^2$ compared to the IMES results in Figure~\ref{fig:fraction_sr_recovered}. This indicates that the non-Gaussianity correction is the dominant factor contributing to the observed SNR loss.

The remaining SNR loss is attributed to factors such as the SES outlier masking, detrending of low frequencies not resolved by the coarse PSD frequency grid, PSD measurement error and overfitting (Appendix~\ref{ap:psd_estimation}), false negatives of the SES vetting, and template mismatch error (Appendix~\ref{ap:templates_shape_losses}). Among these, after the SES outlier rejection for low transit number and large SNR, the dominant factor is associated with PSD issues, especially loss of power at low frequencies.

We emphasize that although the UMES score exhibits substantially less SNR loss, its detection efficiency in the real, non-Gaussian noise is lower than that of the IMES score. This is due to the necessity of raising the detection threshold to account for the noise tails arising when not correcting for non-Gaussianity (Section~\ref{sec:detection_efficiency_real_data}).

%% file: ap_confirmed_mes.tex
\begin{figure}[h!]
    \centering
    \includegraphics[width=0.94\textwidth]
    {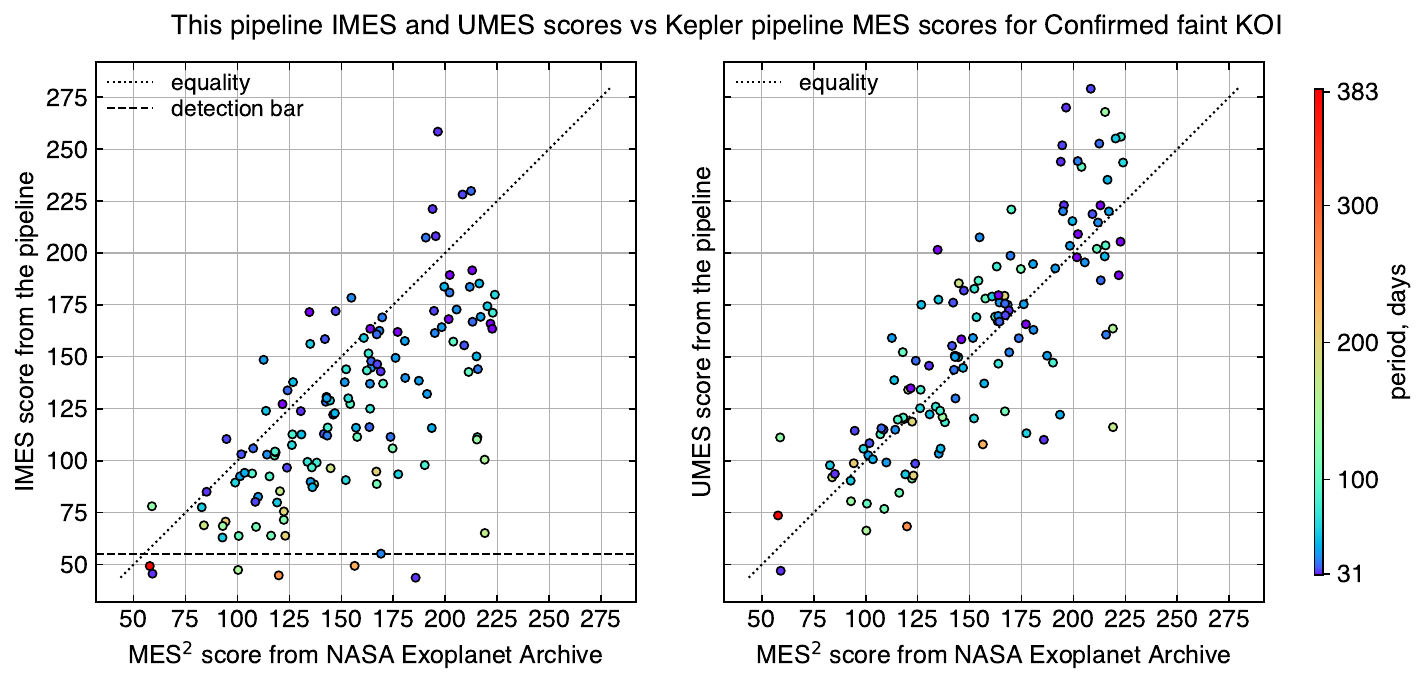}
    \caption{
    Comparison of IMES and UMES scores to \textit{Kepler} MES$^2$ for Confirmed faint KOIs. Points are color-coded by orbital period.
    \textit{Left panel}: IMES score vs \textit{Kepler} MES$^2$. The diagonal line represents the identity, and the dashed horizontal line indicates the empirical detection threshold discussed in Section~\ref{sec:reliability_of_pipeline}.
    \textit{Right panel}: UMES score vs \textit{Kepler} MES$^2$. The detection threshold is not provided because it is harder to define it for the UMES score due to the non-Gaussian background (see Section~\ref{sec:detection_efficiency_real_data}).
    }
    \label{fig:imes_umes_kmes_comparison}
\end{figure}

In this section, we compare the IMES and the UMES scores to the \textit{Kepler} MES score squared using Confirmed \textit{Kepler} planets. The resulting comparison can be seen in Figure~\ref{fig:imes_umes_kmes_comparison}.

For this analysis, we selected Confirmed KOIs and operated the pipeline in the vicinity of their orbital periods. Only faint KOIs with \textit{Kepler} MES$\leq 15$ were included, as they are in the range of interest for our pipeline.

As shown in the right panel of Figure~\ref{fig:imes_umes_kmes_comparison}, the UMES score approximately aligns with \textit{Kepler} MES$^2$, up to statistical noise. This similarity is expected, as both metrics essentially measure the same quantity but employ different methodologies.

In contrast, the left panel of Figure~\ref{fig:imes_umes_kmes_comparison} reveals that the IMES score exhibits SNR loss compared to \textit{Kepler} MES$^2$. This discrepancy is attributed to the non-Gaussianity correction applied to the IMES score. The SNR loss becomes more pronounced at longer orbital periods, where the correction is stricter.

However, the non-Gaussianity correction enables the definition of an approximate detection threshold (discussed in Section~\ref{sec:detection_efficiency_real_data}), which can be used to determine which Confirmed planets exceed it and are therefore detectable by the pipeline. A more detailed analysis of Confirmed planets detection by our pipeline, alongside a discussion of those that did not cross the detection threshold, is provided in Section~\ref{sec:confirmed_koi}.

%% file: ap_p_planet.tex
The $P_{\mathrm{planet}}$ score (Equation~\ref{eq:p_planet_score}), hereafter denoted as $P_p$ for brevity, is to answer the question "Given this trigger, what is the probability that it originates from a planet and not from background noise?" This appendix provides a detailed explanation of how this score is calculated and its interpretation.

\subsection{Conceptual definition}
This score represents the odds ratio for having a planet given that the pipeline returned a trigger with score \mbox{value $\rho^2$}. For this, it is necessary that a transiting planet exists around the target and that the pipeline successfully identifies transits' timing. If no planet is present, or if the transit times indicated by the trigger are incorrect, $\rho^2$ arises from the background noise distribution.

\begin{align}
\begin{split}
    &\text{Pr}\left(\text{planet}|\rho^2\right)
    =
    \frac{\text{Pr}\left(\rho^2|\text{planet}\right)
    \text{Pr}\left(\text{planet}\right)
    \text{Pr}\left(\text{success}\right)}
    {\text{Pr}\left(\rho^2\right)}
    \\&=\frac{\text{Pr}\left(\rho^2|\text{planet}\right)
    \text{Pr}\left(\text{planet}\right)
    \text{Pr}\left(\text{success}\right)}{\text{Pr}\left(\rho^2|\text{planet}\right)
    \text{Pr}\left(\text{planet}\right)
    \text{Pr}\left(\text{success}\right)
    +
    \text{Pr}\left(\rho^2|\text{no planet}\right)
    \left(1 - \text{Pr}\left(\text{planet}\right)
    + \left(1 - \text{Pr}\left(\text{success}\right)\right)
    \text{Pr}\left(\text{planet}\right)
    \right)}.
    \label{eq:p_planet_score_definition}
\end{split}
\end{align}
Here, 
\begin{itemize}[noitemsep, topsep=0pt, left=1pt, label=-]
    \item $\rho^2$ is the value of the detection statistic for the trigger, determined by any statistic of choice. It is associated with a certain orbital period and phase, with all the other parameters undefined.
    \item $\text{Pr}\left(\text{planet}\right)$ is the prior probability of a transiting planet around the target.
    \item $\text{Pr}\left(\text{success}\right)$ is the probability that the pipeline correctly identifies the orbital period and phase. Since this probability is nearly one in our regime (see Section~\ref{sec:trigger_example}), we assume it is unity and omit it in further calculations.
    \item $\text{Pr}\left(\rho^2|\text{planet}\right)$ is the probability density for obtaining $\rho^2$ from a planet.  
    It comes from the hypothetical distribution of the statistic values that the pipeline would return if we could take this star, plant different planets, generate light curves, and run the pipeline.
    \item $\text{Pr}\left(\rho^2|\text{no planet}\right)$ is the probability density for obtaining a statistic value $\rho^2$ in pure noise. That is, what would be the distribution of detection scores if we were able to generate more data for the same star without a planet. Here, we do not consider other origins of triggers, such as eclipsing binaries.
\end{itemize}

\subsection{Prior rate}
\label{ap_sec:p_planet_prior_rate}
Let $\eta$ represent the rate of transiting planets per star. Since the pipeline finds one maximal trigger per star, we need the probability that there is at least one planet in the light curve given the rate $\eta$:
\begin{align}
    \text{Pr}\left(\text{planet}\right) = 1 - \text{Poisson}_{\eta}(0).
\end{align}

Not all the planets can result in a given statistic value $\rho^2$: some of them are too faint or too strong.  The value of the detected statistic is a random variable, but its distribution is still localized: planets with parameters incompatible with $\rho^2$ will not contribute to its probability density.

Denote by $\boldsymbol{\Theta}$ all the planet parameters except for the orbital period and phase. Define $\boldsymbol{\Theta}_1$ as the subset of parameters that potentially can result in a statistic value $\rho^2$, and $\boldsymbol{\Theta}_2$ as those that can not. Then, $\text{Pr}\left(\rho^2
|\boldsymbol{\Theta}_2\right)=0$,  therefore
\begin{align}
    \text{Pr}\left(\rho^2
    |\boldsymbol{\Theta}\right)
    \text{Pr}\left(\boldsymbol{\Theta}\right)
    =\text{Pr}\left(\rho^2
    |\boldsymbol{\Theta}_1\right)
    \text{Pr}\left(\boldsymbol{\Theta}_1\right).
\end{align}
This implies that injection-recovery tests (Section~\ref{sec:injections}) may be omitted for irrelevant parameters $\boldsymbol{\Theta}_1$, reducing the computational effort.
It is enough to select the relevant $\boldsymbol{\Theta}_1$ and calculate the rate $\eta_{\boldsymbol{\Theta}_1}$ for it.

However, the denominator in Equation~\ref{eq:p_planet_score_definition} is related to the probability of encountering any planet and requires the full occurrence rate $\eta$, including $\boldsymbol{\Theta}_2$. The exact value of $\eta$ may be uncertain, but the denominator's dependence on it is weak, due to the small value of transit probability.

\subsection{Background rate}
\label{ap_sec:p_planet_background_rate}
The background distribution $\text{Pr}\left(\rho^2|\text{no planet}\right)$ is a hypothetical distribution of the pipeline scores from the ensemble of this star's possible light curves if it did not host a planet. 

We call this unknown hypothetical distribution the true background and denote its value for 
$\rho^2$ by \mbox{$f^t=\text{Pr}\left(\rho^2|\text{no planet}\right)$}. Aiming to estimate it from the available light curve, we perform a scrambled search, with and without masking the trigger (Section~\ref{sec:scrambling}). We denote the probability density values resulting from these two searches as $f^{sm}$ and $f^{s}$. 

If the star contains no planet, $f^{s}$ is an unbiased estimator of $f^{t}$, meaning that it would converge to $f^{t}$ when averaged over a hypothetical ensemble of possible light curves. The influence of the finite data effect arising from scrambling the same light curve will be discussed below. $f^{sm}$ is biased, as masking the maximal trigger shifts the scrambled scores distribution toward lower values.

If a planet is present, $f^{sm}$ is unbiased because we remove the planet and only look at the distribution of the background signal. 
$f^{s}$ will be a biased estimate of the background because the planet will contaminate the noise distribution, shifting it toward higher values.

We summarize these statements as
\begin{align}
    &\text{E}\left[f^{s}|\text{no planet}\right]=f^{t},\\
    &\text{E}\left[f^{sm}|\text{planet}\right]=f^{t},\\
    &\text{E}\left[f^{sm}|\text{no planet}\right]=f^{t}-\mu_{n},\\
    &\text{E}\left[f^{s}|\text{planet}\right]=f^{t}+\mu_{p},
\end{align}
where $\mu_{p}$ is the bias of the background estimate from an unmasked planet, and $\mu_{n}$ is the bias from masking the maximal noise trigger before scrambling.

It is unknown whether the planet is present in the data, therefore we combine $f^{s}$ and $f^{sm}$ into a composite estimator $\hat{f}$ of the form
\begin{align}
    \hat{f}=\alpha f^{sm}+\left(1-\alpha\right)f^{s},
\end{align}
aiming at selecting the optimal parameter $\alpha$ minimizing the bias. It is not known whether a planet is present, but by definition, given a trigger $\rho^2$, the planet exists with probability $P_{p}$. Therefore, we can calculate the total expectancy of the estimator given that the trigger $\rho^2$,
\begin{align}
    \text{E}\left[\hat{f}|\rho^{2}\right]	=P_{p}\text{E}\left[\hat{f}|\text{planet}\right]+\left(1-P_{p}\right)\text{E}\left[\hat{f}|\text{no planet}\right]
	=f^{t}+\left(1-\alpha\right)P_{p}\mu_{p}-\alpha\left(1-P_{p}\right)\mu_{m},
\end{align}
and require it to be unbiased:
\begin{align}
    \text{E}\left[\hat{f}|\rho^{2}\right] \overset{!}{=} f^t.
    \label{eq:bg_estimator_requirement}
\end{align}
Since we consider triggers close to the detection threshold, we choose an approximation $\mu_{p}=\mu_{m}$. Solving Equation~\ref{eq:bg_estimator_requirement} yields
\begin{align}
    \alpha=\frac{P_{p}\mu_{p}}{P_{p}\mu_{p}+\left(1-P_{p}\right)\mu_{m}}.
\end{align}
Substituting it to Equation~\ref{eq:p_planet_score_definition}, we get
\begin{align}
    P_{p}=\frac{\pi_{p}f^{p}}{\pi_{p}f^{p}+\left(1-\pi_{p}\right)\left(P_{p}f^{sm}+\left(1-P_{p}\right)f^{s}\right)},
\end{align}
which now needs to be solved for $P_p$. We introduce notations
\begin{align}
    g^{p}=\pi_{p}f^{p},\,\,\,g^{s}=\left(1-\pi_{p}\right)f^{s},\,\,\,g^{sm}=\left(1-\pi_{p}\right)f^{sm},
\end{align}
and write the solution as
\begin{align}
    P_{p}=\frac{g^{p}}{\frac{1}{2}\left(g^{p}+g^{s}\right)+\frac{1}{2}\sqrt{\left(g^{p}+g^{s}\right)^{2}-4g^{p}\left(g^{s}-g^{sm}\right)}}.
    \label{eq:p_planet_formula}
\end{align}
This is the formula that we use in the pipeline to calculate $P_{\mathrm{planet}}$.

We note that this formula can be easily generalized to the case when $\mu_{p}\neq\mu_{m}$, and will depend on their ratio.

\subsection{Extrapolating the background distribution}
\label{ap_sec:bg_extrapolation}
If the score $\rho^2$ is sufficiently high, obtaining it from the noise distribution is a very rare event. However, planets are also rare (See discussion of rates in Appendix~\ref{ap:rates_look_elsewhere}). Comparing small rates requires knowing the background distribution value for the rare tail events. It is hard to get the tail from the scrambled search because it requires running the search hundreds of thousands of times and sometimes is limited by entropy (Appendix~\ref{ap:rates_look_elsewhere}). 

If the noise was purely Gaussian, the trigger score would distribute like a maximum of a number of independent $\chi^2(1)$ variables, defined by the search volume per target. Then, its tail would be asymptotically proportional to a $\chi^2(1)$ distribution.

Fortunately, after applying the non-Gaussianity correction~(Section~\ref{sec:score_correction}), the real score distribution tail resembles a scaled $\chi^2(1)$ distribution, as can be observed in Figures~\ref{fig:background_injections_histogram},~\ref{fig:search_triggers_and_scrambled},~\ref{fig:detection_efficiency_scores}. Therefore, it is possible to use extrapolation and estimate what would be the background value if we had more data had were able to run the scrambled search for longer. 
We adopt the following functional form of the distribution tail:
\begin{align}
    f\left(\rho^2\right) = A\sqrt{\frac{a}{\pi \rho^2}}
    e^{-a\rho^2},
\end{align}
where $a$ is the shape parameter allowing to fit the real slope of the distribution, and $A$ is the tail normalization parameter. 

Assume that in the scrambled search we obtained $N$ scores $\left\{\rho^2_1...\rho^2_N\right\}$. From them, we selected only the tail scores that exceed $\rho^2_0$. Assume we got $n$ such scores, $\left\{\rho^2_1...\rho^2_n\right\}$. If the initial $N$ values distribution is normalized to unity, the distribution limited by $\rho^2_0$ will be normalized to $n/N$,
\begin{align}
    \int_{\rho^2_0}^{\infty} d\rho^2 f\left(\rho^2\right) = n/N.
\end{align}
From this normalization,
\begin{align}
    A = \frac{n}{N} \frac{1}{\mathrm{erfc}\left(\sqrt{a\rho_0^2}\right)},
\end{align}
where $\mathrm{erfc}$ is the complementary error function.

We perform a maximum-likelihood estimation for the parameter $a$ using the scores $\left\{\rho^2_1...\rho^2_n\right\}$ obtained from the scrambled search. The log-likelihood for these scores is
\begin{align}
    \mathcal{L}\left(\left\{\rho^2_1...\rho^2_n\right\}\right)
    =\frac{n}{N} \frac{1}{\mathrm{erfc}\left(\sqrt{a\rho_0^2}\right)}
    \left(\frac{a}{\pi}\right)^n
    \frac{1}{\sqrt{\prod_{i=1}^n \rho^2_i}}
    \exp{\left(- a \sum_{i=1}^n \rho^2_i\right)}.
\end{align}
Its extremum
\begin{align}
    \frac{\partial}{\partial a}\mathcal{L}\left(\left\{\rho^2_1...\rho^2_n\right\}\right)
    = 0
\end{align}
yields the equation
\begin{align}
    1+\frac{2\sqrt{a\rho_0^2}}
    {\sqrt{\pi}\mathrm{erfc}\left(\sqrt{a\rho_0^2}\right)}
    e^{-a\rho_0^2}
    -2a\frac{1}{n}\sum_{i=1}^{n}\rho_i^2=0.
\end{align}
We solve it numerically to get the value of $a$. The resulting function is used to extrapolate the background distribution and get the background rate for the trigger. An example is shown in Figure~\ref{fig:background_injections_histogram}.

\subsection{Errors of $P_{\mathrm{planet}}$ score}
\label{ap_sec:p_planet_errors}
The background and foreground rates that are used to compute the $P_{\mathrm{p}}$ score~(Equation~\ref{eq:p_planet_score}) are the estimates of the true rates, subject to errors. Eventually, the goal is to assess the expectancy of the $P_{\mathrm{p}}$ score given these estimates. For instance, if $f^\mathrm{true}$ is the true background rate and $f^\mathrm{meas}$ is the measured rate from the scrambled search, we compute
\begin{align}
    P_{\mathrm{p}}|f^{\mathrm{meas}} = \int df^\mathrm{true} 
    P_{\mathrm{p}}\left(f^\mathrm{true}\right)
    \text{Pr}\left(f^\mathrm{true}
    |f^{\mathrm{meas}}\right),
\end{align}
where $\text{Pr}\left(f^\mathrm{true}
    |f^{\mathrm{meas}}\right)$
is the distribution of the true rate given the measured value. It will be affected by the uncertainty sources of the rates, some of which are described below. Their quantitative effects on candidate planet scores will be discussed in future work~\citep[][]{our_paper_2}.

\paragraph{Finite data effect}
The scrambled search~(Section~\ref{sec:scrambling}) can only be done on one light curve, and scrambling this data repeatedly does not reproduce the distribution that hypothetical new data would provide.

For example, the maximal value in a given light curve remains unchanged in all scrambles. If we obtained new data, its maximal values could be larger (or smaller), and it would populate differently the tail of the distribution. Thus, the estimate of the background density is expected to have an error with respect to the true distribution. 

This error is more pronounced with fewer transits when the smaller entropy contained in the data gets exhausted in the scrambled search. In addition, the variance of the maximal SES score per dataset has a stronger impact on the scrambled scores distribution tail.

\paragraph{Extrapolation error}
The extrapolation of the scrambled scores distribution needed to evaluate the background rate at the location of the trigger~(Appendix~\ref{ap_sec:bg_extrapolation}) also introduces errors. Furthermore, when the finite data effect is dominant, the extrapolation is applied to a biased distribution, giving a wrong result even if it is very precise.

\paragraph{Planet occurrence prior uncertainty}
Errors in the numerator of Equation~\ref{eq:p_planet_score} mainly arise from the uncertainty in the prior occurrence rate of planets. Although this rate also affects the denominator, its impact on the numerator is more significant due to direct proportionality. 

After all the planetary candidates have their $P_{\mathrm{planet}}$ score calculated, it will be possible to provide a self-consistent estimation of the rate which could reduce this error.

\section{Limiting orbital period range for $P_{\mathrm{planet}}$ calculation}
\label{ap_sec:p_planet_period_range}
The definition of the $P_{\mathrm{planet}}$ score (Equation~\ref{eq:p_planet_score}) does not exactly specify the definition of "this trigger", which can be caused either by a planet or by the noise. It can be defined as a "trigger having score $\rho^2_{\mathrm{trigger}}$", meaning that the parameters of the event were maximized or marginalized over. Alternatively, it can refer to a "trigger having score $\rho^2_{\mathrm{trigger}}$ and period $p$", or "trigger having score $\rho^2_{\mathrm{trigger}}$ and parameters $\boldsymbol{\Theta}$".

Trigger parameters are treated differently in the search and have different scientific value:
\begin{itemize}[noitemsep, topsep=0pt, left=1pt, label=-]
    \item Transit duration and shape are marginalized over in the search. After the peak is found, we approximately recover those parameters, but they are not well-measurable.
    \item The first transit phase undergoes maximization and integration around the peak. Its range is bounded, and its prior is flat. The pipeline performance does not depend on it, and it is not physically interesting.
    \item The depth of the transit is not a parameter of the search. It is determined in the end from the trigger score.
    \item The orbital period of the planet is a crucial parameter both technically and physically. It has a physical prior used to find the best trigger; pipeline performance depends on it; it is important for the occurrence rate calculations.
\end{itemize}

In the $P_\mathrm{planet}$ score calculation, we choose to focus on a narrow period range around the trigger period. Below, we justify this approach and show that it does not lead to an additional look-elsewhere effect.

\paragraph{Injections}
Consider a small range of periods $p_{\mathrm{trigger}}\pm \Delta p$. Assuming we are in the regime where the pipeline can recover the injected timing correctly, the detected period will be in the same range as the injected period:
\begin{align}
    \text{Pr}\left(p_{\mathrm{max}}
    =p_{\mathrm{trigger}}\pm \Delta p
    | p_{\mathrm{inj}}
    =p_{\mathrm{trigger}}\pm \Delta p
    \right) = 1.
\end{align}
Therefore limiting the search to this range $p_{\mathrm{trigger}}\pm \Delta p$ does not alter the score distribution:
\begin{align}
    \text{Pr}\left(\rho^2_{\mathrm{max}}
    =\rho^2_{\mathrm{trigger}}
    |p_{\mathrm{max}}
    =p_{\mathrm{trigger}}\pm \Delta p
    \right)
    =
    \text{Pr}\left(\rho^2_{\mathrm{max}}
    =\rho^2_{\mathrm{trigger}}\right)
    \label{eq:period_limitation_doesnt_matter}
\end{align}
The rate of planets in this range is defined by the prior occurrence of transiting planets and the size of the range. 

\paragraph{Background}
Assume that the periodicity search was split into a grid of chunks of size $2\Delta p$ such that the look-elsewhere effect is the same in all the chunks.
If the detection score is prior-weighted, then the probability that the maximal background score is in this period range will also follow the prior, so that
\begin{align}
    \text{Pr}\left(p_{\mathrm{max}}
    =p_{\mathrm{trigger}}\pm \Delta p
    | \text{no planet}
    \right) 
    =
    \text{Pr}\left(p_{\mathrm{max}}
    =p_{\mathrm{trigger}}\pm \Delta p
    | \text{planet}
    \right).
    \label{eq:equal_probability_of_maximal_period}
\end{align}
Since the distributions of $\rho^2$ in all chunks are equivalent, then Equation~\ref{eq:period_limitation_doesnt_matter} also holds for the background distribution.

From equations \ref{eq:period_limitation_doesnt_matter} and \ref{eq:equal_probability_of_maximal_period}, we conclude that the ratio between the foreground rate and the background rate is the same for the full search and for the search restricted to one period chunk. Therefore, in the $P_{\mathrm{planet}}$ score, we can use the distributions obtained for a search limited to a narrow period range around the trigger of interest.

%% file: ap_fp_rates_estimation.tex
This appendix provides theoretical estimations of the expected background rate of the search, the expected number of detectable faint planets, and the score values achievable in a scrambled search.

\paragraph{Expected rate of noise triggers}
In the absence of a planetary signal, the distribution of the IMES scores $f\left(\rho^2\right)$ approximately follows the distribution of the maximum of $N$ Gaussian random variables, where $N$ is the effective number of independent search options for one target. This $N$ can be estimated empirically using the tail of the scrambled search distribution. We verified that a scrambled search on simulated Gaussian data gives similar results to real light curves (for instance, in Figure~\ref{fig:background_injections_histogram}). The distribution of the square of a maximum of $N$ Gaussian random variables for large values of argument asymptotically behaves as
\begin{align}
    f\left(\rho^2\right)
    \underset{\rho^2>2\log N }{\sim }
    \frac{N}{\sqrt{8\pi \rho^2}} \exp\left(-\frac{\rho^2}{2}\right).
\end{align}
From the measured values of $f\left(\rho^2\right)$ for large $\rho^2$, the estimated $N$ for one period grid chunk is $\sim5\cdot 10^5$. 

Using the Gaussian inverse survival function (ISF), we estimated $\rho^2$ expected once per search on one chunk to be $\sim 23$. 
This value is consistent with the analytical estimation considering the overlap of periodic transit models, counting two options having SNR$^2$ of overlap 0.5 as independent.

When extended to all the period chunks, the effective number of options in the search is $N\sim7\cdot 10^8$, predicting a typical $\rho^2$ value expected in a search on one target to be $\sim 37$.

With the $\sim1.5\cdot 10^5$ stars over which the search is run, a noise $\rho^2$ value expected to occur once in a search over all stars is $\sim 60$. The value of $~51$ would occur 100 times per search.

\paragraph{Expected number of planets}
We queried the \textit{NASA Exoplanet Archive}~\citep[][]{akeson_2013_nasa_archive} for Confirmed planets with periods from 50 to 500 days. For MES in ranges [10-12.5] and [12.5-15], there are 35 and 34 planets, respectively. We can assume that in the range [7.5-10], there should also exist a similar number of planets transiting Kepler stars. According to~\citep[][]{zhu_dong_2021}, the occurrence rate does not decline with decreasing planetary radius. 

From this, it follows that per star per $\rho^2$ bin, one expects $\sim 5\cdot 10^{-6}$ planets. For a candidate to reach 50\% probability to be real (or $P_{\mathrm{planet}}$ score 0.5), this rate should be at least the background rate corresponding to the candidate's score. Substituting this rate to the ISF for one-target search gives the approximate expected IMES of $P_{\mathrm{planet}}=0.5$ $\rho^2\approx60$. The assessments of the expected planetary occurrence here are very approximate, and in reality, this number is closer to 55.

\paragraph{Rate achievable with a scrambled search}
A typical light curve duration of $\sim$1400 days and a transit duration of $\sim$0.5 days imply $\sim 3000$ options for selecting an independent transit time. For a scrambled search of $n$-transit events, the best-case scenario involves exploring all independent combinations of transit options.

For a 3-transit event, the ISF for the number of all possible transit combinations provides $\rho^2\sim 40$. This means that the scrambling cannot explore the $\rho^2$ range beyond 40 because there is not enough entropy in the data. For 4 transits, it is 53, for 5 transits, it is 65.

However, not all the combinations of transits are fully independent. For example, two 5-transit events sharing 4 common transits exhibit significant correlation. Correlations further reduce effective independence, particularly when the scrambled search approaches the entropy limit of the data. 

Running $5\cdot 10^4$ scrambled iterations achieves a maximal $\rho^2\sim 43$, as estimated from the ISF. For a 3-transit event, it surpasses the entropy limit, introducing additional challenges for detecting very long periods.

%% file: ap_basic_math.tex
This appendix provides a brief overview of the definitions and derivations of the maximum-likelihood detection statistics.

The likelihoods associated with the data model described in Equation~\ref{eq:data_model}, which represents the data as a Gaussian noise vector with or without a potential signal, are:
\begin{align}
    \mathcal{L}\left(\mathbf{d}|\mathcal{H}_0\right) &= \frac{1}{\sqrt{(2\pi)^N \det C}}
    \exp \left(-\frac{1}{2} 
    \mathbf{d}^T C^{-1} \mathbf{d}\right),
    \\
    \mathcal{L}\left(\mathbf{d}|\mathcal{H}_1\right) &= \frac{1}{\sqrt{(2\pi)^N \det C}}
    \exp \left(-\frac{1}{2} 
    (\mathbf{d}-A\mathbf{h})^T C^{-1} (\mathbf{d}-A\mathbf{h})\right),
    \label{eq:likelihoods_gaussian}
\end{align}
where $\mathcal{H}_0$ denotes the null hypothesis (no signal), 
$\mathcal{H}_1$ denotes the alternative hypothesis (signal present),  $\mathbf{d}$ is the data vector,  $C$ is the covariance matrix, $\mathbf{h}$ is the signal template, and $A$ is the amplitude of the signal.

The log-likelihood ratio test statistic is then given by:
\begin{align}
    \rho^2 = 2\frac{\mathcal{L}\left(\mathbf{d}|\mathcal{H}_1\right)}{\mathcal{L}\left(\mathbf{d}|\mathcal{H}_0\right)}
    =2A\mathbf{d}^TC^{-1}\mathbf{h}-A^2\mathbf{h}^TC^{-1}\mathbf{h}.
    \label{eq:stat_with_free_term}
\end{align}
The second term can be omitted as it is a constant, and the amplitude $A$ can be factored out as a scaling coefficient. This leads to the classical matched-filtering detection statistic 
\begin{align}
    \rho=\mathbf{d}^TC^{-1}\mathbf{h}.
    \label{eq:reminder_stat_raw}
\end{align}
The independence of the statistic on the amplitude is related to it being the Uniformly Most Powerful test~\citep[][]{castella_statistical_inference}.

\paragraph{Amplitude estimator}
The signal amplitude $A$ can be measured using the maximum likelihood estimator
\begin{align}
    \left<A\right>
    =\frac{\mathbf{d}^TC^{-1}\mathbf{h}}{\mathbf{h}^TC^{-1}\mathbf{h}}.
    \label{eq:amplitude_estimator}
\end{align}
It can be verified that it is an unbiased estimator, as
$\mathrm{E}\left[\left<A\right>|\mathcal{H}_1\right] = A$. Substituting this estimator to Equation \ref{eq:stat_with_free_term}, we obtain
\begin{align}
    \rho^2 = \left(\frac{\mathbf{d}^TC^{-1}\mathbf{h}}{\sqrt{\mathbf{h}^TC^{-1}\mathbf{h}}}\right)^2
    \label{eq:stat_snr_2},
\end{align}
which is the matched-filtering detection statistic in units of SNR squared, justifying the notation $\rho^2$. Under $\mathcal{H}_0$ its expected distribution is $\rho^2|\mathcal{H}_0\sim\chi^2(1)$.

\paragraph{Signal-to-noise ratio (SNR)}

The SNR of the signal is defined as
\begin{align}
    \text{SNR}^2 = \left(\frac{
    \text{E}\left[\rho|\mathcal{H}_1\right]
    -\text{E}\left[\rho|\mathcal{H}_0\right]}{
    \text{V}\left[\rho|\mathcal{H}_0\right]}\right)^2
    \label{eq:snr_definition}
\end{align}
Substituting the detection statistic (Equation~\ref{eq:reminder_stat_raw}), the SNR in terms of the signal parameters is:
\begin{align}
    \text{SNR}^2 = A^2\mathbf{h}^TC^{-1}\mathbf{h}
\end{align}

%% file: ap_gaussianization.tex
This appendix derives an alternative way of non-Gaussianity control that was not used in the pipeline but may be useful in other searches. 

A robust way to treat the non-Gaussianity is to apply a transformation to the SES ensuring that it follows a strictly Gaussian distribution. 
This transformation erases the information about the actual values of the SES, preserving only their relative ranking and the temporal ordering. 
It replaces the original values with a Gaussian sequence based on the data length. That means, for example, that the maximal SES would be the value expected to occur once per data length. 

This limitation is destructive for the deep transits that would be limited so as not to exceed this maximum. On the contrary, low SES buried in noise remain almost unaffected. At the same time, the non-Gaussian background is eliminated, which makes the true periodic signal easily detectable. It makes this method efficient for short-period low-SNR planets which would almost not lose SNR and get well-detectable on the strictly Gaussian background.

\paragraph{Derivation}
Let $\mathcal{L}(\rho)$ denote the probability density of SES values $\rho$. A transformation $\tilde{\rho}=f(\rho)$ with some function~$f$ modifies the distribution as
\begin{equation}
    \mathcal{L}\left(f\left(\rho\right)\right)
    =\frac{\mathcal{L}\left(\rho\right)}
    {f'\left(\rho\right)}.
    \label{eq:distr_coordinate_transform}
\end{equation}
The function $f$ can be selected in such a way that the distribution $\mathcal{L}\left(\tilde{\rho}\right)$ is Gaussian. This condition results in a differential equation
\begin{equation}
    \frac{\mathcal{L}\left(\rho\right)}
    {\tilde{\rho}'\left(\rho\right)}
    =\frac{1}{\sqrt{2\pi}}e^{-\frac{1}{2}\tilde{\rho}^{2}}.
\end{equation}
Integrating with bounds of minimal scores $\rho_0$, $\tilde{\rho}_0$ and the scores of consideration, $\rho$, $\tilde{\rho}$ gives
\begin{equation}
    \int_{\tilde{\rho}_{0}}^{\tilde{\rho}}d\tilde{\rho}\frac{1}{\sqrt{2\pi}}e^{-\frac{1}{2}\tilde{\rho}^{2}}=\int_{\rho_{0}}^{\rho}d\rho\,\mathcal{L}\left(\rho\right)
\end{equation}
This equation connects the corrected score $\tilde{\rho}$ to the CDF of the original score $\rho$. If the distribution is two-sided, including both negative and positive parts, then $\rho_{0}=\tilde{\rho}_{0}=-\infty$, so that
\begin{equation}
    \frac{1}{2}\left(1+\text{erf}\left(\frac{\tilde{\rho}}{\sqrt{2}}\right)\right)=
    \int_{-\infty}^{\rho}d\rho\,\mathcal{L}\left(\rho\right),
\end{equation}
and the corrected score is given by
\begin{equation}
    \tilde{\rho}\left(\rho\right)=\sqrt{2}\;\text{erf}^{-1}\left(2\int_{-\infty}^{\rho}d\rho\,\mathcal{L}\left(\rho\right)-1\right).
    \label{eq:score_correction_two_sided}
\end{equation}
If the distribution includes only non-negative values, then $\rho_{0}=\tilde{\rho}_{0}=0$, and
\begin{equation}
    \text{erf}\left(\frac{\tilde{\rho}}{\sqrt{2}}\right)=\int_{0}^{\rho} d\rho\,\mathcal{L}\left(\rho\right),
\end{equation}
so the corrected score yields
\begin{equation}
    \tilde{\rho}\left(\rho\right)=\sqrt{2}\;\text{erf}^{-1}\left(\int_{0}^{\rho}d\rho\,\mathcal{L}\left(\rho\right)\right).
\end{equation}
The CDF of the observed score distribution is not known, but it can be estimated using the rank of the score~\citep[][]{venumadhav_2019_pipeline}:
\begin{equation}
    \int d\rho\,\mathcal{L}\left(\rho\right)=\text{Rank}\left(\rho\right).
    \label{rank_score_formula}
\end{equation}

This Gaussianization erases information about the absolute depth of the SES, only keeping their relative values and timing order. Due to the upper bound set by the data length, this approach is stricter than the non-Gaussianity correction (Section~\ref{sec:score_correction}), making it not suitable for detecting long-period planets. However, for short-period faint planets, this approach can effectively treat the non-Gaussianity, while preserving most of the signal SNR.

Gaussianized score can be interpreted as a lower bound on the likelihood. SES populating the distribution tail are pushed towards smaller values by the transformation, so its derivative $f'\leq 1$, implying $\mathcal{L}(f(\rho))\leq\mathcal{L}(\rho)$. The multiple-transit likelihood $\mathcal{L}\left(\rho_{\mathrm{MES}}\right)$ can be represented as a product of the single-transit likelihoods, assuming they are independent. Then, 
\begin{align}
    \mathcal{L}\left(\rho_{\mathrm{MES}}\right) 
    = \prod_i \mathcal{L}\left(\rho_{\mathrm{SES,}i} \right)
    \leq \prod_i \mathcal{L}\left(\tilde{\rho}_{\mathrm{SES,}i} \right).
\end{align}
Therefore, the MES obtained from the Gaussianized scores corresponds to the lower bound of the true likelihood.